\documentclass[11pt,a4paper,onecolumn,twosided,rotate]{IEEEtran} 
\usepackage{amsmath,amsthm,amssymb,amsfonts,upref,cite,epsf,color}
\usepackage{pst-all,pstricks,pst-node,pst-text,pst-3d}
\usepackage{hyperref}
\usepackage{pstricks-add}
\usepackage{multido}
\usepackage{bm}
\usepackage{graphicx,epsfig,rotating} 
\usepackage{psfrag,texdraw,bm}
\usepackage{nomencl}
\usepackage{amsmath}
\usepackage{amssymb}
\usepackage{subfigure}
\usepackage{xspace}
\usepackage{dsfont}
\usepackage{pifont}
\usepackage{listings,color}   
\usepackage[T1]{fontenc}
\usepackage{textcomp}

\newtheorem{theorem}{Theorem}[section]

\newtheorem{lemma}[theorem]{Lemma}

\newtheorem{corollary}[theorem]{Corollary}

\def\baselinestretch{1}
\renewcommand{\baselinestretch}{1.6}\small\normalsize
\def \expect {\mathsf{E} }
\def \genericfuncRg {r^{(\mathbf{p})}}

\def \twiddle[#1] {e^{-j \frac{2 \pi}{N}  #1 }}
\def \twiddleneg[#1] {e^{j \frac{2 \pi}{N}  #1 }}

\newcommand{\eq}{\,=\,}

\def\llr{\rho_{\mathbf{x}_{0}}}
\def\zz{\mathbf{z}}

\DeclareMathOperator{\supp}{supp}
\DeclareMathOperator*{\argmax}{argmax}
\DeclareMathOperator*{\spark}{spark}
\DeclareMathOperator*{\rank}{rank}
\DeclareMathOperator*{\kernel}{ker}

\DeclareMathOperator*{\argmin}{argmin}

\DeclareMathOperator*{\trace}{tr}

\DeclareMathOperator{\linspan}{span}

\def\ML_est{\hat{\mathbf{x}}_{\text{ML}}}
\newcommand{\CRBfull}{{C}ram\'{e}r--{R}ao bound\xspace}

\newcommand{\minachievvar}{M}
\newcommand{\genericfuncbound}{u}
\newcommand{\bb}{\mathbf{b}}

\newcommand{\RKHSSLGM}{\mathcal{H}_{\text{SLGM},\mathbf{x}_{0}}}
\newcommand{\RKHSSSNM}{\mathcal{H}_{\text{SSNM},\mathbf{x}_{0}}}
\newcommand{\SLGMMinAchVar}{M_{\text{SLGM}}(c(\cdot),\mathbf{x}_{0})}
\newcommand{\SSNMMinAchVar}{M_{\text{SSNM}}(c(\cdot),\mathbf{x}_{0})}

\newcommand{\be}{\begin{equation}}
\newcommand{\ee}{\end{equation}}
\newcommand{\ist}{\hspace*{.2mm}}
\newcommand{\rmv}{\hspace*{-.2mm}}

\newcommand{\detm}[1]{\det(#1)}
\newcommand{\tracem}[1]{\trace(#1)}

\newcommand{\scalarestproblem}{\big(  \mathcal{X}, f(\mathbf{y};\mathbf{x}),g(\cdot) \big)}
\newcommand{\SLGMvectorestproblem}{\big(  \mathcal{X}_{S}, f_{\mathbf{H}}(\mathbf{y};\mathbf{x}),\mathbf{g}(\cdot) \big)}
\newcommand{\SLGMscalarestproblem}{\big(  \mathcal{X}_{S}, f_{\mathbf{H}}(\mathbf{y};\mathbf{x}),g(\cdot) \big)}
\allowdisplaybreaks

\makeatother

\begin{document}

\title{\vspace*{5mm}Minimum Variance Estimation of a Sparse Vector\\[-3.5mm]within the Linear Gaussian Model:\\[-3.5mm]An RKHS Approach \vspace*{3mm}}
\author{\emph{ Alexander Jung\textsuperscript{a} (corresponding author), 
Sebastian Schmutzhard\textsuperscript{$\ist$b}, 
Franz Hlawatsch\textsuperscript{a},\\[-1.5mm] 
Zvika Ben-Haim\textsuperscript{c},
and Yonina C.\ Eldar\textsuperscript{$\ist$d} 
\thanks{This work was supported by the FWF under Grants S10602-N13 (Signal and Information Representation) and S10603-N13 
\vspace{-.8mm}
(Statistical Inference) within the National Research Network SISE, by the WWTF under Grant MA 07-004 (SPORTS), 
\vspace{-.8mm}
by the Israel Science Foundation under Grant 1081/07, and by the European Commission under the FP7 Network of 
\vspace{-.8mm}
Excellence in Wireless Communications NEWCOM++ (contract no. 216715).
Parts of this work were previously presented at the 44th 
\vspace{-.8mm}
Asilomar Conference on Signals, Systems and Computers, Pacific Grove, CA, Nov.\ 2010
and at the 2011 IEEE International 
\vspace{-.8mm}
Symposium on Information Theory (ISIT 2011), Saint Petersburg, Russia, July/Aug.\ 2011.}
} 
\thanks{
Submitted to the IEEE Transactions on Information Theory, \today}
\\[2mm]
{\small\emph{\textsuperscript{a}}Institute of Telecommunications, Vienna University of Technology;
\{ajung,\,fhlawats\}@nt.tuwien.ac.at}\\[-1.7mm]
{\small\emph{\textsuperscript{b}}NuHAG, Faculty of Mathematics, University of Vienna;
sebastian.schmutzhard@univie.ac.at} \\[-1.7mm]
{\small\emph{\textsuperscript{c}}Google, Inc., 
Israel; zvika@google.com}\\[-1.7mm]
{\small\emph{\textsuperscript{d}}Technion---Israel Institute of Technology;
yonina@ee.technion.ac.il} 
}


 \date{Submitted to IEEE Transactions on Information Theory, \today}

\renewcommand{\baselinestretch}{1.4}\small\normalsize

\maketitle

\vspace*{-8mm}

\renewcommand{\baselinestretch}{1.33}\small\normalsize

\begin{abstract}
\vspace*{-1.2mm}
We consider 
minimum variance estimation within the \emph{sparse linear {G}aussian model} (SLGM). A sparse vector is to be estimated 
from a linearly transformed version 
embedded in
Gaussian noise. Our analysis is based on the theory of reproducing kernel Hilbert spaces (RKHS).
After a characterization of the RKHS associated with the SLGM, we derive
novel lower bounds on the minimum variance achievable by estimators with
a prescribed bias function. This includes the 
important case of unbiased estimation. The variance bounds are obtained via an 
orthogonal projection of the prescribed mean function onto a subspace of the RKHS 
associated with the SLGM. 
Furthermore, we specialize our bounds to compressed sensing measurement matrices and express them 
in terms of the restricted isometry and coherence parameters.
For the special case of the SLGM given by the \emph{sparse signal in noise model} (SSNM), we 
derive closed-form expressions of the minimum achievable variance (Barankin bound) and the corresponding locally minimum variance estimator.
We also analyze the 
effects of exact and approximate sparsity information
and show that the minimum achievable variance for
exact sparsity is not a limiting case of that for approximate sparsity.
Finally, we compare our bounds with the variance of three well-known 
estimators,
namely, the maximum-likelihood estimator, the hard-thresholding estimator, and compressive reconstruction using the orthogonal matching 
pursuit.
\end{abstract}

\begin{keywords}
\vspace*{-1.2mm}
Sparsity, compressed sensing, unbiased estimation, denoising, RKHS, Cram\'{e}r--Rao bound, Barankin bound, Ham\-mersley--Chapman--Robbins 
bound, locally minimum variance unbiased estimator.
\vspace*{4mm}
\end{keywords}

\renewcommand{\baselinestretch}{1.5}\small\normalsize

\section{{Introduction}}\label{sec.intro}


We study the problem of estimating the value $\mathbf{g}(\mathbf{x})$ of a known 
vector-valued function $\mathbf{g}(\cdot)$ evaluated at the 
unknown parameter vector $\mathbf{x} \!\in\! \mathbb{R}^{N}\!$. It is known that 
$\mathbf{x}$ is 
$S$-sparse, i.e., at most $S$ of its entries are nonzero, where $S \in [N] \triangleq \{1,\ldots,N\}$ (typically $S \!\ll\! N$). 
While the sparsity degree $S$ is 
known, the set of positions of the nonzero entries of $\mathbf{x}$, i.e., the \emph{support} 
$\supp(\mathbf{x}) \subseteq [N]$, is unknown. 
The estimation of $\mathbf{g}(\mathbf{x})$ is based on an observed random vector $\mathbf{y} = \mathbf{H} \mathbf{x} + \mathbf{n} \in \mathbb{R}^{M}\rmv\rmv$, 
with a known 
system matrix $\mathbf{H} \!\in\! \mathbb{R}^{M \times N}\!$ and independent and identically distributed (i.i.d.)\ 
Gaussian noise 
$\mathbf{n} \sim \mathcal{N}(\mathbf{0},  \sigma^{2} \mathbf{I})$ with known noise variance $\sigma^{2} >0$. 
We assume that the minimum number of linearly dependent columns of $\mathbf{H}$ is larger than $S$.

The data model
described above will be termed the \emph{sparse linear Gaussian model} (SLGM).
The SLGM is relevant, e.g., to sparse channel estimation \cite{Carbonelli06}, where the sparse parameter vector $\mathbf{x}$ represents the tap
coefficients of a linear time-invariant 
channel and the system matrix $\mathbf{H}$ represents the training signal. More generally, the SLGM can be used for any type of sparse deconvolution \cite{MallatBook}.
The special case of the SLGM obtained for $\mathbf{H} \rmv=\rmv \mathbf{I}$ (so that $M \rmv=\rmv N$ and $\mathbf{y} = \mathbf{x} + \mathbf{n}$)
will be referred to as the \emph{sparse signal in noise model} (SSNM). The SSNM can be used, e.g., for sparse channel estimation \cite{Carbonelli06}
employing an orthogonal training signal \cite{OptimalTrainingDong} and for image denoising employing an orthonormal wavelet basis \cite{Donoho94idealspatial}. 


A fundamental question, to be considered in this work, is how to exploit the knowledge of the sparsity degree $S$. 
In contrast to compressed sensing (CS), where the sparsity 
is exploited for compression \cite{Don06,CandesCSTutorial,Can06a}, 
here we investigate how much the sparsity assumption helps us 
improve the accuracy of estimating $\mathbf{g}(\mathbf{x})$. 
Related questions have been previously addressed for the SLGM in 
\cite{Donoho94idealspatial} and \cite{Raskutti2011,VerzelenMinmaxSLM,DonohoJohnstone94,ZvikaCRB,ZvikaSSP,AlexZvikaJournal}.
In \cite{Raskutti2011} and \cite{VerzelenMinmaxSLM}, bounds on the minimax risk
and approximate minimax estimators whose worst-case risk is close to these bounds have been derived for the SLGM. 
An asymptotic analysis of minimax estimation for the SSNM has been given in the seminal work \cite{DonohoJohnstone94,Donoho94idealspatial}.
In the context of 
minimum variance estimation (MVE), which is relevant to our present work, lower bounds on the minimum achievable 
variance for the SLGM have been derived recently. 
In particular, the \CRBfull (CRB) 
for the SLGM has been derived and analyzed in \cite{ZvikaCRB} and \cite{ZvikaSSP}.
Furthermore, in our previous work \cite{AlexZvikaJournal}, we derived lower and upper bounds on the minimum achievable variance of unbiased estimators 
for the SSNM. 


The contributions of the present paper can be summarized as follows. First, we present novel CRB-type lower bounds on the variance of estimators for the SLGM. 
These bounds are derived by an application of the mathematical framework of \emph{reproducing kernel Hilbert spaces} (RKHS) \cite{aronszajn1950,Parzen59,Duttweiler73b}. 
Since they hold for any estimator with a prescribed mean function, they 
are also lower bounds on the minimum achievable variance (also known as \emph{Barankin bound}) for the SLGM.
The bounds are tighter than those presented in \cite{ZvikaCRB,ZvikaSSP}, and they have an appealing form in that they are scaled versions of the 
conventional CRB obtained for the nonsparse case \cite{kay,LC}.
We note that our RKHS approach 
is quite different from the technique used in \cite{AlexZvikaJournal}. Also, a shortcoming
of the lower bounds presented in \cite{ZvikaCRB}, and \cite{AlexZvikaJournal} is the fact that they exhibit 
a discontinuity when passing from the case ${\| \mathbf{x} \|}_{0} \!=\! S$ (i.e., $\mathbf{x}$ has exactly $S$ nonzero values) to the case ${\| \mathbf{x} \|}_{0} \!<\! S$ 
(i.e., $\mathbf{x}$ has less than $S$ nonzero values). 
For unbiased estimation, we 
derive a lower bound that is tighter
than the bounds in \cite{ZvikaCRB,ZvikaSSP,AlexZvikaJournal} and, moreover, 
a continuous function of $\mathbf{x}$. 
In particular, this bound exhibits a smooth transition between the two regimes given by ${\| \mathbf{x} \|}_{0} \!=\! S$ and ${\| \mathbf{x} \|}_{0} \!<\! S$.
Based on the fact that the linear CS recovery problem is
an instance of the SLGM, 
we specialize our lower bounds to system matrices that are CS measurement matrices, and we express
them in terms of the restricted isometry and coherence parameters of these matrices. 

Furthermore, for the 
SSNM, we derive expressions of the minimum achievable variance at a given 
parameter vector $\mathbf{x} = \mathbf{x}_{0}$
and of the \emph{locally minimum variance} (LMV) estimator,
i.e., the estimator 
achieving the minimum variance at $\mathbf{x}_{0}$. Simplified expressions of the minimum achievable variance 
and the LMV estimator are obtained for a certain subclass of ``diagonal'' bias functions (which includes the unbiased case).

Finally, we consider the SLGM with an approximate sparsity constraint and show that the minimum achievable variance under an 
exact sparsity constraint is not a limiting case of the minimum achievable variance under an approximate sparsity constraint.

A central aspect of this paper is the application of the mathematical framework of RKHS  \cite{aronszajn1950}
to the SLGM. The RKHS framework has been previously applied to classical estimation in the seminal work reported in \cite{Parzen59} and \cite{Duttweiler73b}, and our present treatment
is substantially based on that work. However, to the best of our knowledge, the RKHS framework has not been applied to the SLGM or, more generally, to the estimation 
of (functions of) sparse vectors. 
The sparse case is specific in that we are considering 
functions whose domain is the set 
of 
$S$-sparse vectors. For $S < N$, the interior of this set is empty, and thus there do not exist derivatives 
in every possible direction. This lack of a differentiable structure makes the characterization of the RKHS 
a somewhat delicate matter.

The remainder of this paper is organized as follows. 
We begin in Section \ref{SecSLGM} with formal statements of the SLGM and SSNM and continue in 
Section \ref{Sec_MVE} with a 
review of basic elements of MVE.
In Section \ref{sec_basic_toolkit}, we review some fundamentals of
RKHSs and the application of RKHSs to MVE.
In Section \ref{sec_RKHS_MVE_SLGM}, we characterize and discuss the RKHS associated with the SLGM.
For the SLGM, we then use the RKHS framework to present formal characterizations of the class of bias functions allowing for 
finite-variance estimators, of the minimum achievable variance (Barankin bound), and of the LMV estimator. We also 
present a result on the 
shape of the Barankin bound. 
In Section \ref{sec_bounds_SLGM}, we 
reinterpret the sparse CRB of \cite{ZvikaCRB}
from the 
RKHS perspective, and we present two novel lower variance bounds for the SLGM.
In Section \ref{sec_slm_viewpoint_CS}, 
we specialize the bounds of Section \ref{sec_bounds_SLGM} to system matrices that are CS measurement matrices.
The important special case 
given by the SSNM is discussed in Section \ref{sec_rkhs_approach_ssnm}, where we derive closed-form expressions 
of the minimum achievable variance (Barankin bound) and of the corresponding LMV estimator. 
A discussion of the 
effects of exact and approximate sparsity information
from the MVE perspective is presented in Section \ref{sec_strict_sparstiy_SLGM}. 
Finally, in Section \ref{sec_numerical_SLGM}, we present numerical results comparing our theoretical bounds with the actual variance 
of some popular estimation schemes.


\emph{Notation and basic definitions}.\, 
The sets of real, nonnegative real, natural, and nonnegative integer numbers are denoted by 
$\mathbb{R}$, $\mathbb{R}_{+}$, $\mathbb{N} \triangleq \{1,2,\ldotsÊ\}$, and $\mathbb{Z}_{+}\! \triangleq \{0,1,\ldots\}$, respectively.
For $L \in \mathbb{N}$, we define $[L] \triangleq \{1,\ldots,L\}$.
The space of all discrete-argument functions $f[\cdot]: \mathcal{T} \!\rightarrow \mathbb{R}$ (with $\mathcal{T} \!\subseteq\rmv \mathbb{Z}$) 
for which $\sum_{l \in \mathcal{T}} f^2[l] < \infty$ is denoted by $\ell^{2}(\mathcal{T})$, with associated norm
${\| f[\cdot]\|}_{\mathcal{T}} \triangleq  \sqrt{\sum_{l \in \mathcal{T}} f^2[l] }$.
The Kronecker delta $\delta_{k,l}$ is $1$ if $k=l$ and $0$ otherwise.
Given an $N$-tuple of nonnegative integers (a ``multi-index'') $\mathbf{p} = (p_1 \cdots\, p_N)^T\!\rmv \in \mathbb{Z}_{+}^{N}$ \cite{KranzPrimerAnalytic}, we define
$\mathbf{p}! \triangleq \prod_{l \in [N]} p_{l}!$, $| \mathbf{p}|\triangleq \sum_{l \in [N]} p_{l}$, and $\mathbf{x}^{\mathbf{p}} \triangleq \prod_{l \in [N]} (x_{l})^{p_l}$
(for $\mathbf{x} \in \mathbb{R}^{N}$). 
Given two multi-indices $\mathbf{p}_{1}, \mathbf{p}_{2} \in \mathbb{Z}_{+}^{N}$, the inequality $\mathbf{p}_{1} \leq \mathbf{p}_{2}$ is understood to hold
elementwise, 
i.e., $p_{1,l} \leq p_{2,l}$ for all $l \in [N]$. 

Lowercase (uppercase) boldface letters denote column vectors (matrices). 
The superscript $^{T}\rmv$ stands for transposition. 
The $k$th unit vector is denoted by $\mathbf{e}_{k}$, and the identity matrix 
by $\mathbf{I}$.
For a rectangular matrix $\mathbf{H} \rmv\in\rmv \mathbb{R}^{M \times N}\rmv\rmv$, we denote by $\mathbf{H}^{\dagger}$ its Moore-Penrose pseudoinverse \cite{golub96},
by $\kernel(\mathbf{H}) \triangleq \{ \mathbf{x} \rmv\in\rmv \mathbb{R}^{N} |\ist  \mathbf{H} \mathbf{x} \rmv=\rmv \mathbf{0} \}$ its kernel (or null space),
by $\linspan(\mathbf{H}) \triangleq \{ \mathbf{y} \rmv\rmv\in\rmv \mathbb{R}^{M} | \,\exists \mathbf{x} \rmv\rmv\in\rmv \mathbb{R}^{N} \!\rmv: \mathbf{y} \rmv=\rmv \mathbf{H} \mathbf{x} \}$ 
its column span, and by $\rank( \mathbf{H} )$ its rank.
For a square matrix $\mathbf{H} \in \mathbb{R}^{N \times N}\rmv\rmv$, we denote by 
$\tracem{ \mathbf{H}}$, $\detm{\mathbf{H}}$, and $\mathbf{H}^{-1}$ its trace, determinant, and inverse (if it exists), respectively.
The $k$th entry of a vector $\mathbf{x}$ is denoted by ${(\mathbf{x})}_{k} = x_k\ist$, 
and the entry in the $k$th row and $l$th column of a matrix $\mathbf{H}$ 
by ${(\mathbf{H})}_{k,l} = H_{k,l}$.
The support (i.e., set of indices of all nonzero entries) and the number of nonzero entries of a vector 
$\mathbf{x}$ are denoted by $\supp(\mathbf{x})$ and ${\| \mathbf{x} \|}_{0} = | \rmv \supp(\mathbf{x}) |$, respectively.
Given an index set $\mathcal{I} \subseteq [N]$, we denote by $\mathbf{x}^{\mathcal{I}} \!\in \mathbb{R}^{N}\rmv$ 
the vector obtained from $\mathbf{x} \in \mathbb{R}^{N}\rmv$ by zeroing all entries except those indexed by $\mathcal{I}$,
and by $\mathbf{H}_{\mathcal{I}} \in \mathbb{R}^{M \times |\mathcal{I}|}$ the matrix formed by those columns of $\mathbf{H} \in \mathbb{R}^{M \times N}\rmv$ 
that are indexed by $\mathcal{I}$.
The $p$-norm of a vector $\mathbf{x} \rmv\in\rmv \mathbb{R}^{N}\rmv$ 
is defined
as $\| \mathbf{x} \|_{p} \triangleq \big( \rmv\sum_{k \in [N]} x_{k}^{p}\big)^{1/p}$.



\section{The Sparse Linear Gaussian Model}
\label{SecSLGM}

We will first present a more detailed statement of the SLGM. 
Let $\mathbf{x} \!\in\! \mathbb{R}^{N}\rmv$ be an unknown parameter vector that is known to be 
$S$-sparse in the sense that at most $S$ of its entries are nonzero, i.e., ${\| \mathbf{x} \|}_{0} \le S$, with a known sparsity degree $S \in [N]$
(typically $S \!\ll\! N$).
We will express this $S$-sparsity in terms of a parameter set $\mathcal{X}_{S}$, 
\vspace{-1mm}
i.e.,
\be
\label{equ_SLGM_parameter}
\mathbf{x} \rmv\in\rmv  \mathcal{X}_{S} \,, \quad\; \text{with} \;\; 
  \mathcal{X}_{S} \ist\triangleq\ist \big\{\mathbf{x}' \rmv\!\in\rmv \mathbb{R}^{N} \big| \ist {\|\mathbf{x}' \|}_{0} \rmv\leq\rmv S \big\}  \subseteq \mathbb{R}^{N} .
\ee 
In the limiting case where $S$ is equal to the dimension
of $\mathbf{x}$, i.e., $S \rmv=\rmv N$, we have $\mathcal{X}_{S} = \mathbb{R}^{N}\rmv$. 
Note that the support $\supp(\mathbf{x}) \subseteq [N]$ is unknown. We observe a linearly transformed and noisy version of $\mathbf{x}$, 
\be
\label{equ_linear_observation_model}
\mathbf{y} \ist=\ist \mathbf{H} \mathbf{x} + \mathbf{n} \in \mathbb{R}^{M}\ist,
\ee
where $\mathbf{H} \!\in\rmv \mathbb{R}^{M \times N}\rmv$ is a known matrix and $\mathbf{n} \!\in\rmv \mathbb{R}^{M}\rmv$ is i.i.d.\ Gaussian noise, i.e.,
$\mathbf{n} \sim \mathcal{N}(\mathbf{0},  \sigma^{2} \mathbf{I})$, with a known noise variance $\sigma^{2} >0$.
It follows that the probability density function (pdf) of the observation $\mathbf{y}$ for a specific value of $\mathbf{x}$ is given 
by 
\begin{equation} 
\label{equ_def_LGM_stat_model}
f_{\mathbf{H}}(\mathbf{y}; \mathbf{x}) \eq \frac{1}{(2 \pi \sigma^{2})^{M/2} } 
  \,\exp \rmv\rmv\rmv\bigg( \!\!-\rmv\rmv  \frac{1}{2 \sigma^{2}} {\| \mathbf{y} \rmv\rmv-\rmv\rmv \mathbf{H} \mathbf{x} \|}^{2}_{2} \bigg) \,.
\vspace{-1mm}
\end{equation}
We assume 
\vspace{-1mm}
that 
\begin{equation} 
\spark(\mathbf{H}) > S \,, 
\label{equ_spark_cond}
\end{equation} 
where $\rm{spark}(\mathbf{H})$ denotes the minimum number of linearly dependent columns of $\mathbf{H}$ \cite{GreedisGood,DonohoElad2002}.
Note that we also allow $M \!<\rmv N$ 
(this case is relevant to CS methods as discussed in Section \ref{sec_slm_viewpoint_CS});
however, condition \eqref{equ_spark_cond} implies that $M \rmv\ge\rmv S$. 
Condition \eqref{equ_spark_cond} is weaker than the standard condition 
$\spark(\mathbf{H}) > 2S$ \cite{ZvikaCRB}. Still, the standard condition is reasonable since otherwise one can find two different parameter vectors 
$\mathbf{x}_{1}, \mathbf{x}_{2} \in \mathcal{X}_{S}$ for 
which 
$f_{\mathbf{H}}(\mathbf{y}; \mathbf{x}_{1}) = f_{\mathbf{H}}(\mathbf{y}; \mathbf{x}_{2})$ for all $\mathbf{y}$,
which implies that
one cannot distinguish between $\mathbf{x}_{1}$ and $\mathbf{x}_{2}$ based on knowledge of $\mathbf{y}$. 
Finally, we note that the assumption of i.i.d.\ noise in \eqref{equ_linear_observation_model} does not imply a loss of generality. Indeed, consider 
an SLGM $\mathbf{y} = \mathbf{H} \mathbf{x} + \mathbf{n}$ where $\mathbf{n}$ is not i.i.d.\ with some positive definite (hence, nonsingular) covariance matrix $\mathbf{C}$.
Then, the ``whitened observation'' $\tilde{\mathbf{y}} \triangleq \mathbf{C}^{-1/2}\ist\tilde{\mathbf{y}}$ \cite{papoulis91}, where $\mathbf{C}^{-1/2}$ is the 
inverse of the matrix square root $\mathbf{C}^{1/2}\rmv$ \cite{higham86}, can be written as
$\tilde{\mathbf{y}}= \widetilde{\mathbf{H}} \mathbf{x} + \tilde{\mathbf{n}}$, with $\widetilde{\mathbf{H}} \triangleq \mathbf{C}^{-1/2} \ist \mathbf{H}$
and $\tilde{\mathbf{n}} \triangleq \mathbf{C}^{-1/2} \ist \mathbf{n}$. It can be verified that 
$\widetilde{\mathbf{H}}$ also satisfies \eqref{equ_spark_cond} and 
$\tilde{\mathbf{n}}$ is i.i.d.\ with variance $\sigma^{2}=1$, i.e., $\tilde{\mathbf{n}} \sim \mathcal{N}(\mathbf{0},\mathbf{I})$.

The task considered in this paper is 
estimation of the function value $\mathbf{g}(\mathbf{x})$ from the observation 
$\mathbf{y} = \mathbf{H} \mathbf{x} + \mathbf{n}$, where the \emph{parameter function}
$\mathbf{g}(\cdot) \!: \mathcal{X}_{S} \rmv\rightarrow \mathbb{R}^{P}\rmv$ is a known deterministic function.
The \emph{estimate} $\hat{\mathbf{g}} = \hat{\mathbf{g}}(\mathbf{y})\rmv\in\rmv \mathbb{R}^{P}\rmv$
is derived from $\mathbf{y}$ via a deterministic \emph{estimator} $\hat{\mathbf{g}}(\cdot) \!: \mathbb{R}^{M} \rmv\!\rightarrow \mathbb{R}^{P}\rmv\rmv$.
We allow $\hat{\mathbf{g}}\!\in\rmv\rmv \mathbb{R}^{P}\!$ without constraining $\hat{\mathbf{g}}$ to be in 
$\mathbf{g}(\mathcal{X}_{S}) \triangleq \{ \mathbf{g}(\mathbf{x}) \ist|\ist\ist \mathbf{x} \!\in\! \mathcal{X}_{S} \}$, 
even though it is known that $\mathbf{x} \!\in\! \mathcal{X}_{S}$. The reason for not enforcing the sparsity constraint 
$\hat{\mathbf{g}} \in \mathbf{g}(\mathcal{X}_{S})$ is twofold: first, it would complicate the analysis; 
second, it would typically
result in a worse achievable estimator performance (in terms of mean squared error) since it restricts the class of allowed estimators. 
In particular, it has been shown that a sparsity constraint can increase the worst-case risk of the resulting estimators significantly \cite{Leeb2008201}. 

Estimation of the parameter vector $\mathbf{x}$ itself is a special case obtained by choosing the parameter function as the identity mapping, 
i.e., $\mathbf{g}(\mathbf{x}) = \mathbf{x}$, which implies $P \!=\! N$. Again,
we allow $\hat{\mathbf{x}}\rmv\in\rmv \mathbb{R}^{N}\rmv$ and do not constrain 
$\hat{\mathbf{x}}$ to be in $\mathcal{X}_{S}$.

In what follows, it will be convenient to denote the 
SLGM-based estimation problem by the triple
\[
\mathcal{E}_{\text{SLGM}} \ist\triangleq\ist \SLGMvectorestproblem \ist, 
\] 
where $f_{\mathbf{H}}(\mathbf{y}; \mathbf{x})$ is given by \eqref{equ_def_LGM_stat_model} 
and will be referred to as the \emph{statistical model}. A related estimation problem is based on the \emph{linear Gaussian model} (LGM) 
\cite{poorsspbook,kay,scharf91,HeroUniformCRB}, for which $\mathbf{x} \rmv\in\rmv \mathbb{R}^{N}\rmv$ rather than $\mathbf{x} \rmv\in\rmv \mathcal{X}_{S}$; 
this problem will be denoted 
\vspace{-1mm}
by 
\[
\mathcal{E}_{\text{LGM}} \ist\triangleq\ist \big( \ist\mathbb{R}^{N}\!, f_{\mathbf{H}}(\mathbf{y}; \mathbf{x}), \mathbf{g}(\cdot) \big) \ist.
\]
The SLGM shares with the LGM the observation model \eqref{equ_linear_observation_model} and the statistical model \eqref{equ_def_LGM_stat_model}; 
it is obtained from the LGM by restricting
the parameter set $\mathbb{R}^{N}\rmv$ to the set of $S$-sparse vectors, $\mathcal{X}_{S}$. For $S=N$, the SLGM reduces to the LGM.  
Another important special case of the SLGM is given by the SSNM, for which $\mathbf{H} \!=\! \mathbf{I}$, $M \!=\! N$, 
\vspace{-2mm}
and 
\[
\mathbf{y} \ist=\ist \mathbf{x} + \mathbf{n} \,,
\]
where $\mathbf{x} \!\in\! \mathcal{X}_{S}$ and 
$\mathbf{n} \sim \mathcal{N}(\mathbf{0}, \sigma^{2}  \mathbf{I})$ with known variance $\sigma^{2} \!\rmv >\! 0$. 
The SSNM-based estimation problem will be denoted 
\vspace*{-2mm}
as
\[ 
\mathcal{E}_{\text{SSNM}} \ist\triangleq\ist \big( \mathcal{X}_{S}, f_{\mathbf{I}}\rmv(\mathbf{y}; \mathbf{x}), \mathbf{g}(\cdot)  \big) \ist. 
\]

\section{Basic Elements of Minimum Variance Estimation}
\label{Sec_MVE}

Let us 
consider\footnote{This 
introductory section closely parallels \cite[Section II]{RKHSExpFamIT2012}. We include it nevertheless because it constitutes an important basis for our subsequent 
discussion.} 
a general estimation problem $\mathcal{E} = \big(  \mathcal{X}, f(\mathbf{y};\mathbf{x}),\mathbf{g}(\cdot) \big)$ 
based on an arbitrary parameter set $\mathcal{X} \!\subseteq\rmv \mathbb{R}^{N}$ and an arbitrary statistical model $f(\mathbf{y}; \mathbf{x})$.
The general goal in the design of an estimator $\hat{\mathbf{g}}(\cdot)$ is that $\hat{\mathbf{g}}(\mathbf{y})$
should be close to the true value $\mathbf{g}(\mathbf{x})$. A frequently used criterion for assessing the quality 
of an 
estimator $\hat{\mathbf{g}}(\mathbf{y})$ is the mean squared error (MSE) 
defined as 
\[
 \varepsilon \,\triangleq\, \expect_{\mathbf{x}} \big\{ {\|  \hat{\mathbf{g}}(\mathbf{y}) \rmv-\rmv \mathbf{g}(\mathbf{x})\|}^{2}_{2} \big\} 
 \,= \int_{\mathbb{R}^{M}} \! {\|  \hat{\mathbf{g}}(\mathbf{y}) \rmv-\rmv \mathbf{g}(\mathbf{x})\|}^{2}_{2} 
 \, f(\mathbf{y}; \mathbf{x}) \, d \mathbf{y} \,. 
\]
Here, $\expect_{\mathbf{x}} \{ \cdot \}$ denotes the expectation operation with respect to the pdf $f(\mathbf{y}; \mathbf{x})$; the subscript 
in $\expect_{\mathbf{x}}$ indicates the dependence on the parameter vector $\mathbf{x}$ parametrizing $f(\mathbf{y}; \mathbf{x})$.
We will write $\varepsilon(\hat{\mathbf{g}}(\cdot); \mathbf{x})$ to indicate the dependence of the MSE on the estimator $\hat{\mathbf{g}}(\cdot)$ and the parameter vector 
$\mathbf{x}$. In general,
there does not exist an estimator $\hat{\mathbf{g}}(\cdot)$ that minimizes the MSE simultaneously for all 
$\mathbf{x} \rmv\in\rmv \mathcal{X}$ \cite{RethinkingBiasedEldar}. 
This follows from the fact that minimizing the MSE at a given parameter vector $\mathbf{x}_{0}$ always yields zero MSE; this is achieved by the trivial estimator 
$\hat{\mathbf{g}}(\mathbf{y}) \equiv \mathbf{g}(\mathbf{x}_{0})$,
which 
ignores the observation $\mathbf{y}$.

A popular rationale for the design of good estimators is MVE.
The MSE can be decomposed as 
\begin{equation} 
\label{equ_bias_variance_tradeoff}
\varepsilon(\hat{\mathbf{g}}(\cdot); \mathbf{x}) \,=\, {\| \mathbf{b}(\hat{\mathbf{g}}(\cdot); \mathbf{x}) \|}_2^{2} +\ist v(\hat{\mathbf{g}}(\cdot); \mathbf{x}) \,, 
\end{equation}
with the bias $\mathbf{b}(\hat{\mathbf{g}}(\cdot); \mathbf{x}) \triangleq \expect_{\mathbf{x}} \{ \hat{\mathbf{g}}(\mathbf{y}) \} \ist-\ist \mathbf{g}(\mathbf{x})$ and 
the variance $v(\hat{\mathbf{g}}(\cdot); \mathbf{x}) 
\triangleq \expect_{\mathbf{x}} \big\{ \big\| \hat{\mathbf{g}}(\mathbf{y}) - \expect_{\mathbf{x}} \{ \mathbf{g}(\mathbf{y}) \} \big\|^{2}_{2} \big\}$.
In MVE, one fixes the bias on the entire parameter set $\mathcal{X}$, i.e., one requires that
\be
\label{equ_bias_prescription}
\mathbf{b}(\hat{\mathbf{g}}(\cdot); \mathbf{x}) \ist\stackrel{!}{=}\ist \mathbf{c}(\mathbf{x}) \,, \quad\; \text{for all} \;\ist\ist \mathbf{x} \rmv\rmv\in\rmv\rmv \mathcal{X} \ist, 
\ee
with a \emph{prescribed bias function} $\mathbf{c}(\cdot) \rmv\rmv : \mathcal{X} \rmv\rmv\rightarrow \mathbb{R}^{P}\!$, and 
attempts to minimize the variance $v(\hat{\mathbf{g}}(\cdot); \mathbf{x})$ among all estimators with the given bias function $\mathbf{c}(\cdot)$.
Fixing the bias function is equivalent to fixing the estimator's mean function, i.e., $\expect_{\mathbf{x}}\big\{ \hat{\mathbf{g}}(\mathbf{y}) \big\} \stackrel{!}{=} {\bm \gamma}(\mathbf{x})$ 
for all $\mathbf{x} \rmv\in\rmv \mathcal{X}$, with the \emph{prescribed mean function} ${\bm \gamma}(\mathbf{x}) \triangleq \mathbf{c}(\mathbf{x}) + \mathbf{g}(\mathbf{x})$. 
\emph{Unbiased estimation} is obtained as a special case for $\mathbf{c}(\mathbf{x}) \equiv \mathbf{0}$ or equivalently ${\bm \gamma}(\mathbf{x}) \equiv \mathbf{g}(\mathbf{x})$.
Fixing the bias can be viewed
as a kind of ``regularization'' of the set of considered estimators \cite{LC,RethinkingBiasedEldar}, 
since it excludes useless estimators such as $\hat{\mathbf{g}}(\mathbf{y}) \equiv \mathbf{g}(\mathbf{x}_{0})$. 
Another justification for considering a fixed bias function is that under mild conditions, for a large number of i.i.d.\ observations 
${\{\mathbf{y}_i \}}_{i \in [L]}$, the bias term dominates in the decomposition \eqref{equ_bias_variance_tradeoff}. Thus, in order to achieve a small MSE in that case, 
an estimator has to be at least asymptotically unbiased, i.e., one has to require that, for a large number of observations, $\mathbf{b}(\hat{\mathbf{g}}(\cdot); \mathbf{x}) \approx \mathbf{0}$ for all $\mathbf{x} \rmv\in\rmv \mathcal{X}$.


For an estimation problem $\mathcal{E} = \big(  \mathcal{X}, f(\mathbf{y};\mathbf{x}),\mathbf{g}(\cdot) \big)$, a fixed parameter vector $\mathbf{x}_{0} \rmv\in\rmv \mathcal{X}$, 
and a prescribed bias function $\mathbf{c}(\cdot) \rmv\rmv : \mathcal{X} \rmv\rmv\rightarrow \mathbb{R}^{P}\rmv\rmv$, we define the \emph{set of allowed estimators} by 
\[
\mathcal{A}(\mathbf{c}(\cdot),\mathbf{x}_{0}) \,\triangleq\, \big\{ \hat{\mathbf{g}}(\cdot) \,\big|\, 
v(\hat{\mathbf{g}}(\cdot);\mathbf{x}_{0}) < \infty \,, \, \mathbf{b}(\hat{\mathbf{g}}(\cdot);\mathbf{x}) = \mathbf{c}(\mathbf{x}) \,\, \forall \mathbf{x} \rmv\rmv\in\rmv\rmv \mathcal{X} \big\} \,. 
\]
We call a bias function $\mathbf{c}(\cdot)$ \emph{valid} for the estimation problem $\mathcal{E}$ at $\mathbf{x}_{0} \!\in\rmv\rmv \mathcal{X}$ 
if the set $\mathcal{A}(\mathbf{c}(\cdot),\mathbf{x}_{0})$ is nonempty, which means that there is at least one estimator $\hat{\mathbf{g}}(\cdot)$ 
that has finite variance at $\mathbf{x}_{0}$ and whose bias function equals $\mathbf{c}(\cdot)$, i.e., 
$\mathbf{b}(\hat{\mathbf{g}}(\cdot);\mathbf{x}) = \mathbf{c}(\mathbf{x})$ for all $\mathbf{x} \rmv\in\rmv \mathcal{X}$. 
For the SLGM, in particular, this definition trivially entails the following fact:
If a bias function $\mathbf{c}(\cdot) \rmv\rmv : \mathcal{X}_{S} \rmv\rightarrow\rmv \mathbb{R}^{P}$ 
is valid for $S \!=\! N$, it is also valid for $S \rmv<\rmv N$.

It follows from \eqref{equ_bias_variance_tradeoff} that, for a fixed bias function $\mathbf{c}(\cdot)$, minimizing the MSE $\varepsilon(\hat{\mathbf{g}}(\cdot); \mathbf{x}_{0})$ 
is equivalent to minimizing the variance $v(\hat{\mathbf{g}}(\cdot); \mathbf{x}_{0})$. 
Let us denote the minimum (strictly speaking, infimum) variance at $\mathbf{x}_{0}$ for bias function $\mathbf{c}(\cdot)$ 
\vspace{-1mm}
by
\begin{equation}
\label{equ_minimum_variance_problem}
\minachievvar(\mathbf{c}(\cdot),\mathbf{x}_{0}) \,\triangleq \inf_{\hat{\mathbf{g}}(\cdot) \ist\in\ist \mathcal{A}(\mathbf{c}(\cdot),\mathbf{x}_{0})} 
  \rmv v(\hat{\mathbf{g}}(\cdot); \mathbf{x}_{0}) \,.
\vspace{1mm}
\end{equation}
If $\mathcal{A}(\mathbf{c}(\cdot),\mathbf{x}_{0})$ is empty, i.e., if $\mathbf{c}(\cdot)$ is not valid, we set $\minachievvar(\mathbf{c}(\cdot),\mathbf{x}_{0}) \triangleq \infty$.
Any estimator $\hat{\mathbf{g}}^{(\mathbf{c}(\cdot),\mathbf{x}_{0})}(\cdot)  \in \mathcal{A}(\mathbf{c}(\cdot),\mathbf{x}_{0})$ that achieves the infimum in \eqref{equ_minimum_variance_problem}, i.e., 
for which 
\begin{equation} 
\label{equ_def_LMV_estimator}
v \big(\hat{\mathbf{g}}^{(\mathbf{c}(\cdot),\mathbf{x}_{0})}(\cdot);\mathbf{x}_{0} \big) \ist=\ist \minachievvar(\mathbf{c}(\cdot),\mathbf{x}_{0}) \,,
\end{equation}
is called an \emph{LMV estimator} at $\mathbf{x}_{0}$
for bias function $\mathbf{c}(\cdot)$ \cite{LC,Parzen59,Duttweiler73b}. 
The corresponding minimum variance $\minachievvar(\mathbf{c}(\cdot),\mathbf{x}_{0})$ is called the \emph{minimum achievable variance} at $\mathbf{x}_{0}$
for bias function $\mathbf{c}(\cdot)$. The minimization problem defined by \eqref{equ_minimum_variance_problem} is referred to as a \emph{minimum variance problem} (MVP).
From its definition in \eqref{equ_minimum_variance_problem}, it follows that
$\minachievvar(\mathbf{c}(\cdot),\mathbf{x}_{0})$ is a lower bound on the variance at $\mathbf{x}_{0}$ of any estimator with bias function $\mathbf{c}(\cdot)$, i.e., 
\[ 
\hat{\mathbf{g}}(\cdot) \in \mathcal{A}(\mathbf{c}(\cdot),\mathbf{x}_{0}) \;\;\Rightarrow\;\; v(\hat{\mathbf{g}}(\cdot);\mathbf{x}_{0}) \ist\geq\ist \minachievvar(\mathbf{c}(\cdot),\mathbf{x}_{0}) \,.
\]
This is sometimes referred to as the \emph{Barankin bound}; it is the tightest possible lower bound on the variance at $\mathbf{x}_{0}$ of estimators with bias 
function $\mathbf{c}(\cdot)$.

If, for a prescribed bias function $\mathbf{c}(\cdot)$, there exists an estimator that is the LMV estimator \emph{simultaneously} at all $\mathbf{x}_0 \rmv\in\rmv \mathcal{X}$, 
then that estimator is termed
the \emph{uniformly minimum variance} (UMV) estimator for bias function $\mathbf{c}(\cdot)$ \cite{LC,Parzen59,Duttweiler73b}. 
For the SLGM, a UMV estimator does not exist in general \cite{AlexZvikaJournal,JungPHD}. A noteworthy exception is the 
SLGM where
$\mathbf{H}$ has full column rank, $\mathbf{g}(\mathbf{x}) = \mathbf{x}$, $S = N$, and $\mathbf{c}(\cdot) \equiv \mathbf{0}$; here, it is well known \cite{LC}, \cite[Thm. 4.1]{kay} that the 
least squares estimator, $\hat{\mathbf{x}}
= \mathbf{H}^{\dag} \mathbf{y}$, is the UMV estimator.

Finally, let $\hat{g}_{k}(\cdot) \triangleq \big(\hat{\mathbf{g}}(\cdot)\big)_k$ and $c_{k}(\cdot) \triangleq \big(\mathbf{c}(\cdot)\big)_k$.
The 
variance of the vector estimator $\hat{\mathbf{g}}(\cdot)$
can be decomposed as
\vspace{-1mm}
\be
\label{equ_sum_var_scalar}
v(\hat{\mathbf{g}}(\cdot); \mathbf{x}) \,= \sum_{k \in [P]} \! v(\hat{g}_{k}(\cdot); \mathbf{x}) \,,
\vspace{.5mm}
\ee 
where $v(\hat{g}_{k}(\cdot); \mathbf{x}) \triangleq \expect_{\mathbf{x}} \big\{ \big[ \hat{g}_{k}(\mathbf{y}) - \expect_{\mathbf{x}} \{ \hat{g}_{k}(\mathbf{y}) \} \big]^{2} \big\}$
is the variance of the $k$th estimator component $\hat{g}_{k}(\cdot)$.
Furthermore, $\hat{\mathbf{g}}(\cdot) \in \mathcal{A}(\mathbf{c}(\cdot),\mathbf{x}_{0})$ if and only if $\hat{g}_{k}(\cdot) \in \mathcal{A}(c_{k}(\cdot),\mathbf{x}_{0})$ for all $k \rmv\in\rmv [P]$.
This shows that the MVP \eqref{equ_minimum_variance_problem}
can be reduced to $P$ separate \emph{scalar} MVPs
\[
\minachievvar(c_k(\cdot),\mathbf{x}_{0}) \,\triangleq \inf_{\hat{g}_k(\cdot) \ist\in\ist \mathcal{A}(c_k(\cdot),\mathbf{x}_{0})} \rmv v(\hat{g}_k(\cdot); \mathbf{x}_{0}) \,, \quad\; k \rmv\in\rmv [P] \,,
\vspace{1mm}
\]
each requiring the optimization of a single scalar component $\hat{g}_{k}(\cdot)$ of $\hat{\mathbf{g}}(\cdot)$.  
Therefore, without loss of generality, we will hereafter assume that the parameter function $\mathbf{g}(\mathbf{x})$ is scalar-valued, i.e., $P \!=\! 1$ 
and $\mathbf{g}(\mathbf{x}) \rmv=\rmv g(\mathbf{x})$.

\section{RKHS Fundamentals} 
\label{sec_basic_toolkit}

As mentioned in Section \ref{sec.intro}, the existing variance bounds for the SLGM are not maximally tight. Using the theory
of RKHSs will allow us to derive variance bounds which are tighter than the existing bounds. For the SSNM (see Section \ref{sec_rkhs_approach_ssnm}), 
the RKHS approach even yields a precise characterization of the minimum achievable variance (Barankin bound) and of the accompanying LMV estimator.
In this section, we present a review (similar in part to \cite[Section III]{RKHSExpFamIT2012}) of some 
fundamentals of the theory of RKHSs and of the application of RKHSs to MVE. These fundamentals will provide a 
framework 
for our analysis of
the SLGM in later sections.

\subsection{Basic Facts}
\label{SecRKHSbasics}

An RKHS is associated with a \emph{kernel function} $R(\ist\cdot\ist\ist,\cdot) \!: \mathcal{X} \rmv\times\rmv \mathcal{X} \rmv\rightarrow \mathbb{R}$,
where $\mathcal{X}$ is an arbitrary set. The defining properties of a kernel function are (i) symmetry, i.e., 
$R(\mathbf{x}_{1}, \mathbf{x}_{2}) = R(\mathbf{x}_{2}, \mathbf{x}_{1})$ for all $\mathbf{x}_{1}, \mathbf{x}_{2} \in \mathcal{X}$,
and (ii) positive semidefiniteness in the sense that, for every finite set 
$\{\mathbf{x}_{1},\ldots,\mathbf{x}_{D}\} \subseteq \mathcal{X}$, the matrix $\mathbf{R} \in \mathbb{R}^{D \times D}$ with entries
$R_{m,n} = R(\mathbf{x}_{m},\mathbf{x}_{n})$ is positive semidefinite.
A fundamental result \cite[p. 344]{aronszajn1950} states that for any such kernel function $R$, there exists an RKHS
$\mathcal{H}(R)$, which is a Hilbert space equipped with an inner product 
${\langle \cdot\ist\ist, \cdot \rangle}_{\mathcal{H}(R)}$ and satisfying the following two properties:

\begin{itemize}

\vspace{1mm}

\item  
For any $\mathbf{x} \!\in\! \mathcal{X}$, $R(\ist\cdot\ist\ist,\mathbf{x}) \rmv\in\rmv \mathcal{H}(R)$ (here, $R(\ist\cdot\ist\ist,\mathbf{x})$ denotes the function 
$f_{\mathbf{x}}(\mathbf{x}') = R(\mathbf{x}'\!, \mathbf{x})$ for fixed $\mathbf{x} \rmv\in\rmv \mathcal{X}$).

\vspace{1mm}

\item For any function $f(\cdot) \rmv\in\rmv \mathcal{H}(R)$ and any 
\vspace{-1mm}
$\mathbf{x} \!\in\! \mathcal{X}$,
\begin{equation} 
\label{equ_reproducing_property}
\big\langle f(\cdot), R(\ist\cdot\ist\ist,\mathbf{x}) \big\rangle_{\mathcal{H}(R)} = f(\mathbf{x}) \,.
\vspace{-1mm}
\end{equation}

\vspace{.5mm}

\end{itemize}

\noindent The ``reproducing property'' \eqref{equ_reproducing_property} defines the inner product ${\langle f_1, f_2 \rangle}_{\mathcal{H}(R)}$ 
for all $f_1(\cdot), f_2(\cdot) \rmv\in \mathcal{H}(R)$, because any $f(\cdot) \rmv\in\rmv \mathcal{H}(R)$ can be expanded into the 
set of functions ${\{ R(\ist\cdot\ist\ist,\mathbf{x}) \}}_{\mathbf{x} \in \mathcal{X}}$. The induced norm is 
${\| f \|}_{\mathcal{H}(R)} = \sqrt{ {\langle f, f \rangle}_{\mathcal{H}(R)} }\ist\ist$.


For later use, we mention the following result \cite[p.\ 351]{aronszajn1950}.
Consider a kernel function $R(\ist\cdot\ist\ist,\cdot) \!: \mathcal{X} \!\times\! \mathcal{X} \rmv\rightarrow \mathbb{R}$, 
its restriction $R'(\ist\cdot\ist\ist,\cdot) \!: \mathcal{X}' \!\rmv\rmv\times\! \mathcal{X}' \!\rmv\rightarrow \mathbb{R}$ 
to a given subdomain $\mathcal{X}' \!\rmv\rmv\times\! \mathcal{X}'$ with $\mathcal{X}' \!\rmv\subseteq\rmv\rmv\rmv \mathcal{X}$, 
and the corresponding RKHSs $\mathcal{H}(R)$ and $\mathcal{H}(R')$. 
Then, a function $f'(\cdot) \!: \mathcal{X}' \!\rmv\rightarrow\rmv \mathbb{R}$ belongs to $\mathcal{H}(R')$ if and only if there exists a function 
$f(\cdot) \!: \mathcal{X} \!\rightarrow\rmv \mathbb{R}$ belonging to $\mathcal{H}(R)$ whose restriction to $\mathcal{X}'\rmv$, denoted $f(\cdot)\big|_{\mathcal{X}'}$, 
equals $f'(\cdot)$. Thus, $\mathcal{H}(R')$ equals the set of functions 
that is obtained by restricting each function $f(\cdot) \rmv\in\rmv \mathcal{H}(R)$ to the subdomain $\mathcal{X}'\rmv$, i.e.,
\begin{equation} 
\label{equ_restrict}
\mathcal{H}(R') \eq \big\{ f'(\cdot) = f(\cdot)\big|_{\mathcal{X}'} \ist\big| \, f(\cdot) \rmv\in \mathcal{H}(R) \big\} \,.
\end{equation}
Furthermore \cite[p.\ 351]{aronszajn1950}, the norm of a function $f'(\cdot) \rmv\in\rmv \mathcal{H}(R')$ is equal to the minimum of the norms of all functions 
$f(\cdot) \rmv\in\rmv \mathcal{H}(R)$ whose restriction to $\mathcal{X}'\rmv$ equals $f'(\cdot)$, i.e.,
\begin{equation} 
\label{equ_thm_reducing_domain_RKHS}
{\| f'(\cdot) \|}_{\mathcal{H}(R')} \,= \min_{\substack{ \rule{0mm}{2.5mm}f(\cdot) \ist\in\ist \mathcal{H}(R) \\ f(\cdot) \big|_{\mathcal{X}'} =\ist f'(\cdot)}} \!\! {\| f(\cdot) \|}_{\mathcal{H}(R)} \,.
\vspace{-2.5mm}
\end{equation}

\subsection{The RKHS Approach to MVE}
\label{SecRKHSMVE}

RKHS theory provides a powerful mathematical framework for 
MVE 
\cite{Parzen59}. Given an arbitrary estimation problem $\mathcal{E} =\scalarestproblem$ and a parameter vector $\mathbf{x}_{0} \in \mathcal{X}$
for which $f(\mathbf{y};\mathbf{x}_{0}) \not= 0$,  
a kernel function $R_{\mathcal{E},\mathbf{x}_{0}}(\ist\cdot\ist\ist,\cdot)$ and, in turn, an RKHS $\mathcal{H}_{\mathcal{E},\mathbf{x}_{0}}\rmv\rmv$ can be defined as follows. 
We first define the \emph{likelihood ratio} 
\begin{equation} 
\label{equ_def_likelihood}
\llr (\mathbf{y},\mathbf{x}) \,\triangleq\, \frac{f(\mathbf{y};\mathbf{x})}{f(\mathbf{y};\mathbf{x}_{0})} \,, 
\end{equation}
which is considered 
as a random variable (since it is a function of the random vector $\mathbf{y}$) that is parametrized by $\mathbf{x} \rmv\in\rmv \mathcal{X}$. 
Next, we define the Hilbert space $\mathcal{L}_{\mathcal{E}\rmv,\mathbf{x}_{0}}$ as the closure of the linear 
span\footnote{For 
a detailed discussion of the concepts of closure, inner product, orthonormal basis, and linear span in the context of abstract 
Hilbert spaces, see
\cite{Parzen59} and \cite{RudinBook}.} 
of the set of random variables $\big\{ \llr (\mathbf{y},\mathbf{x}) \big\}_{\mathbf{x} 
\in \mathcal{X}}$. The inner product in $\mathcal{L}_{\mathcal{E}\rmv,\mathbf{x}_{0}}$ is defined by 
\[
\big\langle  \llr (\mathbf{y}, \mathbf{x}_{1}) \ist, \llr (\mathbf{y}, \mathbf{x}_{2}) \big\rangle_{\text{RV}} 
\,\triangleq\, \expect_{\mathbf{x}_{0}} \big\{ \llr (\mathbf{y}, \mathbf{x}_{1}) \, \llr (\mathbf{y}, \mathbf{x}_{2}) \big\}
\eq \expect_{\mathbf{x}_{0}} \bigg\{ \frac{ f(\mathbf{y}; \mathbf{x}_{1}) \ist f(\mathbf{y}; \mathbf{x}_{2}) }{ f^2(\mathbf{y}; \mathbf{x}_{0})} \bigg \} \,.
\]
(It can be shown that it is sufficient to define 
${\langle  \cdot \ist, \cdot \rangle}_{\text{RV}}$ 
for the random variables 
$\big\{ \llr(\mathbf{y}, \mathbf{x}) \big\}_{\mathbf{x} \in \mathcal{X}}$ \cite{Parzen59}.)
From now on, we consider only estimation problems $\mathcal{E}=\scalarestproblem$ such that 
$\big\langle  \llr (\mathbf{y}, \mathbf{x}_{1}) \ist, \llr (\mathbf{y}, \mathbf{x}_{2}) \big\rangle_{\text{RV}} < \infty$
for all $\mathbf{x}_{1}, \mathbf{x}_{2} \in \mathcal{X}$, or, equivalently,
\[
\expect_{\mathbf{x}_{0}} \bigg\{ \frac{ f(\mathbf{y}; \mathbf{x}_{1}) \ist f(\mathbf{y}; \mathbf{x}_{2}) }{ f^2(\mathbf{y}; \mathbf{x}_{0})} \bigg\} 
< \infty \,, \quad \text{for all} \;\, 
\mathbf{x}_{1}, \mathbf{x}_{2} \in \mathcal{X} \,.
\] 
Thus, 
${\langle \cdot \ist\ist, \cdot \rangle}_{\text{RV}}$ is well defined.
We can 
interpret the inner product ${\langle \cdot \ist\ist, \cdot \rangle}_{\text{RV}} \!: 
\mathcal{L}_{\mathcal{E}\rmv,\mathbf{x}_{0}} \!\times\rmv \mathcal{L}_{\mathcal{E}\rmv,\mathbf{x}_{0}} \!\rightarrow \mathbb{R}$ as a kernel function 
$R_{\mathcal{E}\rmv,\mathbf{x}_{0}}(\ist\cdot\ist\ist,\cdot) \rmv\rmv: \mathcal{X} \rmv\rmv\rmv\times\rmv\rmv \mathcal{X} \rmv\rmv\rightarrow\rmv \mathbb{R}$:
\be
\label{equ_def_kernel_est_problem}
R_{\mathcal{E}\rmv,\mathbf{x}_{0}}(\mathbf{x}_{1},\mathbf{x}_{2}) \,\triangleq\, \big\langle \llr(\mathbf{y},\mathbf{x}_{1}) \ist, \llr (\mathbf{y}, \mathbf{x}_{2}) \big\rangle_{\text{RV}} 
\eq  \expect_{\mathbf{x}_{0}} \bigg\{  \frac{ f(\mathbf{y}; \mathbf{x}_{1}) \ist f(\mathbf{y}; \mathbf{x}_{2}) }{ f^2(\mathbf{y}; \mathbf{x}_{0})} \bigg\} \,.
\vspace{.5mm}
\ee
The RKHS associated with the estimation problem $\mathcal{E}=\big( Ê\mathcal{X}, f(\mathbf{y}; \mathbf{x}), g(\cdot) \big)$ and the parameter vector 
$\mathbf{x}_{0} \rmv\in\rmv \mathcal{X}$ is then defined to be the RKHS induced by the kernel function $R_{\mathcal{E}\rmv,\mathbf{x}_{0}}(\ist\cdot\ist\ist,\cdot)$. 
We will denote this RKHS as $\mathcal{H}_{\mathcal{E},\mathbf{x}_{0}}$, i.e.,
$\mathcal{H}_{\mathcal{E},\mathbf{x}_{0}} \!\triangleq \mathcal{H}(R_{\mathcal{E}\rmv,\mathbf{x}_{0}})$.
As shown in \cite{Parzen59}, the two Hilbert spaces $\mathcal{L}_{\mathcal{E},\mathbf{x}_{0}}$ and $\mathcal{H}_{\mathcal{E},\mathbf{x}_{0}}$ are isometric,
and a specific congruence, i.e., isometric mapping $\mathsf{J}[\cdot] \rmv\rmv: \mathcal{H}_{\mathcal{E},\mathbf{x}_{0}} \!\rightarrow \mathcal{L}_{\mathcal{E}\rmv,\mathbf{x}_{0}}$
is given by
\[
\mathsf{J}[ R_{\mathcal{E}\rmv,\mathbf{x}_{0}}(\ist\cdot\ist\ist,\mathbf{x})] = \llr (\ist\cdot\ist\ist,\mathbf{x}) \,.  
\]


A fundamental relation of the RKHS $\mathcal{H}_{\mathcal{E},\mathbf{x}_{0}}\rmv\rmv$ with MVE is established by the following central result:

\begin{theorem}[\!\!\cite{Parzen59,Duttweiler73b}]
\label{equ_main_thm_RKHS_MVE}
Consider an estimation problem $\mathcal{E}= \big( \mathcal{X}, f(\mathbf{y}; \mathbf{x}), g(\cdot) \big)$, 
a fixed parameter vector $\mathbf{x}_{0}Ê\rmv\in\rmv \mathcal{X}$, and 
a prescribed bias function $c(\cdot) \!: \mathcal{X} \rmv\rmv\rightarrow \mathbb{R}$, corresponding to the prescribed mean function $\gamma(\cdot) = c(\cdot) + g(\cdot)$.
Then, the following holds:

\begin{enumerate} 

\item The bias function $c(\cdot)$ is valid for $\mathcal{E}$ at $\mathbf{x}_{0}$ if and only if 
$\gamma(\cdot)$ belongs to the RKHS $\mathcal{H}_{\mathcal{E},\mathbf{x}_{0}}$.

\item If the bias function $c(\cdot)$ is valid for $\mathcal{E}$ at $\mathbf{x}_{0}$, the 
minimum achievable variance at $\mathbf{x}_{0}$ (Barankin bound) is given by
\begin{equation}
\label{equ_min_achiev_var_sqared_norm}
\minachievvar(c(\cdot),\mathbf{x}_{0}) \eq {\| \gamma(\cdot) \|}^{2}_{\mathcal{H}_{\mathcal{E},\mathbf{x}_{0}}} \!\rmv- \gamma^{2}(\mathbf{x}_{0}) \,,
\end{equation}
and the LMV estimator at $\mathbf{x}_{0}$ is given 
\vspace{-2.5mm}
by 
\[
\hat{g}^{(c(\cdot),\mathbf{x}_{0})}(\cdot) \ist=\ist \mathsf{J}[\gamma(\cdot)] \,.
\vspace{1mm}
\]

\end{enumerate}

\end{theorem}

Based on Theorem \ref{equ_main_thm_RKHS_MVE}, the following remarks can be made:

\vspace{1mm}

\begin{itemize}

\item The RKHS $\mathcal{H}_{\mathcal{E},\mathbf{x}_{0}}$ can be interpreted as the set of the mean functions $\gamma(\mathbf{x}) = \expect_{\mathbf{x}}Ê\{ \hat{g}(\mathbf{y}) \}$ of 
all estimators $\hat{g}(\cdot)$ with a finite variance at $\mathbf{x}_{0}$, i.e., $v(\hat{g}(\cdot);\mathbf{x}_{0}) < \infty$. 

\vspace{1.5mm}

\item The MVP \eqref{equ_minimum_variance_problem} can be reduced to the computation 
of the squared norm ${\|\gamma(\cdot) \|}^{2}_{\mathcal{H}_{\mathcal{E},\mathbf{x}_{0}}}\!$ and isometric image $\mathsf{J}[\gamma(\cdot)]$ of the prescribed mean 
function $\gamma(\cdot)$, viewed as an element of the RKHS $\mathcal{H}_{\mathcal{E},\mathbf{x}_{0}}$. This theoretical result is especially helpful if a simple characterization of 
$\mathcal{H}_{\mathcal{E},\mathbf{x}_{0}}\rmv$ is available. 
A simple characterization in the sense of \cite{Duttweiler73b} is
given by an orthonormal basis for $\mathcal{H}_{\mathcal{E},\mathbf{x}_{0}}\rmv$ such that the inner products of 
$\gamma(\cdot)$ with the basis functions can be computed easily.

\vspace{1.5mm}

\item 
If a simple characterization of $\mathcal{H}_{\mathcal{E},\mathbf{x}_{0}}\rmv$ is not available, we can still use \eqref{equ_min_achiev_var_sqared_norm}
to establish a large class of lower bounds on the minimum achievable variance $\minachievvar(c(\cdot),\mathbf{x}_{0})$.
Indeed, let $\mathcal{U} \subseteq \mathcal{H}_{\mathcal{E},\mathbf{x}_{0}}\rmv\rmv$ be an arbitrary subspace of $\mathcal{H}_{\mathcal{E},\mathbf{x}_{0}}\rmv\rmv$ 
and let $\mathsf{P}_{\mathcal{U}}Ê\ist\gamma(\cdot)$ denote the orthogonal projection of $\gamma(\cdot)$ onto $\mathcal{U}$. We then have 
${\| \gamma(\cdot) \|}^{2}_{\mathcal{H}_{\mathcal{E},\mathbf{x}_{0}}} \ge {\| \mathsf{P}_{\mathcal{U}} \ist\gamma(\cdot) \|}^{2}_{\mathcal{H}_{\mathcal{E},\mathbf{x}_{0}}}$ 
\cite[Chapter 4]{RudinBook} and thus, from \eqref{equ_min_achiev_var_sqared_norm},
\begin{equation}
\label{equ_lower_bound_min_achiev_var_projection}
 \minachievvar(c(\cdot),\mathbf{x}_{0}) 
 \,\geq\, {\| \mathsf{P}_{\mathcal{U}} \ist\gamma(\cdot) \|}^{2}_{\mathcal{H}_{\mathcal{E},\mathbf{x}_{0}}} \!\rmv-\ist \gamma^{2}(\mathbf{x}_{0}) \,.
\end{equation}
Some well-known lower bounds on the estimator variance, such as the Cram\'{e}r--Rao and Bhattacharya bounds, 
are obtained from \eqref{equ_lower_bound_min_achiev_var_projection} by specific choices of the subspace $\mathcal{U}$ \cite{RKHSExpFamIT2012}.
 
\end{itemize}

\subsection{The RKHS Associated with the LGM} 
\label{SecRKHSLGM}

In our analysis of the SLGM, the RKHS associated with the LGM will play an important role.
Consider $\mathcal{X} \rmv= \mathbb{R}^{N}\rmv\rmv$ and $f(\mathbf{y}; \mathbf{x}) = f_{\mathbf{H}}(\mathbf{y}; \mathbf{x})$
as defined in \eqref{equ_def_LGM_stat_model}, 
where the system matrix $\mathbf{H} \in \mathbb{R}^{M \times N}\rmv\rmv$ is \emph{not} required to satisfy condition \eqref{equ_spark_cond}.
The likelihood ratio \eqref{equ_def_likelihood} for $f(\mathbf{y}; \mathbf{x}) = f_{\mathbf{H}}(\mathbf{y}; \mathbf{x})$ is obtained as
\begin{equation} 
\label{equ_likelihood_LGM}
\rho_{\text{LGM},\mathbf{x}_{0}} (\mathbf{y},\mathbf{x}) \eq \frac{f_{\mathbf{H}}(\mathbf{y};\mathbf{x})}{f_{\mathbf{H}}(\mathbf{y};\mathbf{x}_{0})} 
\eq \exp\bigg(\!\!\rmv-\rmv\frac{1}{2\sigma^{2}} \big[ 2 \mathbf{y}^{T} \mathbf{H}(\mathbf{x}_{0}  \!-\rmv\rmv \mathbf{x}) + \| \mathbf{H} \mathbf{x} \|_{2}^{2} - \| \mathbf{H} \mathbf{x}_{0} \|_{2}^{2} \big] \bigg) \,. 
\end{equation}
Furthermore, from \eqref{equ_def_kernel_est_problem}, the kernel associated with the LGM follows 
\vspace{.5mm}
as
\be
\label{equ_def_kernel_LGM}
R_{\text{LGM},\mathbf{x}_{0}}(\ist\cdot\ist\ist,\cdot) \rmv\rmv: \ist\ist\mathbb{R}^{N} \!\!\times\rmv\rmv \mathbb{R}^{N} \!\rightarrow \mathbb{R} \,; 
  \quad\;\,\ist R_{\text{LGM},\mathbf{x}_{0}}(\mathbf{x}_{1},\mathbf{x}_{2}) \rmv\eq \exp\!\bigg( \frac{1}{\sigma^{2}} (\mathbf{x}_{2} \!-\rmv\rmv \mathbf{x}_{0})^{T} 
    \mathbf{H}^{T} \mathbf{H} \ist (\mathbf{x}_{1} \!-\rmv\rmv \mathbf{x}_{0}) \rmv\bigg) \ist\ist.
\vspace{.5mm}
\ee
 
Let $D \triangleq \rank( \mathbf{H} )$.
We will use the thin singular value decomposition (SVD) of $\mathbf{H}$, i.e., $\mathbf{H} = \mathbf{U} \mathbf{\Sigma} \mathbf{V}^{T}\rmv\rmv$,
where $\mathbf{U} \in \mathbb{R}^{M \times D}\rmv\rmv$ with $\mathbf{U}^{T}\mathbf{U}=\mathbf{I}$, 
$\mathbf{V} \in \mathbb{R}^{N \times D}\rmv\rmv$ with $\mathbf{V}^{T}\mathbf{V}=\mathbf{I}$, and
$\mathbf{\Sigma} \in \mathbb{R}^{D \times D}\rmv\rmv$ is a diagonal matrix with 
positive diagonal entries ${(\mathbf{\Sigma})}_{k,k} > 0$ \cite{golub96}.
The next theorem has been shown in \cite[Sec.\ 5.2]{JungPHD}.

\begin{theorem}
\label{thm_isometry_LGM}
Let $\mathcal{H}_{\text{\emph{LGM}},\mathbf{x}_{0}}$ denote the RKHS associated with the LGM-based estimation problem
$\mathcal{E}_{\text{\emph{LGM}}} = \big(\ist\mathbb{R}^{N}\!,f_{\mathbf{H}}(\mathbf{y}; \mathbf{x}), g(\cdot) \big)$ and the parameter vector $\mathbf{x}_{0} \in \mathbb{R}^{N}\rmv\rmv$,
and let $\widetilde{\mathbf{H}} \triangleq \mathbf{V} \mathbf{ \Sigma}^{-1} \!\rmv \in\rmv \mathbb{R}^{N \times D}\rmv$. 
Then, the following holds:

\begin{enumerate} 

\vspace{1.5mm}

\item Any function $f(\cdot) \in \mathcal{H}_{\emph{LGM},\mathbf{x}_{0}}\rmv\rmv$ is invariant to translations by vectors $\mathbf{x}' \in \mathbb{R}^{N}\rmv\rmv$ 
belonging to the null space of $\mathbf{H}$, i.e., $f(\mathbf{x}) = f(\mathbf{x} + \mathbf{x}')$ for all $\mathbf{x}' \!\in \kernel(\mathbf{H})$ and $\mathbf{x} \rmv\in\rmv \mathbb{R}^{N}\rmv$.  

\vspace{2mm}

\item The RKHS $\mathcal{H}_{\text{\emph{LGM}},\mathbf{x}_{0}}\rmv\rmv$ is isometric to the RKHS $\mathcal{H}(R_\text{\emph{G}})$ whose kernel 
$R_{\text{\emph{G}}}(\ist\cdot\ist\ist,\cdot) \!: \mathbb{R}^{D} \!\times \mathbb{R}^{D} \!\rightarrow \mathbb{R}$ is given 
\vspace{-.5mm}
by  
\[
R_{\text{\emph{G}}}(\zz_{1}, \zz_{2}) \eq \exp \rmv\big( \zz_{1}^{T} \zz_{2} \big) \,, \quad\; \zz_{1}, \zz_{2} \rmv\in\rmv \mathbb{R}^{D} \rmv.
\vspace{-.5mm}
\]
A congruence from $\mathcal{H}(R_{\text{\emph{G}}})$ to $\mathcal{H}_{\text{\emph{LGM}},\mathbf{x}_{0}}\rmv\rmv$
is constituted by the mapping $\mathsf{K}_{\text{\emph{G}}}[\ist\cdot\ist] \!: \mathcal{H}(R_{\text{\emph{G}}}) \rightarrow \mathcal{H}_{\text{\emph{LGM}},\mathbf{x}_{0}}$ given by
\begin{align} 
&\hspace{-10mm}\mathsf{K}_{\text{\emph{G}}}[f(\cdot)] \eq \widetilde{f}(\mathbf{x}) 
\,\triangleq\, f\bigg(\frac{1}{\sigma} \ist\widetilde{\mathbf{H}}^{\dagger} \mathbf{x} \rmv \bigg) \ist \exp\rmv\rmv\bigg( \frac{1}{2 \sigma^{2}} \ist {\| \mathbf{H} \mathbf{x}_{0} \|}^{2}_{2} 
  - \frac{1}{\sigma^{2}} \ist\mathbf{x}^{T} \mathbf{H}^{T}\mathbf{H} \mathbf{x}_{0} \rmv\bigg) \ist\ist, \quad \mathbf{x} \rmv\in\rmv \mathbb{R}^{N} \rmv, \nonumber\\[-2mm]
& \rule{100mm}{0mm}    \text{for all} \;\, f(\cdot) \rmv\in\rmv \mathcal{H}(R_{\text{\emph{G}}}) \,,
\label{equ_def_isometry_LGM}
\end{align} 
and a congruence from $\mathcal{H}_{\text{\emph{LGM}},\mathbf{x}_{0}}\rmv\rmv$ to $\mathcal{H}(R_{\text{\emph{G}}})$ is constituted by the inverse mapping 
$\mathsf{K}^{-1}_{\text{\emph{G}}}[\ist\cdot\ist] \!: \mathcal{H}_{\text{\emph{LGM}},\mathbf{x}_{0}} \!\rightarrow \mathcal{H}(R_{\text{\emph{G}}})$ given by
\begin{align} 
&\hspace{-10mm}\mathsf{K}^{-1}_{\text{\emph{G}}}[\widetilde{f}(\cdot)] \eq f(\zz) 
\eq \widetilde{f}\big(\sigma \widetilde{\mathbf{H}} \zz \big) \ist\exp\rmv\rmv\bigg( \!\!-\rmv\frac{1}{2 \sigma^{2}} \ist {\| \mathbf{H} \mathbf{x}_{0} \|}^{2}_{2} 
  \ist+\ist \frac{1}{\sigma} \ist\zz^{T}  \widetilde{\mathbf{H}}^{\dagger} \mathbf{x}_{0} \rmv\bigg)  \ist\ist, \quad \zz \rmv\in\rmv \mathbb{R}^{D} \rmv, \nonumber\\[-2mm]
& \rule{100mm}{0mm} \text{for all} \;\, \widetilde{f}(\cdot) \rmv\in\rmv \mathcal{H}_{\text{\emph{LGM}},\mathbf{x}_{0}} \,.  
\label{equ_def_isometry_LGM_inverse}
\end{align}  

\vspace{-1mm}

\end{enumerate}

\end{theorem}


The congruence $\mathsf{K}_{\text{G}}$ reduces the characterization of the RKHS $\mathcal{H}_{\text{LGM},\mathbf{x}_{0}}$ to that of the RKHS $\mathcal{H}(R_{\text{G}})$.
A simple characterization (in the sense of an orthonormal basis) of the RKHS $\mathcal{H}(R_{\text{G}})$ can be obtained 
by noting that the kernel $R_{\text{G}}(\ist\cdot\ist\ist,\cdot)$ is infinitely often differentiable and applying the results for RKHSs with differentiable kernels
presented in \cite{zhou_jfaa}. This leads to the following theorem \cite{JungPHD,zhou_jfaa}.

\begin{theorem}
\label{thm_entire_character_H_R_g}

\rule{1mm}{0mm}

\begin{enumerate} 

\vspace{1.5mm}

\item For any $\mathbf{p} \in \mathbb{Z}_+^{D}$, the RKHS $\mathcal{H}(R_{\text{\emph{G}}})$ contains the function $\genericfuncRg(\cdot) \!: \mathbb{R}^{D} \!\rightarrow\rmv \mathbb{R}$ 
given by
\[ 
\genericfuncRg (\zz) \,\triangleq\ist \frac{1}{\sqrt{\mathbf{p}!}} \ist \frac{ \partial^{\mathbf{p}} 
R_{\text{\emph{G}}}(\zz, \zz_{2})}{\partial \zz_{2}^{\mathbf{p}}}\bigg|_{\zz_{2} = \mathbf{0}} 
\rmv=\, \frac{1}{\sqrt{\mathbf{p}!}} \ist\ist \zz^{\mathbf{p}} \ist.
\]

\vspace{1.5mm}

\item The inner product of an arbitrary function $f(\cdot) \in \mathcal{H}(R_{\text{\emph{G}}})$ with $\genericfuncRg(\cdot)$ is given by 
\begin{equation} 
\label{equ_inner_prod_part_der_RKHS_Rg}
\big\langle f (\cdot), \genericfuncRg (\cdot) \big\rangle_{\mathcal{H}(R_{\text{\emph{G}}})} 
  \eq \frac{1}{\sqrt{\mathbf{p}!}} \ist \frac{ \partial^{\mathbf{p}} f(\zz)} {\partial \zz^{\mathbf{p}}} \bigg|_{\zz = \mathbf{0}} . 
\end{equation} 

\vspace{1.5mm}

\item The set of functions $\big\{ \genericfuncRg(\cdot) 
\big\}_{\mathbf{p} \in \mathbb{Z}_{+}^{D}}$ is an orthonormal basis for $\mathcal{H}(R_{\text{\emph{G}}})$. 

\vspace{2mm}

\end{enumerate}

\end{theorem} 


In particular, because of result 3, a function $f(\cdot) \!: \mathbb{R}^{D} \!\rmv\rmv\rightarrow\rmv\rmv \mathbb{R}$ belongs to $\mathcal{H}(R_{\text{G}})$
if and only if it can be written pointwise as 
\begin{equation}
\label{equ_series_repr_RKHS_R_g}
f(\zz) \,= \sum_{\mathbf{p} \,\in \mathbb{Z}_{+}^{D}} \! a[\mathbf{p}] \, \genericfuncRg(\zz) 
  \,= \sum_{\mathbf{p} \in \mathbb{Z}_{+}^{D}} \rmv\rmv \frac{a[\mathbf{p}]}{\sqrt{\mathbf{p}!}} \ist\ist \zz^{\mathbf{p}} \ist,
\end{equation} 
with a unique coefficient sequence $a[\mathbf{p}] \in \ell^{2}(\mathbb{Z}_{+}^{D})$. The coefficient $a[\mathbf{p}]$ is given by \eqref{equ_inner_prod_part_der_RKHS_Rg}, 
\vspace{1mm}
i.e.,
\begin{equation}
\label{equ_series_repr_RKHS_R_g_coeff}
a[\mathbf{p}] \eq \frac{1}{\sqrt{\mathbf{p}!}} \ist \frac{ \partial^{\mathbf{p}} f(\zz)} {\partial \zz^{\mathbf{p}}} \bigg|_{\zz = \mathbf{0}} .
\vspace{1mm}
\end{equation}
Expression \eqref{equ_series_repr_RKHS_R_g} implies
that any $f(\zz) \in \mathcal{H}(R_{\text{G}})$ is infinitely often differentiable and, because of \eqref{equ_series_repr_RKHS_R_g_coeff}, fully determined by 
its partial derivatives at $\zz \!=\! \mathbf{0}$, i.e., 
$\frac{\partial^{\mathbf{p}} f(\zz)}{\partial \zz^{\mathbf{p}} } \big|_{\zz= \mathbf{0}}$ for $\mathbf{p} \in \mathbb{Z}_{+}^{D}$. 
Furthermore, since according to \eqref{equ_def_isometry_LGM} any function $\widetilde{f}(\cdot) \rmv\in\rmv \mathcal{H}_{\text{LGM},\mathbf{x}_{0}}\rmv\rmv$ 
is the image of a function $f(\cdot) \rmv\in\rmv \mathcal{H}(R_{\text{G}})$ under the congruence $\mathsf{K}_{\text{G}}[\ist\cdot\ist]$, 
it follows that also any $\widetilde{f}(\cdot) \rmv\in\rmv \mathcal{H}_{\text{LGM},\mathbf{x}_{0}}\rmv\rmv$ is infinitely often differentiable
and fully determined by its partial derivatives at $\mathbf{x} \!=\! \mathbf{0}$, i.e., 
$\frac{\partial^{\mathbf{p}} \widetilde{f}(\mathbf{x})}{\partial \mathbf{x}^{\mathbf{p}} } \big|_{\mathbf{x}= \mathbf{0}}$ for $\mathbf{p} \in \mathbb{Z}_{+}^{N}$. 
(The latter fact holds because the partial derivatives of $\widetilde{f}(\cdot)$ uniquely determine the partial derivatives of $f(\cdot)= \mathsf{K}^{-1}_{\text{G}}[\widetilde{f}(\cdot)]$ 
via \eqref{equ_def_isometry_LGM_inverse} and the generalized Leibniz rule for the differentiation of a product of functions.)
This agrees with the well-known result \cite[Lemma 2.8]{FundmentExpFamBrown} that for a statistical model of the exponential family type, the mean function of any finite-variance estimator 
is analytic, and thus fully determined by its partial derivatives at zero. 
(To appreciate the connection with the mean function of finite-variance estimators, recall from the discussion
following Theorem \ref{equ_main_thm_RKHS_MVE} that the elements of $\mathcal{H}_{\text{LGM},\mathbf{x}_{0}}\rmv\rmv$ are 
the mean functions of all finite-variance estimators for the LGM, which is a special case of an exponential family.)


\section{RKHS-based Analysis of Minimum Variance Estimation for the SLGM}
\label{sec_RKHS_MVE_SLGM}

In this section, we apply the RKHS framework to
the SLGM-based estimation problem $\mathcal{E}_{\text{SLGM}}=$\linebreak 
$\SLGMscalarestproblem$.
Thus, the parameter set is the set of $S$-sparse vectors, $\mathcal{X} \rmv=\rmv \mathcal{X}_{S} \rmv\subseteq\rmv \mathbb{R}^{N}\rmv$ in \eqref{equ_SLGM_parameter}, 
and the statistical model is given by
$f(\mathbf{y}; \mathbf{x}) \rmv=\rmv f_{\mathbf{H}}(\mathbf{y}; \mathbf{x})$ in \eqref{equ_def_LGM_stat_model}.
More specifically, we consider SLGM-based MVE at a given parameter vector $\mathbf{x}_{0} \!\in\! \mathcal{X}_{S}$, for a prescribed bias function 
$c(\cdot) \rmv: \mathcal{X}_{S} \rmv\rightarrow\rmv \mathbb{R}$. We recall that the set of allowed estimators, $\mathcal{A}(c(\cdot),\mathbf{x}_{0})$, consists of
all estimators $\hat{g}(\cdot)$ with finite variance at $\mathbf{x}_{0}$, i.e., $v(\hat{g}(\cdot);\mathbf{x}_{0}) < \infty$, whose bias function equals $c(\cdot)$, 
i.e., $b(\hat{g}(\cdot); \mathbf{x}) = c(\mathbf{x})$ for all $\mathbf{x} \in \mathcal{X}_{S}$.

Our results can be summarized as follows. We characterize the RKHS associated with the SLGM and employ it to analyze
SLGM-based MVE. 
Using this characterization together with Theorem \ref{equ_main_thm_RKHS_MVE}, we provide conditions on the prescribed bias function $c(\cdot)$ such that the minimum achievable variance 
is finite, i.e., we characterize the set of valid bias functions (cf.\ Section \ref{Sec_MVE}).
Furthermore, we present expressions of the minimum achievable variance (Barankin bound) $\minachievvar_{\text{SLGM}}(c(\cdot),\mathbf{x}_{0})$ 
and of the associated LMV estimator $\hat{g}^{(c(\cdot),\mathbf{x}_{0})}(\cdot)$ for 
an arbitrary valid bias function $c(\cdot)$. Since these expressions 
are difficult to evaluate in general, we finally derive lower bounds on the minimum achievable variance.
These lower bounds
are also lower bounds on the variance 
of any estimator with the prescribed bias function.

\subsection{The RKHS Associated with the SLGM} 
\label{sec_RKHS_SLGM}

Let us consider the SLGM-based estimation problem
$\mathcal{E}_{\text{SLGM}}=\SLGMscalarestproblem$ and the corresponding LGM-based estimation problem
$\mathcal{E}_{\text{LGM}} \rmv\rmv=\! \big(\mathbb{R}^{N}\!,f_{\mathbf{H}}(\mathbf{y}; \mathbf{x}), g(\cdot) \big)$ with the same 
system matrix $\mathbf{H} \!\in\! \mathbb{R}^{M \times N}\rmv\rmv$ satisfying condition \eqref{equ_spark_cond} and with the same noise variance $\sigma^2\rmv$.
For an $S$-sparse parameter vector $\mathbf{x}_{0} \!\in\! \mathcal{X}_S$, let
$\mathcal{H}_{\text{SLGM},\mathbf{x}_{0}}\rmv\rmv$ and $\mathcal{H}_{\text{LGM},\mathbf{x}_{0}}\rmv\rmv$ denote the RKHSs associated with the estimation problems
$\mathcal{E}_{\text{SLGM}}$ and $\mathcal{E}_{\text{LGM}}$, respectively. Using \eqref{equ_def_kernel_est_problem} and \eqref{equ_def_LGM_stat_model},
the kernel underlying $\mathcal{H}_{\text{SLGM},\mathbf{x}_{0}}\rmv\rmv$ is obtained as
\be 
R_{\text{SLGM},\mathbf{x}_{0}}(\ist\cdot\ist\ist,\cdot) \rmv\rmv: \mathcal{X}_{S} \rmv\rmv\times\! \mathcal{X}_{S} \rightarrow \mathbb{R} \,; 
  \quad\; R_{\text{SLGM},\mathbf{x}_{0}}(\mathbf{x}_{1},\mathbf{x}_{2}) \rmv\eq \exp\!\bigg( \frac{1}{\sigma^{2}} (\mathbf{x}_{2} \!-\rmv\rmv \mathbf{x}_{0})^{T} 
    \mathbf{H}^{T} \mathbf{H} \ist (\mathbf{x}_{1} \!-\rmv\rmv \mathbf{x}_{0}) \rmv\bigg) \ist\ist .
\label{equ_def_kernel_SLGM} 
\ee
Comparing with the kernel $R_{\text{LGM},\mathbf{x}_{0}}(\ist\cdot\ist\ist,\cdot)$ underlying $\mathcal{H}_{\text{LGM},\mathbf{x}_{0}}\ist$, 
which was presented in \eqref{equ_def_kernel_LGM},
we conclude that $R_{\text{SLGM},\mathbf{x}_{0}}(\ist\cdot\ist\ist,\cdot)$ is the restriction of 
$R_{\text{LGM},\mathbf{x}_{0}}(\ist\cdot\ist\ist,\cdot)$ to the subdomain 
$\mathcal{X}_{S} \rmv\rmv\times\! \mathcal{X}_{S} \subseteq \mathbb{R}^{N} \!\!\times\rmv\rmv \mathbb{R}^{N}\rmv$.

The characterization of $\mathcal{H}_{\text{LGM},\mathbf{x}_{0}}\rmv\rmv$ provided by Theorems \ref{thm_isometry_LGM} and \ref{thm_entire_character_H_R_g}
is also relevant to $\mathcal{H}_{\text{SLGM},\mathbf{x}_{0}}$.
This is due to the following 
application of the ``RKHS restriction result'' in Section \ref{SecRKHSbasics} 
(see \eqref{equ_restrict} and \eqref{equ_thm_reducing_domain_RKHS}):

\begin{corollary}
\label{thm_relation_RKHS_LGM_SLM}
The RKHS $\mathcal{H}_{\text{\emph{SLGM}},\mathbf{x}_{0}}\rmv\rmv$ consists of the restrictions of all functions $f(\cdot)\rmv\rmv: \mathbb{R}^{N} \!\rmv\rightarrow \mathbb{R}$
contained in 
$\mathcal{H}_{\emph{\text{LGM}},\mathbf{x}_{0}}\rmv\rmv$ to the subdomain $\mathcal{X}_{S} \rmv\subseteq \mathbb{R}^{N}\rmv$, i.e., 
\[
\mathcal{H}_{\text{\emph{SLGM}},\mathbf{x}_{0}} \eq \big\{ f'(\cdot) = f(\cdot)\big|_{\mathcal{X}_{S}} \ist\big| \, f(\cdot) \rmv\in \mathcal{H}_{\emph{\text{LGM}},\mathbf{x}_{0}} \big\} \,.
\]
Furthermore, the norm of a function $f'(\cdot) \rmv\in\rmv \mathcal{H}_{\text{\emph{SLGM}},\mathbf{x}_{0}}\rmv\rmv$ is equal to the minimum of the norms of all functions 
$f(\cdot) \rmv\in\rmv \mathcal{H}_{\emph{\text{LGM}},\mathbf{x}_{0}}\rmv\rmv$ whose restriction to $\mathcal{X}_{S}$ equals $f'(\cdot)$, i.e.,
\begin{equation} 
\label{equ_relation_norm_SLM_norm_LGM}
{\| f'(\cdot) \|}_{\mathcal{H}_{\text{\emph{SLGM}},\mathbf{x}_{0}}} 
  \rmv\rmv=\rmv \min_{\substack{ \rule{0mm}{2.5mm}f(\cdot) \ist\in\ist \mathcal{H}_{\text{\emph{LGM}},\mathbf{x}_{0}} \\ f(\cdot) \big|_{\mathcal{X}_{S}} =\ist f'(\cdot)}} \!\! 
    {\| f(\cdot) \|}_{\mathcal{H}_{\text{\emph{LGM}},\mathbf{x}_{0}}} .
\vspace{1mm}
\end{equation}
\end{corollary}

An immediate consequence of Corollary \ref{thm_relation_RKHS_LGM_SLM} is the obvious\footnote{Indeed, prescribing the bias for all $\mathbf{x} \in \mathbb{R}^{N}$ (as is done within the LGM), instead of prescribing it only for the sparse vectors $\mathbf{x} \in \mathcal{X}_{S}$ (as is done within the SLGM) can only result in a higher (or equal) minimum achievable variance.} fact that the minimum achievable variance for the SLGM can never exceed that 
for the LGM (if the prescribed bias 
function for the SLGM is the restriction of the prescribed bias 
function for the LGM). Indeed, letting $c(\cdot)\rmv\rmv: \mathbb{R}^{N} \!\rmv\rightarrow \mathbb{R}$ be the prescribed bias function for the LGM 
and $\gamma(\cdot)= c(\cdot) + g(\cdot)$ the corresponding mean function, and recalling that $\mathbf{x}_{0} \!\in\! \mathcal{X}_{S}$, we 
\vspace{1mm}
have
\[ 
\minachievvar_{\text{SLGM}}\big(c(\cdot) \big|_{\mathcal{X}_{S}}, \mathbf{x}_{0} \big)
\ist\ist\stackrel{\eqref{equ_min_achiev_var_sqared_norm}}{=}\ist\ist\ist
\big\| \gamma(\cdot) \big|_{\mathcal{X}_{S}} \big\|^{2}_{ \mathcal{H}_{\text{{SLGM}},\mathbf{x}_{0}}} 
   \!\!- \gamma^{2}(\mathbf{x}_{0})
\ist\ist\stackrel{\eqref{equ_relation_norm_SLM_norm_LGM}}{\leq}\ist\ist
{\| \gamma(\cdot) \|}^{2}_{ \mathcal{H}_{\text{{LGM}},\mathbf{x}_{0}}} \!\!- \gamma^{2}(\mathbf{x}_{0})
   \ist\ist\stackrel{\eqref{equ_min_achiev_var_sqared_norm}}{=}\ist\ist 
\minachievvar_{\text{LGM}}(c(\cdot), \mathbf{x}_{0}) \,.
\vspace{1mm}
\] 

Thus, in the precise sense of Corollary \ref{thm_relation_RKHS_LGM_SLM}, $\mathcal{H}_{\text{SLGM},\mathbf{x}_{0}}\rmv\rmv$ is
the restriction of $\mathcal{H}_{\text{LGM},\mathbf{x}_{0}}\rmv\rmv$ to the set $\mathcal{X}_S$ of $S$-sparse parameter vectors, and the characterization of 
$\mathcal{H}_{\text{LGM},\mathbf{x}_{0}}\rmv\rmv$ provided by Theorems \ref{thm_isometry_LGM} and \ref{thm_entire_character_H_R_g} can also be used for a
characterization of $\mathcal{H}_{\text{SLGM},\mathbf{x}_{0}}$.
In what follows, we will employ this principle for developing an RKHS-based analysis of MVE for the SLGM. 
Proofs of the presented results can be found in \cite{JungPHD}.
As before, we will use the thin SVD of the system matrix $\mathbf{H}$, i.e., 
$\mathbf{H} = \mathbf{U} \mathbf{\Sigma} \mathbf{V}^{T}\!$, as well as the shorthand notations $\widetilde{\mathbf{H}} = \mathbf{V} \mathbf{ \Sigma}^{-1}\rmv$
and $D = \rank(\mathbf{H})$.
 

\subsection{The Class of Valid Bias Functions} 
\label{sec_RKHS_basic_facts_SLGM}

The class of valid bias functions for the SLGM-based estimation problem 
$\mathcal{E}_{\text{SLGM}}=\SLGMscalarestproblem$ at $\mathbf{x}_{0} \!\in\! \mathcal{X}_{S}$ is characterized by
the following result \cite[Thm. 5.3.1]{JungPHD}:
 
\begin{theorem} 
\label{thm_condition_gamma_valid_SLM_fourier_series_R_g}
A bias function $c(\cdot) \rmv\rmv: \mathcal{X}_{S} \rmv\rightarrow \mathbb{R}$ is valid for $\mathcal{E}_{\emph{\text{SLGM}}}=\SLGMscalarestproblem$
at $\mathbf{x}_{0}\!\in\! \mathcal{X}_{S}$ if and only if it 
can be expressed 
\vspace{.5mm}
as
\begin{equation} 
\label{equ_condition_gamma_valid_SLM_fourier_series_R_g}
c(\mathbf{x}) \eq \exp\rmv\rmv\bigg( \frac{1}{2 \sigma^{2}} \ist {\| \mathbf{H} \mathbf{x}_{0} \|}^{2}_{2} 
  - \frac{1}{\sigma^{2}} \ist\mathbf{x}^{T} \mathbf{H}^{T}\mathbf{H} \mathbf{x}_{0} \rmv\bigg)
\rmv \sum_{\mathbf{p} \in \mathbb{Z}_{+}^{D}} \rmv\rmv \frac{a[\mathbf{p}]}{\sqrt{\mathbf{p}!}} \ist\ist \bigg( \frac{1}{\sigma} \widetilde{\mathbf{H}}^{\dagger} \mathbf{x} \rmv\bigg)^{\!\mathbf{p}} 
  \!- g(\mathbf{x}) \,, \quad\; \mathbf{x} \!\in\! \mathcal{X}_{S} \,,
\vspace{-.5mm}
\end{equation} 
with some coefficient sequence $a[\mathbf{p}] \in \ell^{2}(\mathbb{Z}_{+}^{D})$.
\end{theorem} 

Theorem \ref{thm_condition_gamma_valid_SLM_fourier_series_R_g} implies that the mean function 
$\gamma(\cdot) = c(\cdot) + g(\cdot)$ corresponding to a bias function $c(\cdot)$ that is valid for $\mathcal{E}_{\text{SLGM}}$ at $\mathbf{x}_{0} \!\in\! \mathcal{X}_{S}$ is of the 
\vspace{.5mm}
form 
\begin{equation} 
\label{equ_condition_gamma_valid_SLM_expression_analytic}
\gamma(\mathbf{x}) \eq \exp\rmv\rmv\bigg( \frac{1}{2 \sigma^{2}} \ist {\| \mathbf{H} \mathbf{x}_{0} \|}^{2}_{2} 
  - \frac{1}{\sigma^{2}} \ist\mathbf{x}^{T} \mathbf{H}^{T}\mathbf{H} \mathbf{x}_{0} \rmv\bigg)
\rmv \sum_{\mathbf{p} \in \mathbb{Z}_{+}^{D}} \rmv\rmv \frac{a[\mathbf{p}]}{\sqrt{\mathbf{p}!}} \ist\ist \bigg( \frac{1}{\sigma} \widetilde{\mathbf{H}}^{\dagger} \mathbf{x} \rmv\bigg)^{\!\mathbf{p}}, 
  \quad\; \mathbf{x} \!\in\! \mathcal{X}_{S} \,,
\vspace{-1mm}
\end{equation} 
with some coefficient sequence $a[\mathbf{p}] \in \ell^{2}(\mathbb{Z}_{+}^{D})$. 
The function on the right-hand side in \eqref{equ_condition_gamma_valid_SLM_expression_analytic} is \emph{analytic} on the domain $\mathcal{X}_{S}$
in the 
sense\footnote{Note 
that a function with domain $\mathcal{X}_{S}$, with $S \!<\! N$, cannot be analytic in the conventional sense
since the domain of an analytic function has to be open by definition 
\cite[Definition 2.2.1]{KranzPrimerAnalytic}.} 
that it can be locally represented at any point $\mathbf{x} \!\in\! \mathcal{X}_{S}$ by a convergent power series.
Thus, in particular, the mean function $\gamma(\mathbf{x}) = \expect_{\mathbf{x}} \{ \hat{g}(\mathbf{y}) \}$ of any finite-variance estimator $\hat{g}(\mathbf{y})$ is necessarily 
an ``analytic'' function. Again, this agrees with the general result about the mean function of estimators for exponential families presented in \cite[Lemma 2.8]{FundmentExpFamBrown}.
(Note that the statistical model of the SLGM is a special case of an exponential family.) 

In the special case where $g(\mathbf{x}) = x_{k}$ for some $k \!\in\! [N]$,
a sufficient condition on a bias function to be valid is stated as follows \cite[Thm. 5.3.4]{JungPHD}:

\begin{theorem} 
\label{thm_suff_cond_valid_erervwhere_SLM_power_series}
The function
\begin{equation}
\label{equ_prescr_bias_finite_order_polynom_SLM_112}
c(\mathbf{x}) \eq \exp \rmv\rmv \big( \mathbf{x}_{1}^{T}  \widetilde{\mathbf{H}}^{\dagger} \mathbf{x} \big) 
  \rmv\sum_{\mathbf{p} \in \mathbb{Z}_{+}^{D}} \! \frac{a[\mathbf{p}]}{\mathbf{p}!}   
  \bigg( \frac{1}{\sigma} \widetilde{\mathbf{H}}^{\dagger} \mathbf{x}\rmv \bigg)^{\!\mathbf{p}} 
  \!-\ist x_{k} \,, \quad\; \mathbf{x} \!\in\! \mathcal{X}_{S} \,,
\vspace{-.7mm}
\end{equation}
with an arbitrary $\mathbf{x}_{1} \!\in\rmv \mathbb{R}^{D}$ and coefficients $a[\mathbf{p}]$ satisfying 
$| a[\mathbf{p}] | \leq C^{|\mathbf{p}|}$ with an arbitrary constant $C \rmv\in\rmv \mathbb{R}_{+}\ist\ist$, is a valid bias function for 
$\mathcal{E}_{\emph{SLGM}} = \big(\mathcal{X}_{S},f_{\mathbf{H}}(\mathbf{y}; \mathbf{x}),g(\mathbf{x}) \!=\! x_{k} \big)$ at any 
$\mathbf{x}_{0} \!\in\! \mathcal{X}_{S}$. In particular, for $\mathbf{H} \! = \! \mathbf{I}$, the unbiased case (i.e., $c(\mathbf{x}) \equiv 0$) is 
obtained for $\mathbf{x}_{1} = \mathbf{0}$, $a[\mathbf{e}_{k}] = \sigma$, and $a[\mathbf{p}] = 0$ for all other $\mathbf{p} \in  \mathbb{Z}_{+}^{D}$.
\end{theorem}

Note that the difference of the factors in \eqref{equ_prescr_bias_finite_order_polynom_SLM_112} compared to the factors in \eqref{equ_condition_gamma_valid_SLM_fourier_series_R_g}
(i.e., $\frac{a[\mathbf{p}]}{\mathbf{p}!}$ instead of $\frac{a[\mathbf{p}]}{\sqrt{\mathbf{p}!}}$) is in accordance with the different
condition on the coefficient sequence $a[\mathbf{p}]$ 
(i.e., $| a[\mathbf{p}] | \leq C^{|\mathbf{p}|}\rmv\rmv$ instead of $a[\mathbf{p}] \in \ell^{2}(\mathbb{Z}_{+}^{D})$).

\subsection{Minimum Achievable Variance (Barankin Bound) and LMV Estimator} 
\label{sec_Barankin_LMV}

Let us consider the MVP \eqref{equ_minimum_variance_problem} at a given parameter vector $\mathbf{x}_{0} \in \mathcal{X}_{S}$
for an SLGM-based estimation problem $\mathcal{E}_{\text{SLGM}} \ist\triangleq\ist \SLGMscalarestproblem$ and 
for a prescribed bias function $c(\cdot)\rmv\rmv : \mathcal{X}_{S} \rmv\rightarrow\rmv \mathbb{R}$, which is known to be valid. 
Then, the minimum achievable variance (Barankin bound) at $\mathbf{x}_{0}$, denoted $\SLGMMinAchVar$ (cf.\ \eqref{equ_minimum_variance_problem}), 
and the corresponding LMV estimator $\hat{g}^{(c(\cdot),\mathbf{x}_{0})}(\cdot)$ (cf.\ \eqref{equ_def_LMV_estimator}) are characterized by the following 
\vspace{.5mm}
theorem \cite[Thm. 5.3.1]{JungPHD}.

\begin{theorem}
\label{thm_Barankin_bound_LMV}


Consider an SLGM-based estimation problem $\mathcal{E}_{\text{\emph{SLGM}}} = \SLGMscalarestproblem$
and a valid prescribed bias function $c(\cdot)\rmv\rmv : \mathcal{X}_{S} \rmv\rightarrow\rmv \mathbb{R}\ist$. Then:

\begin{enumerate} 

\vspace{1.5mm}

\item The minimum achievable variance at $\mathbf{x}_{0} \!\in\! \mathcal{X}_{S}$ 
is given 
by
\be
\minachievvar_{\emph{SLGM}}(c(\cdot),\mathbf{x}_{0}) \,= \min_{ a[\cdot] \ist\in\ist \mathcal{C}(c)} \rmv\rmv
{\| a[\cdot] \|}^{2}_{\ell^{2}(\mathbb{Z}_{+}^{D})} \ist-\ist \gamma^2(\mathbf{x}_{0}) \,,
\label{equ_M_min_a_gamma}
\ee 
where $\gamma(\cdot) = c(\cdot) + g(\cdot)\ist$, ${\| a[\cdot] \|}^{2}_{\ell^{2}(\mathbb{Z}_{+}^{D})} \rmv\rmv\triangleq\rmv \sum_{\mathbf{p} \in \mathbb{Z}_{+}^{D}} a^2[\mathbf{p}]\ist$, and
$\mathcal{C}(c) \!\subseteq\rmv\rmv \ell^{2}(\mathbb{Z}_{+}^{D})$ denotes the set of coefficient sequences $a[\mathbf{p}] \in \ell^{2}(\mathbb{Z}_{+}^{D})$ that are consistent with 
\eqref{equ_condition_gamma_valid_SLM_fourier_series_R_g}.

\vspace{2mm}

\item 
The function $\hat{g}(\cdot) \rmv\rmv : \mathbb{R}^{M} \!\rightarrow\rmv \mathbb{R}$ given by 
\begin{equation} 
\label{equ_est_arb_coeffs_SLM}
\hat{g}(\mathbf{y}) \eq \exp\rmv\rmv\bigg( \!\!-\rmv \frac{1}{2 \sigma^{2}} \ist {\| \mathbf{H} \mathbf{x}_{0} \|}^{2}_{2} \rmv\bigg)  
\rmv \sum_{\mathbf{p} \in \mathbb{Z}_{+}^{D}} \rmv\rmv \frac{a[\mathbf{p}]}{\sqrt{\mathbf{p}!}} \, \chi_{\mathbf{p}}(\mathbf{y}) \,, 
\vspace{-1.5mm}
\end{equation}
with an arbitrary coefficient sequence $a[\cdot] \rmv\rmv\in\rmv\rmv \mathcal{C}(c)$ and
\vspace{1.5mm}
\[
\chi_{\mathbf{p}}(\mathbf{y}) \,\triangleq\, \frac{ \partial^{\mathbf{p}} \big[ \rho_{\emph{LGM},\mathbf{x}_{0}}(\mathbf{y},\sigma \widetilde{\mathbf{H}} \mathbf{z}) 
  \ist\exp\rmv\rmv \big( \frac{1}{\sigma} \ist \mathbf{x}_{0}^{T} \mathbf{H}^{T} \mathbf{H}
  \widetilde{\mathbf{H}} \mathbf{z} \big) \big]}{ \partial \mathbf{z}^{\mathbf{p}}} \bigg|_{\mathbf{z} = \mathbf{0}} \,,
\vspace{1mm}
\]
where \emph{$\rho_{\text{LGM},\mathbf{x}_{0}} (\mathbf{y},\mathbf{x})$} is given by \eqref{equ_likelihood_LGM},
is an allowed estimator at $\mathbf{x}_{0}$ for 
$c(\cdot)$, i.e., $\hat{g}(\cdot) \in \mathcal{A}(c(\cdot), \mathbf{x}_{0})$.

\vspace{2mm}

\item The LMV estimator at $\mathbf{x}_{0}$,
$\hat{g}^{(c(\cdot),\mathbf{x}_{0})}(\cdot)$, is given by
\eqref{equ_est_arb_coeffs_SLM} using the specific coefficient sequence $a_{0}[\mathbf{p}] = \argmin_{a[\cdot] \in \mathcal{C}(c)} \rmv\rmv {\| a[\cdot] \|}_{\ell^{2}(\mathbb{Z}_{+}^{D})}\ist$.

\vspace{1.5mm}

\end{enumerate} 
 
\end{theorem} 

The kernel $R_{\text{SLGM},\mathbf{x}_{0}}(\ist\cdot\ist\ist,\cdot)$ given by \eqref{equ_def_kernel_SLGM} is pointwise continuous with respect to the parameter $\mathbf{x}_{0}$, i.e., 
$\lim_{\mathbf{x}'_{0} \rightarrow \mathbf{x}_{0}} R_{\text{SLGM},\mathbf{x}_{0}'}(\mathbf{x}_{1}, \mathbf{x}_{2}) 
= R_{\text{SLGM},\mathbf{x}_{0}}(\mathbf{x}_{1}, \mathbf{x}_{2})$ for all $\mathbf{x}_{0}, \mathbf{x}_{1}, \mathbf{x}_{2} \in \mathcal{X}_{S}$.
Therefore, applying \cite[Thm. 4.3.6]{JungPHD} or \cite[Thm. IV.6]{RKHSExpFamIT2012} to the SLGM yields the following result.

\begin{corollary}
\label{cor_lower_semi_cont_SLGM}
Consider the SLGM with parameter function $g(\mathbf{x}) \!=\! x_{k}\ist$ 
and a prescribed bias function $c(\cdot) \rmv\rmv : \mathcal{X}_{S} \rmv\rightarrow\rmv \mathbb{R}$ that is valid for  
$\mathcal{E}_{\emph{SLGM}} = \big(\mathcal{X}_{S},f_{\mathbf{H}}(\mathbf{y}; \mathbf{x}), g(\mathbf{x}) \rmv\rmv=\rmv\rmv x_{k} \big)$ 
at each parameter vector $\mathbf{x}_{0} \in \mathcal{X}_{S}$. Then if $c(\cdot)$ is continuous, 
the minimum achievable variance $\minachievvar_{\emph{SLGM}}(c(\cdot), \mathbf{x}_{0})$ is a lower 
semi-continuous\footnote{A definition of lower semi-continuity can be found in \cite{RudinBookPrinciplesMatheAnalysis}.} 
function of $\mathbf{x}_{0}$. 
\end{corollary}

From Corollary \ref{cor_lower_semi_cont_SLGM}, we can conclude that the sparse CRB derived in \cite{ZvikaCRB} is not tight, i.e., 
it is not equal to the minimum achievable variance $\minachievvar_{\text{SLGM}}(c(\cdot),\mathbf{x}_{0})$. Indeed, the sparse CRB 
is in general a strictly upper semi-continuous function of the parameter vector $\mathbf{x}_{0}$, 
whereas the minimum achievable variance $\minachievvar_{\text{SLGM}}(c(\cdot),\mathbf{x}_{0})$ is lower semi-continuous according to Corollary \ref{cor_lower_semi_cont_SLGM}. Since a function cannot be simultaneously strictly upper semi-continuous and lower semi-continuous, the sparse CRB cannot be equal to $\minachievvar_{\text{SLGM}}(c(\cdot),\mathbf{x}_{0})$ in general.


\section{Lower Variance Bounds for the SLGM}
\label{sec_bounds_SLGM}

While Theorem \ref{thm_Barankin_bound_LMV} provides a mathematically complete characterization of 
the minimum achievable variance and the LMV estimator, the corresponding expressions are somewhat  
difficult to evaluate in general. Therefore, we will next derive lower bounds on the minimum achievable variance $\SLGMMinAchVar$ for the 
estimation problem $\mathcal{E}_{\text{SLGM}} = \big(\mathcal{X}_{S},f_{\mathbf{H}}(\mathbf{y}; \mathbf{x}),g(\mathbf{x}) \!=\! x_{k} \big)$ with some $k \!\in\! [N]$
and for a prescribed bias function $c(\cdot)$.
These bounds are easier to evaluate. As mentioned before, they
are also lower bounds on the variance 
of any estimator having the prescribed bias function.
Our assumption that $g(\mathbf{x}) \!=\! x_{k}$ is no restriction because, according to \cite[Thm. 2.3.1]{JungPHD}, 
the MVP for a given parameter function $g(\mathbf{x})$ and prescribed bias function $c(\mathbf{x})$ is 
equivalent to the MVP 
for parameter function $g'(\mathbf{x}) = x_{k}$ and prescribed bias function $c'(\mathbf{x}) = c(\mathbf{x}) + g(\mathbf{x}) - x_{k}$.
In particular,\footnote{Indeed, 
if 
$c'(\mathbf{x})$ is valid at $\mathbf{x}_{0}$ for the MVP with parameter function $x_{k}$, 
there exists a finite-variance estimator $\hat{g}(\cdot)$ with mean function $\expect_{\mathbf{x}} \{ \hat{g}(\mathbf{y}) \} = c'(\mathbf{x})+x_{k}$. 
For the MVP with parameter function $g(\cdot)$, that estimator $\hat{g}(\cdot)$ has the bias function
\[
b(\hat{g}(\cdot), \mathbf{x}) \eq \expect_{\mathbf{x}} \{ \hat{g}(\mathbf{y}) \} - g(\mathbf{x}) 
  \eq c'(\mathbf{x})+x_{k} - g(\mathbf{x}) 
  \eq c(\mathbf{x}) \,.
\]
Thus, there exists a finite-variance estimator with bias function $c(\mathbf{x}) = c'(\mathbf{x}) - g(\mathbf{x}) + x_{k}$, which implies that the bias function $c(\cdot)$ is 
valid for the MVP with parameter function 
$g(\cdot)$.} 
if 
$c'(\mathbf{x})$ is valid for the MVP with parameter function $g'(\mathbf{x}) = x_{k}$, then $c(\mathbf{x}) = c'(\mathbf{x}) - g(\mathbf{x}) + x_{k}$ 
is valid for the MVP with parameter function $g(\mathbf{x})$.
Therefore, any MVP can be reduced to an equivalent MVP with $g(\mathbf{x}) = x_{k}$ and an appropriately modified prescribed bias function. 

We assume that the prescribed bias function $c(\cdot)$ is valid for $\mathcal{E}_{\text{SLGM}} = \big(\mathcal{X}_{S},f_{\mathbf{H}}(\mathbf{y}; \mathbf{x}),g(\mathbf{x}) \!=\! x_{k} \big)$.
This validity assumption is no real restriction either, since our lower bounds 
are finite and therefore are lower bounds also if $\SLGMMinAchVar =\infty$, which, by our definition in Section \ref{Sec_MVE}, is the case if $c(\cdot)$ is not valid.

The 
lower bounds to be presented are 
based on the generic lower bound \eqref{equ_lower_bound_min_achiev_var_projection}, i.e., they are of the form
\begin{equation}
\label{equ_projection_theorem_lower_bound_min_ach_variance}
\SLGMMinAchVar \,\geq\, {\| \mathsf{P}_{\mathcal{U}} \ist\gamma(\cdot) \|}^{2}_{\RKHSSLGM} \!\!-\ist \gamma^{2}(\mathbf{x}_{0}) \,,
\end{equation}
for some subspace $\mathcal{U} \subseteq \RKHSSLGM$.
Here, the prescribed mean function $\gamma(\cdot) \rmv\rmv: \mathcal{X}_{S} \rmv\rightarrow\rmv \mathbb{R}$, given by $\gamma(\mathbf{x}) = c(\mathbf{x}) + x_k$,
is an element of $\RKHSSLGM\rmv\rmv$ since $c(\cdot)$ is assumed valid (recall Theorem \ref{equ_main_thm_RKHS_MVE}).

\subsection{The Sparse CRB} 
\label{sec_RKHS_Lower_Bounds_SLGM}

The first bound is an adaptation of the CRB \cite{kay,LC,scharf91,RKHSExpFamIT2012} to the sparse setting and has been previously derived in a slightly different form in \cite{ZvikaCRB}.

\begin{theorem}
\label{thm_CRB_SLM}
Consider the estimation problem $\mathcal{E}_{\emph{SLGM}} = \big(\mathcal{X}_{S},f_{\mathbf{H}}(\mathbf{y}; \mathbf{x}),g(\mathbf{x}) \!=\! x_{k} \big)$ 
with a system matrix $\mathbf{H} \in \mathbb{R}^{M \times N}\rmv\rmv$ satisfying \eqref{equ_spark_cond}. Let $\mathbf{x}_{0} \rmv\in\rmv \mathcal{X}_{S}$.
If the prescribed bias function $c(\cdot) \rmv\rmv: \mathcal{X}_{S} \rmv\rightarrow\rmv \mathbb{R}$ is such that the partial derivatives 
$\frac{\partial c(\mathbf{x})}{\partial x_{l}} \big|_{\mathbf{x} = \mathbf{x}_{0}}$ exist for all $l \rmv\in\rmv [N]$, then 
\be 
\minachievvar_{\emph{SLGM}}(c(\cdot),\mathbf{x}_{0}) \,\geq\ist \begin{cases}
  \sigma^{2} \ist\ist \mathbf{b}^{T} \rmv ( \mathbf{H}^{T} \mathbf{H} )^{\dagger} \ist \mathbf{b} \,, 
    & \mbox{if}\,\, {\| \mathbf{x}_{0} \|}_{0} \le S\!-\!1 \\[-1mm]Ê
  \sigma^{2} \ist\ist \mathbf{b}_{\mathbf{x}_{0}}^{T} \rmv ( \mathbf{H}_{\mathbf{x}_{0}}^{T} \mathbf{H}_{\mathbf{x}_{0}} )^{\dagger} \ist \mathbf{b}_{\mathbf{x}_{0}} \,, 
    & \mbox{if}\,\, {\| \mathbf{x}_{0} \|}_{0} = S \,. 
  \end{cases}
\label{equ_CRB_SLM}
\ee 
Here, in the case ${\| \mathbf{x}_{0} \|}_{0} \rmv<\rmv S$, $\mathbf{b} \rmv\in\rmv \mathbb{R}^{N}\rmv\rmv$ is given 
by $b_{l} \triangleq \delta_{k,l} + \frac{\partial c(\mathbf{x})}{\partial x_{l}} \big|_{\mathbf{x} = \mathbf{x}_{0}}$, $l \rmv\in\rmv [N]$, 
and in the case ${\| \mathbf{x}_{0} \|}_{0} \rmv=\rmv S$,
$\mathbf{b}_{\mathbf{x}_{0}} \rmv\!\in\rmv \mathbb{R}^{S}\rmv\rmv$ and $\mathbf{H}_{\mathbf{x}_{0}} \rmv\!\in\rmv \mathbb{R}^{M \times S}\rmv\rmv$ consist of those entries of 
$\mathbf{b}$ and columns of $\mathbf{H}$, respectively that are indexed by $\supp(\mathbf{x}_{0}) \equiv \{k_{1},\ldots,k_{S}\}$, i.e., 
${(\mathbf{b}_{\mathbf{x}_{0}})}_{i} \rmv=\rmv b_{k_{i}}$ and ${(\mathbf{H}_{\mathbf{x}_{0}} )}_{m,i} \rmv=\rmv {(\mathbf{H})}_{m,k_{i}}$, $i \rmv\in\rmv [S]$. 
\end{theorem} 

A proof of this theorem is given in \cite[Thm. 5.4.1]{JungPHD}. There, it is shown that the bound \eqref{equ_CRB_SLM} for ${\| \mathbf{x}_{0} \|}_{0} \rmv<\rmv S$ 
is obtained from the generic bound \eqref{equ_projection_theorem_lower_bound_min_ach_variance} using the subspace 
$\mathcal{U}  = \linspan\rmv\rmv \big\{ \genericfuncbound_{0}(\cdot), {\{ \genericfuncbound_{l}(\cdot) \}}_{l \in [N]}Ê\big\}$, where
\[
\genericfuncbound_{0} (\cdot) \ist\ist\triangleq\ist\ist R_{\text{SLGM},\mathbf{x}_{0}} (\ist\cdot\ist\ist, \mathbf{x}_{0}) \,, \qquad\;
\genericfuncbound_{l}(\cdot) \ist\ist\triangleq\ist\ist \frac{ \partial R_{\text{SLGM},\mathbf{x}_{0}} (\ist\cdot\ist\ist, \mathbf{x}_{2})}  
  { \partial {(\mathbf{x}_{2})}_{l} } \bigg|_{\mathbf{x}_{2} = \mathbf{x}_{0}} \rmv , \quad l \!\in\! [N] \,,
\] 
with $R_{\text{SLGM},\mathbf{x}_{0}}(\ist\cdot\ist\ist,\cdot)$ given by \eqref{equ_def_kernel_SLGM}, 
and the bound \eqref{equ_CRB_SLM} for ${\| \mathbf{x}_{0} \|}_{0} \rmv=\rmv S$ is obtained from \eqref{equ_projection_theorem_lower_bound_min_ach_variance} 
using the subspace $\mathcal{U} = \linspan\rmv\rmv \big\{ \genericfuncbound_{0}(\cdot), {\{ \genericfuncbound_{l}(\cdot) \}}_{l \in\ist \supp(\mathbf{x}_{0})}Ê\big\}$.
This establishes a new, RKHS-based interpretation of the bound in \cite{ZvikaCRB} in terms of
the projection of the prescribed mean function 
$\gamma(\mathbf{x}) = c(\mathbf{x}) + x_{k}$ onto an RKHS-related subspace $\mathcal{U}$.
We note that the bound in \cite{ZvikaCRB} was formulated as a bound on the variance $v(\hat{\mathbf{x}}(\cdot); \mathbf{x}_{0})$ of a vector-valued estimator 
$\hat{\mathbf{x}}(\cdot)$ of $\mathbf{x}$ (and not only of the $k$th entry $x_{k}$). Consistent with \eqref{equ_sum_var_scalar}, that bound can be reobtained by summing our 
bound in \eqref{equ_CRB_SLM} (with $c(\cdot) = c_k(\cdot)$) over all $k \rmv\in\rmv [N]$. Thus, the two bounds are equivalent.

An important aspect of Theorem \ref{thm_CRB_SLM} is that the lower variance bound 
in \eqref{equ_CRB_SLM} is not a continuous function of $\mathbf{x}_{0}$ on $\mathcal{X}_{S}$ in general.
Indeed, for the case $\mathbf{H} \rmv=\rmv \mathbf{I}$ and $c(\cdot) \rmv\equiv\rmv 0$, which has been considered in \cite{AlexZvikaJournal}, 
it can be verified that the bound 
is a strictly upper semi-continuous function of $\mathbf{x}_{0}$:
for example, for $M \rmv=\rmv N \rmv=\rmv 2$, $\mathbf{H} \rmv=\rmv \mathbf{I}$, $c(\cdot) \rmv\equiv\rmv 0$,
$S \rmv=\rmv 1$, $k \rmv=\rmv 2$,  
and 
$\mathbf{x}_{0} = a \rmv\rmv\cdot\rmv\rmv (1,0)^{T}\!$ with $a \rmv\in\rmv \mathbb{R}_{+}$, the bound 
is equal to $1$ for $a \rmv=\rmv 0$ (case of ${\| \mathbf{x}_{0} \|}_{0} \rmv<\rmv S$)
but equal to $0$ for all $a \rmv>\rmv 0$ (case of ${\| \mathbf{x}_{0} \|}_{0} \rmv=\rmv S$).
However, by Corollary \ref{cor_lower_semi_cont_SLGM}, the minimum achievable variance $\SLGMMinAchVar$ is
a lower semi-continuous function of $\mathbf{x}_{0}$.
It thus follows that the bound in \eqref{equ_CRB_SLM} cannot be tight, i.e., it cannot be equal to $\SLGMMinAchVar$ for all $\mathbf{x}_{0} \rmv\in\rmv \mathcal{X}_{S}$,
which means that we have a strict inequality in \eqref{equ_CRB_SLM} at least for some 
$\mathbf{x}_{0} \rmv\in\rmv \mathcal{X}_{S}$. 

Let us finally consider the special case where $M \ge N$ and $\mathbf{H} \in \mathbb{R}^{M \times N}\rmv\rmv$ has full 
rank, i.e., $\rank(\mathbf{H}) = N$. The least-squares (LS) estimator \cite{kay,scharf91} of $x_k$ is given by 
$\hat{x}_{\text{LS},k} (\mathbf{y}) = \mathbf{e}_{k}^{T} \mathbf{H}^{\dag} \mathbf{y}$; it is unbiased and its variance 
\vspace{-2mm}
is  
\begin{equation} 
\label{equ_var_ord_LS_est}
v(\hat{x}_{\text{LS},k}(\cdot); \mathbf{x}_{0}) 
 \eq \sigma^{2} \ist\ist \mathbf{e}_{k}^{T} ( \mathbf{H}^{T} \mathbf{H} )^{-1} \mathbf{e}_{k} \,.
\end{equation} 
On the other hand, for unbiased estimation, i.e., $c(\cdot) \!\equiv\! 0$, our lower bound for ${\| \mathbf{x}_{0} \|}_{0} \!<\! S$ in \eqref{equ_CRB_SLM} becomes
$\minachievvar_{\text{SLGM}}(c(\cdot) \!\equiv\! 0, \mathbf{x}_{0}) \ge \sigma^{2} \ist\ist \mathbf{b}^{T} ( \mathbf{H}^{T} \mathbf{H} )^{\dagger} \ist \mathbf{b}
= \sigma^{2} \ist\ist \mathbf{e}_{k}^{T} ( \mathbf{H}^{T} \mathbf{H} )^{-1} \mathbf{e}_{k}$. Comparing with \eqref{equ_var_ord_LS_est}, we conclude that
our bound is tight
and the minimum achievable variance is in fact
\[
\minachievvar_{\text{SLGM}}(c(\cdot) \rmv\rmv\equiv\rmv\rmv 0, \mathbf{x}_{0}) \eq \sigma^{2} \ist\ist \mathbf{e}_{k}^{T} ( \mathbf{H}^{T} \mathbf{H} )^{-1} \mathbf{e}_{k} \,,
\]
which is achieved by the LS estimator. Thus, for $M \ge N$ and $\rank(\mathbf{H}) = N$,
the LS estimator is 
the\footnote{If 
an LMV estimator exists, it is unique 
\cite{LC}.} 
LMV unbiased estimator for the SLGM at each parameter vector $\mathbf{x}_{0} \rmv\in\rmv \mathcal{X}_{S}$ with ${\| \mathbf{x}_{0} \|}_{0} \rmv<\rmv S$.
It is interesting to note that the LS estimator does 
not exploit the sparsity information
expressed by the parameter set $\mathcal{X}_{S}$, i.e., the knowledge that ${\|\mathbf{x} \|}_{0} \rmv\le\rmv S$,
and that it has the constant variance \eqref{equ_var_ord_LS_est} for each $\mathbf{x}_{0} \rmv\in\rmv \mathcal{X}_{S}$ (in fact, even for $\mathbf{x}_{0} \rmv\in\rmv \mathbb{R}^N$). 
We also note that the LS estimator is not an LMV unbiased estimator 
for the case ${\| \mathbf{x}_{0} \|}_{0} \rmv=\rmv S$; therefore, it is not a UMV unbiased estimator on $\mathcal{X}_{S}$ (i.e., an unbiased estimator with 
minimum variance at each $\mathbf{x}_{0} \rmv\in\rmv \mathcal{X}_{S}$). In fact, as shown in \cite{AlexZvikaJournal}, and \cite{JungPHD}, there does not exist a 
UMV unbiased estimator for the SLGM in general.

\subsection{A Novel CRB-Type Lower Variance Bound} 
\label{sec_RKHS_Lower_Bounds_SLGM_improved}

A novel
lower bound on $\SLGMMinAchVar$ is stated in the following theorem
\cite{RKHSAsilomar2010}.

\begin{theorem} 
\label{thm_bound_asilomar}
Consider the estimation problem $\mathcal{E}_{\emph{SLGM}} = \big(\mathcal{X}_{S},f_{\mathbf{H}}(\mathbf{y}; \mathbf{x}),g(\mathbf{x}) \!=\! x_{k} \big)$ 
with a system matrix $\mathbf{H} \!\in\rmv\rmv \mathbb{R}^{M \times N}\!$ satisfying \eqref{equ_spark_cond}. Let $\mathbf{x}_{0} \!\in\! \mathcal{X}_{S}$,
and consider an arbitrary index set $\mathcal{K} \rmv\rmv=\rmv\rmv \{k_1,\ldots,k_{|\mathcal{K}|} \} \rmv\rmv\subseteq\rmv\rmv [N]$ consisting 
of no more than $S$ indices, i.e., $|\mathcal{K}| \leq S$.
If the prescribed bias function $c(\cdot) \rmv\rmv: \mathcal{X}_{S} \rmv\rightarrow\rmv \mathbb{R}$ is such that the partial derivatives 
$\frac{\partial c(\mathbf{x})}{\partial x_{k_i}} \big|_{\mathbf{x} = \mathbf{x}_{0}}$ exist for all $k_i \in \mathcal{K}$, 
then\footnote{Note 
that $\big( \mathbf{H}_{\mathcal{K}}^{T} \mathbf{H}_{\mathcal{K}} \big)^{\rmv-1}\rmv\rmv$ exists because 
of \eqref{equ_spark_cond}.} 
\be 
\minachievvar_{\emph{SLGM}}(c(\cdot),\mathbf{x}_{0}) \,\geq\, \exp\! \bigg( \!\!-\rmv\rmv \frac{1}{\sigma^{2}} \ist 
  {\| (\mathbf{I} \!-\! \mathbf{P}) \mathbf{H} \mathbf{x}_{0} \|}^{2}_{2} \bigg) 
  \big[ \sigma^{2} \ist \bb_{\mathbf{x}_{0}}^{T} \rmv\rmv\big( \mathbf{H}_{\mathcal{K}}^{T} \mathbf{H}_{\mathcal{K}} \big)^{\rmv-1} \bb_{\mathbf{x}_{0}} 
  \rmv+\ist \gamma^{2}(\widetilde{\mathbf{x}}_{0}) \big] -\ist \gamma^{2}(\mathbf{x}_{0}) \,.
\label{equ_bound_asilomar_1}
\ee 
Here, $\mathbf{P} \rmv\rmv\triangleq\rmv \mathbf{H}_{\mathcal{K}} (\mathbf{H}_{\mathcal{K}})^{\dagger} \!\in\rmv\rmv \mathbb{R}^{M \times M}\rmv$, 
$\bb_{\mathbf{x}_{0}} \!\rmv\rmv\in\rmv\rmv \mathbb{R}^{|\mathcal{K}|}$ is defined elementwise as 
${(\bb_{\mathbf{x}_{0}})}_i \rmv\rmv\triangleq\rmv \delta_{k,k_i} \!+ \frac{\partial c(\mathbf{x})}{\partial x_{k_i}} \big|_{\mathbf{x} = \widetilde{\mathbf{x}}_{0}}$ 
for $i \rmv\rmv\in\rmv\rmv [\ist |\mathcal{K}| \ist ]$, 
$\widetilde{\mathbf{x}}_{0} \rmv\rmv\in\rmv\rmv \mathbb{R}^{N}$ is defined as the unique (due to \eqref{equ_spark_cond}) 
vector with $\supp(\widetilde{\mathbf{x}}_{0}) \rmv\rmv\subseteq\rmv\rmv \mathcal{K}$ solving 
$\mathbf{H} \widetilde{\mathbf{x}}_{0} = \mathbf{P} \mathbf{H} \mathbf{x}_{0}$, and $\gamma(\mathbf{x}) = c(\mathbf{x})+x_{k}$.
\end{theorem}

According to \cite[Thm. 5.4.3]{JungPHD}, the bound in \eqref{equ_bound_asilomar_1} follows from the generic bound \eqref{equ_projection_theorem_lower_bound_min_ach_variance} 
by using the subspace 
$\mathcal{U}  = \linspan\rmv\rmv \big\{ \tilde{\genericfuncbound}_{0}(\cdot), {\{ \tilde{\genericfuncbound}_{l}(\cdot) \}}_{l \in \mathcal{K}}Ê\big\}$, where
\[
\tilde{\genericfuncbound}_{0} (\cdot) \ist\ist\triangleq\ist\ist R_{\text{SLGM},\mathbf{x}_{0}} (\ist\cdot\ist\ist, \widetilde{\mathbf{x}}_{0}) \,, \qquad\;
\tilde{\genericfuncbound}_{l}(\cdot) \ist\ist\triangleq\ist\ist \frac{ \partial R_{\text{SLGM},\mathbf{x}_{0}} (\ist\cdot\ist\ist, \mathbf{x}_{2})}  
  { \partial {(\mathbf{x}_{2})}_{l} } \bigg|_{\mathbf{x}_{2} = \widetilde{\mathbf{x}}_{0}} , \quad l \!\in\! \mathcal{K} \,.
\] 
We note that the bound presented in \cite{RKHSAsilomar2010} is obtained by maximizing \eqref{equ_bound_asilomar_1} with respect to the index set $\mathcal{K}$;
this gives the tightest possible bound of the type \eqref{equ_bound_asilomar_1}. 

For the special case given by the SSNM, i.e., $\mathbf{H} \rmv=\rmv \mathbf{I}$, and unbiased estimation, i.e., $c(\cdot) \rmv\equiv\rmv 0$, 
the bound \eqref{equ_bound_asilomar_1} is a continuous function of $\mathbf{x}_{0}$ on $\mathcal{X}_{S}$.
This is an important difference from the bound given in Theorem \ref{thm_CRB_SLM} and, also, from the bound to be given in Theorem \ref{thm_HCRB_SLM}.
Furthermore, still for $\mathbf{H} \rmv=\rmv \mathbf{I}$ and $c(\cdot) \rmv\equiv\rmv 0$, the bound \eqref{equ_bound_asilomar_1} 
can be shown \cite{RKHSAsilomar2010},\cite[p.\ 106]{JungPHD} to be tighter (higher) than the bounds in Theorem \ref{thm_CRB_SLM} and Theorem \ref{thm_HCRB_SLM}. 

The matrix $\mathbf{P}$ appearing in \eqref{equ_bound_asilomar_1} is the orthogonal projection matrix \cite{golub96} on the subspace 
$\mathcal{H}_{\mathcal{K}} \rmv\triangleq \linspan(\mathbf{H}_{\mathcal{K}})$\linebreak 
$\subseteq \mathbb{R}^{M}\rmv\rmv$, i.e., the subspace spanned by those columns
of $\mathbf{H}$ whose indices are in $\mathcal{K}$. 
Consequently, $\mathbf{I} \rmv\rmv-\rmv\rmv \mathbf{P}$ is the orthogonal projection matrix on the orthogonal complement 
of $\mathcal{H}_{\mathcal{K}}$, and the norm ${\|(\mathbf{I} \!-\! \mathbf{P}) \mathbf{H} \mathbf{x}_{0} \|}_{2}$
thus represents the distance 
between the point $\mathbf{H} \mathbf{x}_{0}$ and the subspace $\mathcal{H}_{\mathcal{K}}$ \cite{RudinBook}. 
Therefore, the factor
$\exp\rmv\rmv\big( \!\rmv-\rmv\! \frac{1}{\sigma^{2}} \ist {\| (\mathbf{I} \!-\! \mathbf{P}) \mathbf{H} \mathbf{x}_{0} \|}^{2}_{2} \big)$ 
appearing in the bound \eqref{equ_bound_asilomar_1} can be interpreted as a measure 
of the distance between 
$\mathbf{H} \mathbf{x}_{0}$ and 
$\mathcal{H}_{\mathcal{K}}$. 
In general, the bound \eqref{equ_bound_asilomar_1} is tighter (i.e., higher) if $\mathcal{K}$ is chosen such that 
the distance ${\|(\mathbf{I} \!-\! \mathbf{P}) \mathbf{H} \mathbf{x}_{0} \|}_{2}$ is smaller. 

A slight modification in the derivation of 
\eqref{equ_bound_asilomar_1}
yields the following alternative bound:
\begin{equation} 
\label{equ_bound_asilomar_2}
\minachievvar_{\text{SLGM}}(c(\cdot),\mathbf{x}_{0}) \,\geq\, \exp\! \bigg( \!\!-\rmv\rmv \frac{1}{\sigma^{2}} \ist 
  {\| (\mathbf{I} \!-\! \mathbf{P}) \mathbf{H} \mathbf{x}_{0} \|}^{2}_{2} \bigg)
\, \sigma^{2} \ist \bb_{\mathbf{x}_{0}}^{T} \rmv\rmv\big( \mathbf{H}_{\mathcal{K}}^{T} \mathbf{H}_{\mathcal{K}} \big)^{\rmv-1} \bb_{\mathbf{x}_{0}} \ist\ist.
\end{equation} 
As shown in \cite[Thm. 5.4.4]{JungPHD}, this bound follows from the generic lower bound \eqref{equ_projection_theorem_lower_bound_min_ach_variance}
by using the subspace $\mathcal{U}  = \linspan\rmv\rmv \big\{ \genericfuncbound_{0}(\cdot), {\{ \tilde{\genericfuncbound}_{l}(\cdot) \}}_{l \in \mathcal{K}}Ê\big\}$,
with $\genericfuncbound_{0}(\cdot) = R_{\text{SLGM},\mathbf{x}_{0}} (\ist\cdot\ist\ist, \mathbf{x}_{0})$ and
$\tilde{\genericfuncbound}_{l}(\cdot) = \frac{ \partial R_{\text{SLGM},\mathbf{x}_{0}} (\ist\cdot\ist\ist, \mathbf{x}_{2})}  
{ \partial {(\mathbf{x}_{2})}_{l} } \big|_{\mathbf{x}_{2} = \widetilde{\mathbf{x}}_{0}}$ as defined previously.
Note that this subspace deviates from 
the subspace underlying the bound 
\eqref{equ_bound_asilomar_1}
only by the use of $\genericfuncbound_{0}(\cdot)$ instead of $\tilde{\genericfuncbound}_{0}(\cdot)$.
The difference of
the bounds \eqref{equ_bound_asilomar_2} and \eqref{equ_bound_asilomar_1} is 
\begin{equation} 
\label{equ_diff_asilomar_bounds}
\Delta_{\eqref{equ_bound_asilomar_2}-\eqref{equ_bound_asilomar_1}} \eq \gamma^{2}(\mathbf{x}_{0}) \ist-\ist\ist \exp\! \bigg( \!\!-\rmv\rmv \frac{1}{\sigma^{2}} \ist 
  {\| (\mathbf{I} \!-\! \mathbf{P}) \mathbf{H} \mathbf{x}_{0} \|}^{2}_{2} \bigg) \ist\ist \gamma^{2}(\widetilde{\mathbf{x}}_{0}) \,. 
\end{equation} 
This depends on the choice of the index set $\mathcal{K}$ (via $\mathbf{P}$ and $\widetilde{\mathbf{x}}_{0}$). 
If, for some 
$\mathcal{K}$ and 
$c(\cdot)$, 
$\gamma^{2}(\widetilde{\mathbf{x}}_{0}) Ê\approx \gamma^{2}(\mathbf{x}_{0})$, then $\Delta_{\eqref{equ_bound_asilomar_2}-\eqref{equ_bound_asilomar_1}}$
is approximately nonnegative since $\exp\rmv\rmv\big( \!\rmv- \! \frac{1}{\sigma^{2}} \ist {\| (\mathbf{I} \!-\! \mathbf{P}) \mathbf{H} \mathbf{x}_{0} \|}^{2}_{2} \big) \rmv\leq\rmv 1$.
Hence, in that case, the bound \eqref{equ_bound_asilomar_2} is tighter (higher) than the bound \eqref{equ_bound_asilomar_1}.
We note that one sufficient condition for $\gamma^{2}(\widetilde{\mathbf{x}}_{0}) Ê\approx \gamma^{2}(\mathbf{x}_{0})$ is that 
the columns of $\mathbf{H}_{\mathcal{K}}$ are nearly orthonormal and $c(\cdot)Ê\equiv 0$, i.e., 
unbiased estimation.

The bounds \eqref{equ_bound_asilomar_1} and \eqref{equ_bound_asilomar_2} have an intuitively appealing interpretation in terms of a scaled CRB for an LGM. 
Indeed, the quantity $\sigma^{2} \ist \bb_{\mathbf{x}_{0}}^{T} \rmv\rmv\big( \mathbf{H}_{\mathcal{K}}^{T} \mathbf{H}_{\mathcal{K}} \big)^{\rmv-1} \bb_{\mathbf{x}_{0}}\rmv\rmv$ 
appearing in \eqref{equ_bound_asilomar_1} and \eqref{equ_bound_asilomar_2} can be interpreted as the 
CRB \cite{kay} for the LGM with parameter dimension $N \rmv\rmv=\rmv |\mathcal{K}|$, 
parameter function $g(\mathbf{x}) \rmv=\rmv x_{k}$, and prescribed bias function $c(\cdot)$. 
For a discussion of the scaling factor $\exp\rmv\rmv\big( \!\rmv-\! \frac{1}{\sigma^{2}} \ist {\| (\mathbf{I} \rmv-\rmv \mathbf{P}) \mathbf{H} \mathbf{x}_{0} \|}^{2}_{2} \big)$, 
we will consider the following two complementary cases:

\begin{enumerate}

\vspace{1.5mm}

\item For the case where either $k \in \supp(\mathbf{x}_{0})$ or
${\|Ê\mathbf{x}_{0}\|}_{0} \rmv<\rmv S$ (or both), the factor 
$\exp\rmv\rmv\big( \!\rmv-\! \frac{1}{\sigma^{2}} \ist {\| (\mathbf{I} \rmv-\rmv \mathbf{P}) \mathbf{H} \mathbf{x}_{0} \|}^{2}_{2} \big)$
can be made equal to $1$ by choosing $\mathcal{K} = \supp(\mathbf{x}_{0}) \cup \{ k \}$. 

\vspace{1.5mm}

\item On the other hand, consider the complementary 
case where $k \notin \supp(\mathbf{x}_{0})$ and ${\| \mathbf{x}_{0}\|}_{0} \rmv=\rmv S$. 
Choosing $\mathcal{K} = \mathcal{L} \cup\rmv \{k\}$, 
where $\mathcal{L}$ comprises the indices of the $S \rmv-\rmv 1$ largest (in magnitude) entries of $\mathbf{x}_{0}$, 
we obtain ${\| (\mathbf{I} \!-\! \mathbf{P}) \mathbf{H} \mathbf{x}_{0} \|}^{2}_{2} 
= \xi_{0}^{2} \ist {\| ( \mathbf{I} \!-\! \mathbf{P}) \mathbf{H} \mathbf{e}_{j_{0}} \|}^{2}_{2}$, 
where $\xi_{0}$ and $j_{0}$ denote the value and index, respectively, of the smallest (in magnitude) nonzero entry of $\mathbf{x}_{0}$. 
Typically,\footnote{Note 
that, for the case $k \notin \supp(\mathbf{x}_{0})$ and ${\| \mathbf{x}_{0}\|}_{0} \rmv=\rmv S$ considered,
$j_{0} \rmv\notin\rmv \mathcal{K}$ with $|\mathcal{K}| \rmv\leq\rmv S$.
For a system matrix $\mathbf{H}$ satisfying \eqref{equ_spark_cond}, 
we then have $\| ( \mathbf{I} \!-\! \mathbf{P}) \mathbf{H} \mathbf{e}_{j_{0}} \|^{2}_{2} >\rmv 0$ if and only if the submatrix $\mathbf{H}_{\mathcal{K} \cup \{j_{0}\}}$ has full column 
rank.} 
${\| ( \mathbf{I} \!-\! \mathbf{P}) \mathbf{H} \mathbf{e}_{j_{0}} \|}^{2}_{2} > 0$ and 
therefore, as $\xi_{0}$ becomes larger (in magnitude), the bound \eqref{equ_bound_asilomar_2} 
transitions from a ``low signal-to-noise ratio (SNR)''
regime, where $\exp\rmv\rmv\big( \!\rmv-\! \frac{1}{\sigma^{2}} \ist {\| (\mathbf{I} \rmv-\rmv \mathbf{P}) \mathbf{H} \mathbf{x}_{0} \|}^{2}_{2} \big) \rmv\approx\rmv 1$,
to a ``high-SNR'' regime, where 
$\exp\rmv\rmv\big( \!\rmv-\! \frac{1}{\sigma^{2}} \ist {\| (\mathbf{I} \rmv-\rmv \mathbf{P}) \mathbf{H} \mathbf{x}_{0} \|}^{2}_{2} \big) \rmv\approx\rmv 0$.
In the low-SNR regime, the bound \eqref{equ_bound_asilomar_2} is approximately equal to 
$\sigma^{2} \ist \bb_{\mathbf{x}_{0}}^{T} \rmv\rmv\big( \mathbf{H}_{\mathcal{K}}^{T} \mathbf{H}_{\mathcal{K}} \big)^{\rmv-1} \bb_{\mathbf{x}_{0}}\rmv\rmv$,
i.e., to the CRB for the LGM with $N \!=\rmv |\mathcal{K}|$. In the high-SNR regime, the bound becomes approximately equal to $0$; 
this suggests that the zero entries $x_{k}$ with $k \notin \supp(\mathbf{x})$ can be estimated with small variance. Note that for increasing $\xi_{0}$, 
the transition from the low-SNR regime to the high-SNR regime exhibits
an exponential decay. 

\end{enumerate}

\section{The SLGM View of Compressed Sensing} 
\label{sec_slm_viewpoint_CS}

The lower bounds of Section \ref{sec_bounds_SLGM} are also relevant to the linear CS recovery problem, which can be viewed as an instance of the SLGM-based estimation problem.
In this section, we express one of these lower bounds in terms of the restricted isometry constant of the system matrix 
(CS measurement matrix) $\mathbf{H}$.

\subsection{CS Fundamentals}
\label{SecCSfund}

The compressive measurement process within a CS problem
is often
modeled as 
\cite{JustRelax,GreedisGood,Can06a,ZvikaCoherenceTSP,MallatBook}
\begin{equation}
\label{equ_measurment_equ_CS}
\mathbf{y} \ist=\ist \mathbf{H} \mathbf{x} + \mathbf{n} \,. 
\end{equation} 
Here, $\mathbf{y} \rmv\in\rmv \mathbb{R}^{M}\rmv$ denotes the compressive measurements; 
$\mathbf{H} \rmv\in\rmv \mathbb{R}^{M \times N}\rmv\rmv$, where $M \rmv\le\rmv N$ and typically $M \rmv\ll\rmv N$, denotes the CS measurement matrix; 
$\mathbf{x} \rmv\in\rmv \mathcal{X}_{S} \rmv\subseteq\rmv \mathbb{R}^{N}\rmv$ is an unknown $S$-sparse signal or parameter vector, 
with known sparsity degree $S$ (typically $S \rmv\ll\rmv N$); 
and $\mathbf{n}$ represents additive measurement noise.
We assume that $\mathbf{n} \sim \mathcal{N}(\mathbf{0}, \sigma^{2} \mathbf{I})$ and that the columns $\{ \mathbf{h}_{j} \}_{j \in [N]}$ of $\mathbf{H}$ are normalized, i.e., $\| \mathbf{h}_{j} \|_{2} = 1$ for all $j \in [N]$. The CS measurement model
\eqref{equ_measurment_equ_CS} is 
then identical to the SLGM observation model \eqref{equ_linear_observation_model}.  
Any CS recovery method,\footnote{A comprehensive overview is provided at \htmladdnormallink{http://dsp.rice.edu/cs}{http://dsp.rice.edu/cs}.} 
such as the Basis Pursuit (BP) \cite{JustRelax,Chen98atomicdecomposition} or the Orthogonal Matching Pursuit (OMP) 
\cite{GreedisGood,TroppGilbertOMP}, can be interpreted as an estimator $\hat{\mathbf{x}}(\mathbf{y})$ that estimates the sparse vector $\mathbf{x}$ from 
the observation $\mathbf{y}$.

Due to the typically large dimension of the measurement matrix $\mathbf{H}$, a complete characterization of the 
properties of $\mathbf{H}$
(e.g., via its 
SVD) is often infeasible.
Useful incomplete characterizations are provided by the (mutual) coherence and the restricted isometry property \cite{JustRelax,GreedisGood,Can06a,ZvikaCoherenceTSP}.
The \emph{coherence} of a matrix $\mathbf{H} \rmv\in\rmv \mathbb{R}^{M \times N}\rmv$ is defined 
\vspace{-1.5mm}
as 
\[ 
\mu(\mathbf{H}) \ist\ist\triangleq\ist\ist\ist \max\limits_{i \neq j} | \mathbf{h}_{j}^{T} \mathbf{h}_{i} | \,.
\] 
Furthermore, a matrix $\mathbf{H} \rmv\in\rmv \mathbb{R}^{M \times N}\rmv$ is said to satisfy the \emph{restricted isometry property} (RIP) of order $K$ 
if for every index set $\mathcal{I}Ê\rmv\subseteq\rmv [N]$ of size 
$|\mathcal{I}| \rmv=\rmv K$ there is a constant $\delta'_{K} \rmv\rmv\in\rmv \mathbb{R}_{+}\rmv$ such 
\vspace{-1.5mm}
that 
\begin{equation} 
\label{equ_RIP_condition}
(1\rmv\rmv-\rmv \delta'_{K}) \ist {\| \mathbf{z} \|}^{2}_{2} \,\leq\, {\| \mathbf{H}_{\mathcal{I}} \ist \mathbf{z} \|}^{2}_{2} \,\leq\, (1 \rmv +\delta'_{K}) \ist {\| \mathbf{z} \|}^{2}_{2} \ist\,,
  \quad\; \text{for all} \;\ist \mathbf{z} \rmv\in\rmv \mathbb{R}^{K} .
\end{equation}
The smallest $\delta'_{K}$ for which \eqref{equ_RIP_condition} holds---hereafter denoted $\delta_{K}$---is called the \emph{RIP constant} of $\mathbf{H}$.
Condition \eqref{equ_spark_cond} is necessary for a matrix $\mathbf{H}$ to have the RIP of order $S$ with a RIP constant 
$\delta_{S} \rmv<\rmv 1$.\footnote{Indeed, 
assume that $\spark(\mathbf{H}) \rmv\leq\rmv S$. This means that 
there exists an index set $\mathcal{I} \rmv\subseteq\rmv [N]$ consisting of $S$ indices
such that the columns of $\mathbf{H}_{\mathcal{I}}$ are linearly dependent. This, in turn, implies that 
there is a nonzero coefficient vector $\mathbf{z} \rmv\in\rmv \mathbb{R}^{S}$ 
such that $\mathbf{H}_{\mathcal{I}} \ist \mathbf{z} \rmv=\rmv \mathbf{0}$ and consequently ${\|\mathbf{H}_{\mathcal{I}} \ist \mathbf{z}Ê\|}^{2}_{2} \rmv=\rmv 0$. Therefore, 
there cannot exist a constant $\delta'_{K} \rmv<\rmv 1$ satisfying \eqref{equ_RIP_condition} for all 
$\mathbf{z} \rmv\in\rmv \mathbb{R}^{S}\rmv$.} 
It can be easily verified that $\delta_{K'} \!\geq\rmv \delta_{K}$ for $K' \rmv\rmv\geq\rmv K$.
The coherence $\mu(\mathbf{H})$ provides a coarser description of the matrix $\mathbf{H}$ than the RIP constant $\delta_{K}$ but can be calculated more easily.
The two parameters are related according to $\delta_{K} \leq (K \!-\! 1) \ist \mu(\mathbf{H})$ \cite{ZvikaCoherenceTSP}.

\subsection{A Lower Variance Bound}
\label{SecCSbound}

We now specialize the bound \eqref{equ_bound_asilomar_2} 
on the minimum achievable variance for $\mathcal{E}_{\text{SLGM}}$ to the CS scenario, i.e., to the SLGM with sparsity degree $S$ and 
a system matrix $\mathbf{H}$ that is a CS measurement matrix (i.e., $M \rmv\le\rmv N$) with known RIP constant $\delta_{S}<1$. 
Note that 
$\delta_{S} <1$ implies that 
condition \eqref{equ_spark_cond} is satisfied. The following result was presented in \cite[Thm. 5.7.2]{JungPHD}.

\begin{theorem}
\label{thm_CS_measur_matrix_asilomar_bound}
Consider the SLGM-based estimation problem $\mathcal{E}_{\emph{SLGM}} = \big(\mathcal{X}_{S},f_{\mathbf{H}}\rmv\rmv(\mathbf{y}; \mathbf{x}), g(\mathbf{x}) \!=\! x_{k} \big)$, where
$\mathbf{H} \rmv\in\rmv \mathbb{R}^{M \times N}\rmv\rmv$ with $M \rmv\le\rmv N\rmv$ satisfies the RIP of order $S$ with RIP constant $\delta_{S} \rmv<\rmv 1$.
Let $\mathbf{x}_{0} \rmv\in\rmv \mathcal{X}_{S}$, and consider an arbitrary index set $\mathcal{K} 
\rmv\subseteq\rmv [N]$ 
consisting of no more than $S$ indices, i.e., $|\mathcal{K}| \leq S$. If the first-order partial derivatives 
$\frac{\partial c(\mathbf{x})}{\partial x_{l}} \big|_{\mathbf{x} = \mathbf{x}_{0}}\rmv\rmv$ of the 
prescribed bias function $c(\cdot) \rmv\rmv: \mathcal{X}_{S} \rmv\rightarrow\rmv \mathbb{R}\ist$ 
exist for all $l \rmv\in\rmv \mathcal{K}$, then
\vspace*{-.5mm}
\begin{equation}
\label{equ_CS_measur_matrix_asilomar_bound}
\minachievvar_{\emph{SLGM}}(c(\cdot),\mathbf{x}_{0})  \,\geq\, \exp\! \bigg( \!\!-\rmv\rmv \frac{1 \rmv\rmv+\rmv \delta_{S}}{\sigma^{2}} \ist 
  \big\| \mathbf{x}_{0}^{\supp(\mathbf{x}_{0}) \setminus \mathcal{K}} \big\|^{2}_{2} \bigg)
    \,\sigma^{2} \ist \bb_{\mathbf{x}_{0}}^{T} \rmv\rmv\big( \mathbf{H}_{\mathcal{K}}^{T} \mathbf{H}_{\mathcal{K}} \big)^{-1} \bb_{\mathbf{x}_{0}} \,, 
\end{equation} 
with $\bb_{\mathbf{x}_{0}} \!\in\rmv \mathbb{R}^{|\mathcal{K}|}$ as defined in Theorem \ref{thm_bound_asilomar}.
\end{theorem}

Using the inequality $\delta_{S} \leq (S \!-\! 1) \ist \mu(\mathbf{H})$, we obtain from \eqref{equ_CS_measur_matrix_asilomar_bound} the coherence-based bound 
\[
\minachievvar_{\text{SLGM}}(c(\cdot),\mathbf{x}_{0})  \,\geq\, \exp\! \bigg( \!\!-\rmv\rmv \frac{1 \rmv\rmv+\rmv (S \!-\! 1) \ist \mu(\mathbf{H})}{\sigma^{2}} \ist 
  \big\| \mathbf{x}_{0}^{\supp(\mathbf{x}_{0}) \setminus \mathcal{K}} \big\|^{2}_{2} \bigg)
    \,\sigma^{2} \ist \bb_{\mathbf{x}_{0}}^{T} \rmv\rmv\big( \mathbf{H}_{\mathcal{K}}^{T} \mathbf{H}_{\mathcal{K}} \big)^{-1} \bb_{\mathbf{x}_{0}} \,. 
\]

If we want to compare the actual variance behavior of a given CS recovery scheme (or, estimator) $\hat{x}_{k}(\cdot)$ 
with the bound on the 
minimum achievable variance in \eqref{equ_CS_measur_matrix_asilomar_bound}, 
then we have to ensure that the first-order partial derivatives of the estimator's bias function 
$\mathsf{E}_{\mathbf{x}} \{ \hat{x}_{k}(\mathbf{y}) \} - x_k$ 
exist. The following lemma states that this is indeed the case
under mild conditions. Moreover, the lemma gives an explicit expression of these partial derivatives.

\begin{lemma}[\hspace*{-1mm}{\cite[Cor. 2.6]{FundmentExpFamBrown}}]
\label{lem_cond_exist_CS_recovery_partial_der}
Consider the SLGM-based estimation problem $\mathcal{E}_{\emph{SLGM}} = \big(\mathcal{X}_{S},f_{\mathbf{H}}\rmv\rmv(\mathbf{y}; \mathbf{x}), g(\mathbf{x}) \!=\! x_{k} \big)$ 
and an estimator $\hat{x}_{k}(\cdot) \rmv\rmv: \mathbb{R}^{M} \!\rightarrow\rmv \mathbb{R}$. If the mean function 
$\gamma(\mathbf{x}) = \expect_{\mathbf{x}} \{ \hat{x}_{k}(\mathbf{y}) \}$ 
exists for all
$\mathbf{x} \rmv\in\rmv \mathcal{X}_{S}$, then also the partial derivatives
$\frac{\partial c(\mathbf{x})}{\partial x_{l}}$, $l \rmv\in\rmv [N]$ exist for all $\mathbf{x} \rmv\in\rmv \mathcal{X}_{S}$ and are given by
\be
\label{equ_expr_par_der_CS_recovery_scheme}
\frac{\partial c(\mathbf{x})}{\partial x_{l}}
\eq \delta_{k,l} \ist+ \frac{1}{\sigma^{2}} \ist\ist \mathsf{E}_{\mathbf{x}} \big\{ \hat{x}_{k}(\mathbf{y}) \ist (\mathbf{y} \rmv\rmv-\rmv\rmv \mathbf{H}\mathbf{x})^{T} \mathbf{H} \ist \mathbf{e}_{l} \big\} \,. 
\vspace{-2mm}
\ee
\end{lemma}

\subsection{The Case $\delta_{S} \rmv\approx\rmv 0$}
\label{SecCScase}

For CS applications, 
measurement matrices $\mathbf{H}$ with RIP constant close to zero, i.e., $\delta_{S} \rmv\approx\rmv 0$, are generally preferable 
\cite{Can06a,ROMP_stab,AnalysisOMPRIP,CoSAMP,ZvikaCoherenceTSP}. For $\delta_{S} \rmv=\rmv 0$, the bound in \eqref{equ_CS_measur_matrix_asilomar_bound} becomes
\begin{equation}
\label{equ_CS_measur_matrix_asilomar_bound_1}
\minachievvar_{\text{SLGM}}(c(\cdot),\mathbf{x}_{0})  \,\geq\, \exp\! \bigg( \!\!-\rmv\rmv \frac{1}{\sigma^{2}} \ist 
  \big\| \mathbf{x}_{0}^{\supp(\mathbf{x}_{0}) \setminus \mathcal{K}} \big\|^{2}_{2} \bigg)
    \,\sigma^{2} \ist \bb_{\mathbf{x}_{0}}^{T} \rmv\rmv\big( \mathbf{H}_{\mathcal{K}}^{T} \mathbf{H}_{\mathcal{K}} \big)^{-1} \bb_{\mathbf{x}_{0}} \,. 
\end{equation} 
This 
is equal to the bound \eqref{equ_CS_measur_matrix_bound_altern_1} for the SSNM (i.e., $\mathbf{H} \rmv=\rmv \mathbf{I}$) except that the factor 
$\bb_{\mathbf{x}_{0}}^{T} \rmv\big( \mathbf{H}_{\mathcal{K}}^{T} \mathbf{H}_{\mathcal{K}} \big)^{-1} \bb_{\mathbf{x}_{0}}$ 
in \eqref{equ_CS_measur_matrix_asilomar_bound_1} is replaced by ${\| \bb_{\mathbf{x}_{0}} \|}_2^2$ in \eqref{equ_CS_measur_matrix_bound_altern_1}.
For a ``good'' CS measurement matrix, i.e., with 
$\delta_{S} \rmv\approx\rmv 0$,
we have $\bb_{\mathbf{x}_{0}}^{T} \rmv\big( \mathbf{H}_{\mathcal{K}}^{T} \mathbf{H}_{\mathcal{K}} \big)^{-1} \bb_{\mathbf{x}_{0}} 
\rmv\rmv\approx {\| \bb_{\mathbf{x}_{0}} \|}^{2}_{2}$ for any index set $\mathcal{K} \rmv\subseteq\rmv [N]$ of size
$|\mathcal{K}| \rmv\leq\rmv S$. Thus, the bound in \eqref{equ_CS_measur_matrix_asilomar_bound_1} is very close
to \eqref{equ_CS_measur_matrix_bound_altern_1}.
This means that, conversely, in terms of a lower bound on the achievable estimation accuracy, 
relative to the SSNM (case $\mathbf{H} \rmv=\rmv \mathbf{I}$), no loss of information is incurred by multiplying $\mathbf{x}$ by the CS measurement matrix 
$\mathbf{H} \rmv\in\rmv \mathbb{R}^{M \times N}\rmv\rmv$ and thereby reducing the signal dimension from $N$ to $M$, where typically 
$M \rmv\ll\rmv N$.
This agrees with the fact that if $\delta_{S} \approx 0$,
one can recover---e.g., by using the BP---the sparse parameter vector $\mathbf{x} \rmv\in\rmv \mathcal{X}_{S}$ 
from the compressed observation $\mathbf{y} = \mathbf{H} \mathbf{x} + \mathbf{n}$ 
up to an 
error that is typically very small (and whose norm is almost independent of $\mathbf{H}$ and solely determined by the 
measurement noise $\mathbf{n}$
\cite{Can06a,DantzigCandes}).

\section{RKHS-based Analysis of Minimum Variance Estimation for the SSNM} 
\label{sec_rkhs_approach_ssnm}



Next, we specialize our RKHS-based MVE analysis to the SSNM, i.e., to the special case 
given by $\mathbf{H} \!=\! \mathbf{I}$ (which implies $M \!=\! N$ and $\mathbf{y} = \mathbf{x} + \mathbf{n}$). For the SSNM-based estimation problem
$\mathcal{E}_{\text{SSNM}} = \big(\mathcal{X}_{S},f_{\mathbf{I}}\rmv(\mathbf{y}; \mathbf{x}), g(\cdot) 
\big)$ with $k \rmv\rmv\in\rmv\rmv [N]$, we will analyze 
the minimum achievable variance $\minachievvar_{\text{SSNM}}(c(\cdot), \mathbf{x}_{0})$ and the corresponding LMV estimator.
We 
note that the SLGM 
with a system matrix $\mathbf{H} \rmv\in\rmv \mathbb{R}^{M \times N}\rmv\rmv$ having orthonormal columns, i.e., satisfying $\mathbf{H}^{T} \mathbf{H} = \mathbf{I}$, 
is equivalent to the SSNM \cite{AlexZvikaJournal}.

Specializing the kernel $R_{\text{SLGM},\mathbf{x}_{0}}(\ist\cdot\ist\ist,\cdot)$ (see \eqref{equ_def_kernel_SLGM}) to 
the system matrix $\mathbf{H} \rmv=\rmv \mathbf{I}$, we obtain 
\be
R_{\text{SSNM},\mathbf{x}_{0}}(\mathbf{x}_{1},\mathbf{x}_{2}) 
\eq \exp \rmv\rmv\bigg(  \frac{1}{\sigma^{2}} (\mathbf{x}_{2} \rmv\rmv-\rmv\rmv \mathbf{x}_{0})^{T} (\mathbf{x}_{1} \rmv\rmv-\rmv\rmv \mathbf{x}_{0}) \rmv\rmv\bigg) 
  \,, \quad\; \mathbf{x}_{0}, \mathbf{x}_{1}, \mathbf{x}_{2} \in \mathcal{X}_{S} \,.
\label{equ_R_SSNM_exp}
\ee
The corresponding RKHS, $\mathcal{H}(R_{\text{SSNM},\mathbf{x}_{0}})$, will be briefly denoted by $\RKHSSSNM$.

\subsection{Valid Bias Functions, Minimum Achievable Variance, and LMV Estimator} 
\label{sec_lmv_ssnm}
 
Since the SSNM is a special case of the SLGM, we can characterize the class of valid bias functions, the minimum achievable variance (Barankin bound), 
and the corresponding LMV estimator by Theorems \ref{thm_condition_gamma_valid_SLM_fourier_series_R_g} and 
\ref{thm_Barankin_bound_LMV} specialized to $\mathbf{H} \rmv=\rmv \mathbf{I}$, as stated in the following corollary.

\begin{corollary}
\label{cor_condition_gamma_Barankin_bound_LMV}
Consider the SSNM-based estimation problem $\mathcal{E}_{\emph{\text{SSNM}}}=\big(\mathcal{X}_{S},f_{\mathbf{I}}\rmv(\mathbf{y}; \mathbf{x}), g(\cdot) 
\big)$ with $k \rmv\in\rmv [N]$.

\begin{enumerate} 

\vspace{1.5mm}

\item 
A bias function $c(\cdot) \rmv\rmv: \mathcal{X}_{S} \rmv\rightarrow \mathbb{R}$ is valid for 
$\mathcal{E}_{\emph{\text{SSNM}}}$ at $\mathbf{x}_{0}\!\in\! \mathcal{X}_{S}$ if and only if it 
can be expressed 
\vspace{.5mm}
as
\begin{equation} 
\label{equ_condition_gamma_valid_SSNM_fourier_series_R_g}
c(\mathbf{x}) \eq \exp\rmv\rmv\bigg( \frac{1}{2 \sigma^{2}} \ist {\| \mathbf{x}_{0} \|}^{2}_{2} 
  - \frac{1}{\sigma^{2}} \ist\mathbf{x}^{T} \mathbf{x}_{0} \rmv\bigg)
\rmv \sum_{\mathbf{p} \in \mathbb{Z}_{+}^{D}} \rmv\rmv \frac{a[\mathbf{p}]}{\sqrt{\mathbf{p}!}} \ist \bigg( \frac{1}{\sigma}  \mathbf{x} \rmv\bigg)^{\!\mathbf{p}} 
  \!- g(\mathbf{x}) \,, \quad\; \mathbf{x} \!\in\! \mathcal{X}_{S} \,,
\vspace{-1.5mm}
\end{equation} 
with some coefficient sequence $a[\mathbf{p}] \in \ell^{2}(\mathbb{Z}_{+}^{D})$.

\vspace{2mm}

\item Let $c(\cdot)\rmv\rmv : \mathcal{X}_{S} \rmv\rightarrow\rmv \mathbb{R}$ be a valid prescribed bias function. Then: 

\begin{enumerate} 

\vspace{1.5mm}

\item The minimum achievable variance at $\mathbf{x}_{0} \!\in\! \mathcal{X}_{S}$, $\minachievvar_{\emph{SSNM}}(c(\cdot),\mathbf{x}_{0})$, 
is given by \eqref{equ_M_min_a_gamma}, in which 
$\mathcal{C}(c) \rmv\subseteq\rmv \ell^{2}(\mathbb{Z}_{+}^{D})$ denotes the set of coefficient sequences $a[\mathbf{p}] \in \ell^{2}(\mathbb{Z}_{+}^{D})$ that are consistent with 
\eqref{equ_condition_gamma_valid_SSNM_fourier_series_R_g}.

\vspace{1.5mm}

\item The function $\hat{g}(\cdot) \rmv\rmv : \mathbb{R}^{M} \!\rightarrow\rmv \mathbb{R}$ given by 
\begin{equation} 
\label{equ_est_arb_coeffs_SSNM}
\hat{g}(\mathbf{y}) \eq \exp\rmv\rmv\bigg( \!\!-\rmv \frac{1}{2 \sigma^{2}} \ist {\| \mathbf{x}_{0} \|}^{2}_{2} \rmv\bigg)  
\rmv \sum_{\mathbf{p} \in \mathbb{Z}_{+}^{D}} \rmv\rmv \frac{a[\mathbf{p}]}{\sqrt{\mathbf{p}!}} \, \chi_{\mathbf{p}}(\mathbf{y}) \,, 
\vspace{-1.5mm}
\end{equation}
with an arbitrary coefficient sequence $a[\cdot] \rmv\rmv\in\rmv\rmv \mathcal{C}(c)$ and
\vspace{1.5mm}
\[
\chi_{\mathbf{p}}(\mathbf{y}) \,\triangleq\, \frac{ \partial^{\mathbf{p}} \big[ \rho_{\emph{LGM},\mathbf{x}_{0}}(\mathbf{y},\sigma \mathbf{x}) 
  \ist\exp\rmv\rmv \big( \frac{1}{\sigma} \ist \mathbf{x}_{0}^{T} \rmv \mathbf{x} \big) \big]}{ \partial \mathbf{x}^{\mathbf{p}}} \bigg|_{\mathbf{x} = \mathbf{0}} \,,
\vspace{1mm}
\]
is an allowed estimator at $\mathbf{x}_{0}$ for 
$c(\cdot)$, i.e., $\hat{g}(\cdot) \rmv\in\rmv \mathcal{A}(c(\cdot), \mathbf{x}_{0})$.

\vspace{1.5mm}

\item The LMV estimator at $\mathbf{x}_{0}$,
$\hat{g}^{(c(\cdot),\mathbf{x}_{0})}(\cdot)$, is given by
\eqref{equ_est_arb_coeffs_SSNM} using the specific coefficient sequence $a_{0}[\mathbf{p}] = \argmin_{a[\cdot] \in \mathcal{C}(c)} \rmv\rmv {\| a[\cdot] \|}_{\ell^{2}(\mathbb{Z}_{+}^{D})}\ist$.

\end{enumerate} 

\vspace{1.5mm}

\end{enumerate}  
\end{corollary}

However, a more convenient characterization can be obtained by exploiting the specific structure of $\RKHSSSNM$ 
that is induced by the choice $\mathbf{H} \rmv=\rmv \mathbf{I}$. 
We omit the technical details, which can be found in \cite[Sec. 5.5]{JungPHD}, and 
just present the main
results regarding 
MVE
\cite[Thm. 5.5.2]{JungPHD}.

\vspace{1mm}

\begin{theorem} 
\label{thm_general_min_achiev_var_LMV_SSNM}

Consider the SSNM-based estimation problem $\mathcal{E}_{\text{\emph{SSNM}}} 
= \big(\mathcal{X}_{S},f_{\mathbf{I}}\rmv(\mathbf{y}; \mathbf{x}), g(\cdot)
\big)$ with $k \rmv\in\rmv [N]$.

\begin{enumerate} 

\vspace{1.5mm}

\item
A prescribed bias function 
$c(\cdot) \rmv\rmv: \mathcal{X}_{S} \rmv\rightarrow\rmv \mathbb{R}$ is valid for $\mathcal{E}_{\text{\emph{SSNM}}}$ at $\mathbf{x}_{0} \!\in\! \mathcal{X}_{S}$ if and only if 
the associated prescribed mean function $\gamma(\cdot) = c(\cdot) + g(\cdot)$ can be expressed as
\[
\gamma(\mathbf{x}) \eq \frac{1}{\nu_{\mathbf{x}_{0}}\rmv(\mathbf{x})} \rmv \sum_{\mathbf{p} \ist\in\ist \mathbb{Z}_{+}^{N} \cap \ist\mathcal{X}_{S}} 
\!\rmv\rmv \frac{a[\mathbf{p}]}{\sqrt{\mathbf{p}!}} \ist \bigg( \rmv \frac{\mathbf{x}}{\sigma} \rmv \bigg)^{\!\mathbf{p}} , \quad\; \mathbf{x} \rmv\in\rmv \mathcal{X}_{S} \,,
\vspace{-1mm}
\] 
with
\vspace{1mm}
\[
\nu_{\mathbf{x}_{0}}\rmv(\mathbf{x}) \,\triangleq\, \exp\rmv\rmv\bigg( \!\!-\rmv \frac{1}{2 \sigma^{2}} \ist {\|\mathbf{x}_{0} \|}^{2}_{2} \ist\ist+ \frac{1}{\sigma^{2}} \ist \mathbf{x}^{T} \rmv\mathbf{x}_{0} \rmv\rmv\bigg)
\]
and with a 
coefficient sequence $a[\mathbf{p}] \in \ell^{2}(\mathbb{Z}_{+}^{N} \cap \mathcal{X}_{S})$. This coefficient sequence is unique for a given $c(\cdot)$.

\vspace{2mm}

\item Let $c(\cdot)\rmv\rmv : \mathcal{X}_{S} \rmv\rightarrow\rmv \mathbb{R}$ be a valid prescribed bias function. Then: 

\begin{enumerate} 

\vspace{1.5mm}

\item The minimum achievable variance at $\mathbf{x}_{0} \rmv\in\rmv \mathcal{X}_{S}$ is given 
by
\be
\label{equ_ssnm_expr_min_ach_var}
\minachievvar_{\emph{SSNM}}(c(\cdot),\mathbf{x}_{0}) \,= \sum_{\mathbf{p} \ist\in\ist \mathbb{Z}_{+}^{N} \cap \ist\mathcal{X}_{S}} \!\! a_{\mathbf{x}_{0}}^2[\mathbf{p}] 
\ist-\ist \gamma^{2}(\mathbf{x}_{0}) \,,
\vspace{-1mm}
\ee 
with
\vspace{1mm}
\[ 
a_{\mathbf{x}_{0}}[\mathbf{p}] \,\triangleq\, \frac{1}{\sqrt{\mathbf{p}!}} \,\frac{ \partial^{\mathbf{p}} 
  \big( \gamma( \sigma \mathbf{x}) \ist\ist \nu_{\mathbf{x}_{0}}\rmv(\sigma \mathbf{x}) \big)}{\partial \mathbf{x}^{\mathbf{p}}} \bigg|_{\mathbf{x}=\mathbf{0}} \ist . 
\] 

\vspace{1.5mm}

\item The LMV estimator at $\mathbf{x}_{0}$ is given by
\begin{equation}
\label{equ_ssnm_general_expr_LMV}
\hat{g}^{(c(\cdot),\mathbf{x}_{0})}(\mathbf{y})
\,= \sum_{\mathbf{p} \ist\in\ist \mathbb{Z}_{+}^{N} \cap \ist\mathcal{X}_{S}} \!\! \frac{a_{\mathbf{x}_{0}}[\mathbf{p}]}{\sqrt{\mathbf{p}!}} \,
  \frac{\partial^{\mathbf{p}}\psi_{\mathbf{x}_{0}} (\mathbf{x},\mathbf{y})}{ \partial \mathbf{x}^{\mathbf{p}}} \bigg|_{\mathbf{x}= \mathbf{0}} \ist,
\vspace{-1mm}
\ee 
with
\vspace{1mm}
\[
\psi_{\mathbf{x}_{0}} (\mathbf{x},\mathbf{y}) \ist\ist\triangleq\, \exp \rmv\rmv \bigg( \frac{\mathbf{y}^{T} \rmv (\sigma \mathbf{x} \!-\! \mathbf{x}_{0})}{\sigma^{2}} \ist  
+ \frac{\mathbf{x}_{0}^{T} \rmv\mathbf{x}}{\sigma} - \frac{{\|\mathbf{x}\|}^{2}_{2}}{2} \bigg) \,. 
\vspace{2.5mm}
\]

\end{enumerate} 

\end{enumerate} 

\end{theorem}

Note that the statement of Theorem \ref{thm_general_min_achiev_var_LMV_SSNM} 
is stronger than that of Corollary \ref{cor_condition_gamma_Barankin_bound_LMV},
because it contains explicit expressions of the minimum achievable variance $\SSNMMinAchVar$ and the corresponding LMV estimator 
$\hat{g}^{(c(\cdot),\mathbf{x}_{0})}(\mathbf{y})$.

The expression \eqref{equ_ssnm_expr_min_ach_var} nicely shows the influence of the sparsity constraints on the minimum achievable variance.
Indeed, consider a 
prescribed bias $c(\cdot) \!: \mathbb{R}^{N} \!\!\rightarrow\rmv\rmv \mathbb{R}$ 
that is valid for the SSNM with $S \!=\! N$, and 
therefore
also for the SSNM with $S \rmv<\rmv N$.
Let us denote by $\minachievvar_{N}$ and $\minachievvar_{S}$ the minimum achievable variance $\minachievvar(c(\cdot),\mathbf{x}_{0})$ for the 
degenerate SSNM without sparsity ($S \rmv=\rmv N$) and for the SSNM with sparsity ($S \rmv<\rmv N$), respectively.
Note that in the nonsparse case $S \rmv=\rmv N$, the SSNM coincides with the LGM with system matrix $\mathbf{H} \rmv=\rmv \mathbf{I}$. 
It then follows from \eqref{equ_ssnm_expr_min_ach_var} that 
$\minachievvar_{N} = \sum_{ \mathbf{p} \ist\in\ist \mathbb{Z}_{+}^{N} } a_{\mathbf{x}_{0}}^2[\mathbf{p}] - \gamma^{2}(\mathbf{x}_{0})$ and
\be
\label{equ_diff_min_achiev_var_SSNM}
\minachievvar_{N}-\minachievvar_{S} \,=
   \sum_{ \mathbf{p} \ist\in\ist \mathbb{Z}_{+}^{N} \setminus \mathcal{X}_{S} } \!\! a_{\mathbf{x}_{0}}^2[\mathbf{p}] \,.
\ee
Clearly, if $\mathbf{x}$ is more sparse, i.e., if the sparsity degree $S$ is smaller, the number of (nonnegative) terms in the above sum is larger.
This implies a larger difference $\minachievvar_{N}-\minachievvar_{S}$
and, thus, a stronger reduction of the minimum achievable variance due to the sparsity information.

We mention the obvious fact that a UMV estimator for $\mathcal{E}_{\text{SSNM}}=\big(\mathcal{X}_{S},f_{\mathbf{I}}\rmv(\mathbf{y};\mathbf{x}),g(\cdot)
\big)$
and prescribed bias function $c(\cdot)$ exists if and only if the LMV estimator $\hat{g}^{(c(\cdot),\mathbf{x}_{0})}(\cdot)$ given 
by \eqref{equ_ssnm_general_expr_LMV} does not depend on $\mathbf{x}_{0}$. 

Finally,
consider the SSNM with parameter function $g(\mathbf{x}) \rmv=\rmv x_{k}$, i.e., 
$\mathcal{E}_{\text{SSNM}}=\big(\mathcal{X}_{S},f_{\mathbf{I}}\rmv(\mathbf{y};\mathbf{x}),g(\mathbf{x}) \!=\! x_{k} \big)$, 
for some $k \rmv\in\rmv [N]$.
Because the specific estimator $\hat{g}(\mathbf{y}) \rmv=\rmv y_{k}$
has finite variance and zero bias at each $\mathbf{x} \rmv\in\rmv \mathcal{X}_{S}$, 
the bias function $c_{u}(\mathbf{x}) \equiv 0$ must be valid for $\mathcal{E}_{\text{SSNM}}$
at each $\mathbf{x}_{0} \rmv\in\rmv \mathcal{X}_{S}$. 
Therefore, according to Corollary \ref{cor_lower_semi_cont_SLGM},
the minimum achievable variance for unbiased estimation within the SSNM with parameter function $g(\mathbf{x}) = x_{k}$, $\minachievvar_{\text{SSNM}}(c_{u}(\cdot), \mathbf{x}_{0})$, is a lower semi-continuous function 
of $\mathbf{x}_{0}$ on its domain, i.e., on 
$\mathcal{X}_{S}$. (Note that this remark is not related to Theorem \ref{thm_general_min_achiev_var_LMV_SSNM}.)

\subsection{Diagonal Bias Functions}
\label{sec_diag_ssnm}

In this subsection, we 
consider the 
SSNM-based estimation 
problem\footnote{We 
recall that the assumption $g(\mathbf{x}) \rmv=\rmv x_{k}$ is no restriction, because the MVP for any given parameter function 
$g(\cdot)$ is equivalent to the MVP for the parameter function $g'(\mathbf{x}) = x_{k}$ and the modified prescribed bias function 
$c'(\mathbf{x}) = c(\mathbf{x}) + g(\mathbf{x}) - x_{k}$.} 
$\mathcal{E}_{\text{SSNM}}=\big(\mathcal{X}_{S},f_{\mathbf{I}}\rmv(\mathbf{y};\mathbf{x}),g(\mathbf{x}) \!=\! x_{k} \big)$, for some $k \rmv\in\rmv [N]$, and
we study
a specific class of bias functions.
Let us call a bias function $c(\cdot) \rmv\rmv: \mathcal{X}_{S} \rmv\rightarrow\rmv \mathbb{R}$ \emph{diagonal} 
if $c(\mathbf{x})$ depends only on the $k$th entry of the parameter vector $\mathbf{x}$, i.e., 
the specific scalar parameter $x_{k}$ to be estimated.
That is, $c(\mathbf{x}) \rmv=\rmv \tilde{c}(x_{k})$, with some function $\tilde{c}(\cdot) \rmv\rmv: \mathbb{R} \rmv\rightarrow\rmv \mathbb{R}$ 
that may depend on $k$. Similarly, we say that an estimator $\hat{x}_{k}(\mathbf{y})$ 
is diagonal if it depends only on the $k$th entry of $\mathbf{y}$, i.e., $\hat{x}_{k}(\mathbf{y}) = \hat{x}_{k}(y_{k})$ (with an abuse of notation). 
Clearly, the bias function $b(\hat{x}_{k}(\cdot);\mathbf{x})$ of a diagonal estimator $\hat{x}_{k}(\cdot)$ 
is diagonal, i.e., $b(\hat{x}_{k}(\cdot);\mathbf{x})=b(\hat{x}_{k}(\cdot);x_k)$. 
Well-known examples of diagonal estimators 
are the hard- and soft-thresholding estimators described in \cite{MallatBook,Donoho98minimaxestimation}, and \cite{DonohoJohnstone94} 
and the LS estimator, $\hat{x}_{\text{LS},k}(\mathbf{y}) = y_{k}$.
The maximum likelihood estimator for the SSNM is not diagonal, and its bias function is not diagonal either \cite{AlexZvikaJournal}. 

The following theorem \cite[Thm. 5.5.4]{JungPHD}, which can be regarded as a specialization of Theorem \ref{thm_general_min_achiev_var_LMV_SSNM} to the case of diagonal bias functions, provides a 
characterization of the class of valid diagonal bias functions, 
as well as of the minimum achievable variance and LMV estimator for a prescribed diagonal bias function.
In the theorem, we will use the $l$th order (probabilists') Hermite polynomial $H_{l}(\cdot) \rmv: \mathbb{R} \rmv\rightarrow\rmv \mathbb{R}$ defined as \cite{AbramowitzStegun}
\[
H_{l}(x) \,\triangleq\, (-1)^l \ist e^{x^2/2} \ist \frac{d^l}{dx^l} \, e^{-x^2/2} \ist. 
\] 
Furthermore, in the case ${\| \mathbf{x}_0 \|}_{0} = S$,
the 
support of $\mathbf{x}_{0}$ will be denoted as 
$\supp(\mathbf{x}_{0})=\{ k_{1}, \ldots, k_{S} \}$.


\begin{theorem} 
\label{thm_diag_bias_min_achiev_var_LMV}

Consider the SSNM-based estimation problem $\mathcal{E}_{\text{\emph{SSNM}}} 
= \big(\mathcal{X}_{S},f_{\mathbf{I}}\rmv(\mathbf{y}; \mathbf{x}), g(\mathbf{x}) \!=\! x_{k} \big)$, $k \rmv\in\rmv [N]$, at $\mathbf{x}_{0} \rmv\in\rmv \mathcal{X}_{S}$.
Furthermore consider a prescribed bias function $c(\cdot) \rmv\rmv: \mathcal{X}_{S} \rmv\rightarrow\rmv \mathbb{R}$ that is diagonal
and such that the prescribed mean function $\gamma(\mathbf{x}) = c(\mathbf{x}) + x_{k}$ can be written as a convergent power series 
centered at $\mathbf{x}_{0}$, i.e.,
\begin{equation}
\label{equ_prescr_mean_diag_bias}
\gamma(\mathbf{x}) \,= \sum_{l \in\ist \mathbb{Z}_{+}} \!\frac{m_{l}}{l!} \ist (x_{k} \rmv-\rmv x_{0,k})^{l} \ist,
\vspace{0mm}
\end{equation}
with suitable coefficients $m_{l}$. (Note, in particular, that $m_{0} = \gamma(\mathbf{x}_0)$.)
In what follows, let 
\[ 
B_c \,\triangleq\rmv \sum_{l \in\ist \mathbb{Z}_{+}} \!\frac{m_{l}^{2} \ist \sigma^{2l}}{l!} \,.
\]

\begin{enumerate} 

\vspace{1.5mm}

\item
The bias function $c(\cdot)$ is valid at $\mathbf{x}_{0}$ if and only if 
$B_c \rmv<\rmv \infty$. 

\vspace{1.5mm}

\item Assume that $B_c \rmv<\rmv \infty$, i.e., $c(\cdot)$ is valid. Then:

\begin{enumerate} 

\vspace{1.5mm}

\item
The minimum achievable variance at $\mathbf{x}_{0}$ is given by 
\[
\minachievvar_{\emph{SSNM}}(c(\cdot), \mathbf{x}_{0})  \eq B_c \, \phi(\mathbf{x}_{0}) \ist-\ist \gamma^{2}(\mathbf{x}_{0}) \,,
\vspace{-1.5mm}
\] 
with
\begin{equation} 
\phi(\mathbf{x}_{0}) \,\triangleq\ist \begin{cases}
1 \,, & \mbox{if}\;\, | \rmv\supp(\mathbf{x}_{0}) \cup \{k\}| \le S
\\
\displaystyle \sum_{i \in [S]} \rmv\exp\rmv\rmv \bigg( \!\!-\rmv\rmv \frac{x_{0,k_{i}}^{2}}{\sigma^{2}} \rmv\rmv\bigg) 
\!\prod_{j \in [i-1]} \rmv\rmv\bigg[ 1 \rmv-\ist \exp\rmv\rmv \bigg( \!\!-\rmv\rmv \frac{x_{0,k_{j}}^{2}}{\sigma^{2}} \rmv\rmv\bigg) \bigg] < 1 \,,
& \mbox{if}\;\, | \rmv\supp(\mathbf{x}_{0}) \cup \{k\}| = S \rmv+\rmv\rmv 1 \,.
  \end{cases} 
\label{equ_factor_ISIT_SSNM_min_achiev_var}
\vspace{-2.5mm}
\end{equation}
(Recall that $\supp(\mathbf{x}_{0}) = {\{ k_i\}}_{i=1}^S$ in the case $| \rmv\supp(\mathbf{x}_{0}) \cup \{k\}| = S \rmv+\rmv\rmv 1$.)

\vspace{1.5mm}

\item
The LMV estimator at $\mathbf{x}_{0}$ is given by 
\[
\hat{x}_{k}^{(c(\cdot),\mathbf{x}_{0})}(\mathbf{y})  \eq \psi(\mathbf{y}, \mathbf{x}_{0}) 
  \sum_{l\in\ist \mathbb{Z}_{+}} \!\frac{ m_{l}\ist \sigma^{l}}{l!} \ist\ist H_{l} \bigg( \rmv\frac{ y_k \rmv\rmv-\rmv\rmv x_{0,k} }{\sigma}\rmv\rmv \bigg) \,,
\vspace{-1.5mm}
\] 
with 
\be 
\label{equ_factor_ISIT_SSNM_LMV}
\psi(\mathbf{y}, \mathbf{x}_{0}) \,\triangleq\ist \begin{cases}
1 \,, & \hspace*{-0mm} \mbox{if}\;\, | \rmv\supp(\mathbf{x}_{0}) \cup \{k\}| \le S 
\\[.5mm]
\displaystyle\sum_{i \in [S]} \rmv\exp\rmv\rmv \bigg( \!\!-\rmv\rmv \frac{x_{0,k_i}^2 \!+ 2\ist y_{k_i} x_{0,k_i}}{2 \ist\sigma^{2}} \rmv\rmv\bigg) 
& \\[-1mm]
\displaystyle \hspace*{10mm} \times \!\prod_{j \in [i-1]} \rmv\rmv\bigg[ 1 \rmv-\ist \exp\rmv\rmv \bigg( \!\!-\rmv\rmv \frac{x_{0,k_j}^2 \!+ 2\ist y_{k_j} x_{0,k_j}}{2 \ist\sigma^{2}} \rmv\rmv\bigg) \bigg] \,, 
  &\hspace*{-0mm} \mbox{if}\;\, | \rmv\supp(\mathbf{x}_{0}) \cup \{k\}| = S \rmv+\rmv\rmv 1 \,.
  \end{cases}
\vspace{-2mm}
\ee

\end{enumerate} 

\end{enumerate} 

\end{theorem} 

Regarding the case distinction in Theorem \ref{thm_diag_bias_min_achiev_var_LMV}, we note that
$| \rmv\supp(\mathbf{x}_{0}) \cup \{k\}| \le S$ either if
${\| \mathbf{x} \|}_{0} \rmv\rmv<\rmv S$ or if both ${\| \mathbf{x} \|}_{0} \rmv\rmv=\rmv S$ and 
$k \rmv\in \supp(\mathbf{x}_{0})$, and $| \rmv\supp(\mathbf{x}_{0}) \cup \{k\}| = S \rmv+\rmv\rmv 1$ if both
${\| \mathbf{x} \|}_{0} \rmv\rmv=\rmv S$ and 
$k \rmv\not\in \supp(\mathbf{x}_{0})$.

If the prescribed bias function $c(\cdot)$ is the actual bias function $b(\hat{x}'_{k}(\cdot); \mathbf{x})$ of some
diagonal estimator $\hat{x}'_{k}(\mathbf{y}) = \hat{x}'_{k}(y_{k})$ with 
finite variance at $\mathbf{x}_{0}$, the coefficients $m_{l}$ appearing in Theorem \ref{thm_diag_bias_min_achiev_var_LMV} have a particular interpretation. 
For a discussion of this interpretation, we 
need the following lemma \cite{SzegoOrthogPoly}.

\begin{lemma}
\label{lem_def_diag_finite_var_est_hilbert_space}
Consider the SSNM-based estimation problem $\mathcal{E}_{\text{\emph{SSNM}}} 
= \big(\mathcal{X}_{S},f_{\mathbf{I}}\rmv(\mathbf{y}; \mathbf{x}), g(\mathbf{x}) \!=\! x_{k} \big)$, $k \rmv\in\rmv [N]$, at $\mathbf{x}_{0} \rmv\in\rmv \mathcal{X}_{S}$.
Furthermore consider the 
Hilbert space $\mathcal{P}_{\emph{SSNM}}$ consisting of all finite-variance estimator functions $\hat{g}(\cdot) \rmv\rmv: \mathbb{R}^N \!\to\rmv  \mathbb{R}$, i.e., 
$\mathcal{P}_{\emph{SSNM}} \triangleq \{ \hat{g}(\cdot) \ist|\ist v(\hat{g}(\cdot); \mathbf{x}_{0}) \!<\! \infty \}$,
and endowed with the inner product 
\[ 
\big\langle \hat{g}_{1}(\cdot), \hat{g}_{2}(\cdot) \big\rangle_{\emph{RV}} \ist=\,
 \expect_{\mathbf{x}_{0}} \rmv\big\{ \hat{g}_{1}(\mathbf{y}) \ist \hat{g}_{2}(\mathbf{y}) \big\}
 \ist=\ist \frac{1}{(2 \pi \sigma^{2})^{N/2}} \int_{\mathbb{R}^N} \rmv\hat{g}_{1}(\mathbf{y}) \ist\ist \hat{g}_{2}(\mathbf{y}) 
  \ist\exp \rmv\rmv\rmv\bigg( \!\!-\rmv\rmv \frac{1}{2 \sigma^{2}} \ist {\| \mathbf{y} \rmv\rmv-\rmv\rmv \mathbf{x}_{0} \|}^{2}_{2} \bigg) \ist\ist dÊ\mathbf{y} \,. 
\]
Then, the subset $\mathcal{D}_{\emph{SSNM}} \subseteq \mathcal{P}_{\emph{SSNM}}$ consisting 
of all diagonal estimators $\hat{g}(\mathbf{y}) = \hat{g}(y_{k})$ is a subspace of $\mathcal{P}_{\emph{SSNM}}\ist$, with induced inner product
\[ 
\big\langle \hat{g}_{1}(\cdot), \hat{g}_{2}(\cdot) \big\rangle_{\mathcal{D}_{\emph{SSNM}}} \ist=\, 
  \frac{1}{ \sqrt{2 \pi} \sigma} \int_{\mathbb{R}} \hat{g}_{1}(y) \ist\ist \hat{g}_{2}(y)
 \ist\exp \rmv\rmv\rmv\bigg( \!\!-\rmv\rmv \frac{1}{2 \sigma^{2}} \ist (y \rmv-\rmv x_{0,k} )^{2} \rmv\bigg) \ist\ist dÊy \,. 
\]
An orthonormal basis for $\mathcal{D}_{\emph{SSNM}}$ is constituted by ${\{ h^{(l)}(\cdot) \}}_{l \in \mathbb{Z}_{+}}$, with 
$h^{(l)}(\cdot) \rmv\rmv: \mathbb{R}^{N} \!\rightarrow\rmv \mathbb{R}$ given by 
\begin{equation}
\label{equ_ONB_hermite_functions}
h^{(l)}(\mathbf{y}) \eq \frac{1}{\sqrt{l!}} \ist\ist H_{l} \bigg( \rmv\frac{ y_k \rmv\rmv-\rmv x_{0,k}}{\sigma}\rmv\rmv \bigg) \,.
\vspace{2.5mm}
\end{equation}

\end{lemma}

Combining Theorem \ref{thm_diag_bias_min_achiev_var_LMV} with Lemma \ref{lem_def_diag_finite_var_est_hilbert_space} yields 
the following result \cite[Cor. 5.5.7]{JungPHD}. 

\begin{corollary} 
\label{cor_diag_est_min_achiev_var_LMV_ISIT}
Consider the SSNM-based estimation problem $\mathcal{E}_{\text{\emph{SSNM}}} 
= \big(\mathcal{X}_{S},f_{\mathbf{I}}\rmv(\mathbf{y}; \mathbf{x}), g(\mathbf{x}) \!=\! x_{k} \big)$, $k \rmv\in\rmv [N]$, at $\mathbf{x}_{0} \rmv\in\rmv \mathcal{X}_{S}$.
Furthermore consider a prescribed diagonal bias function $c(\cdot) \rmv\rmv: \mathcal{X}_{S} \rmv\rightarrow\rmv \mathbb{R}$ that is the actual 
bias function of a diagonal estimator $\hat{x}_{k}(\mathbf{y}) = \hat{x}_{k}(y_{k})$, i.e., $c(\mathbf{x}) = b(\hat{x}_{k}(\cdot); \mathbf{x})$. The estimator $\hat{x}_{k}(\cdot)$ is 
assumed to have finite variance at $\mathbf{x}_{0}$, $v(\hat{x}_{k}(\cdot); \mathbf{x}_{0}) < \infty$, and hence 
$\hat{x}_{k}(\mathbf{y}) \in \mathcal{D}_{\emph{SSNM}}$ and, also, $c(\cdot)$ is valid.

\begin{enumerate} 

\vspace{1.5mm}

\item
The prescribed mean function $\gamma(\mathbf{x}) = c(\mathbf{x}) + x_{k}= \mathsf{E}_{\mathbf{x}} \{ \hat{x}_{k}(\mathbf{y}) \}$ 
can be written as a convergent power series \eqref{equ_prescr_mean_diag_bias}, 
with coefficients 
given by 
\begin{align} 
m_{l}& \eq \frac{\sqrt{l!}}{\sigma^{l}} \ist\ist \big\langle \hat{x}_{k}(\cdot), h^{(l)}(\cdot) \big\rangle_{\mathcal{D}_{\emph{SSNM}}} 
\label{equ_coeff_mean} \\[1mm]Ê
&\eq \frac{1}{\sqrt{2 \pi} \sigma^{l+1}} \int_{\mathbb{R}} \hat{x}_{k}(y) \,H_{l} \bigg( \rmv\frac{ y \rmv-\rmv x_{0,k}}{\sigma}\rmv\rmv \bigg)
 \ist\exp \rmv\rmv\rmv\bigg( \!\!-\rmv\rmv \frac{1}{2 \sigma^{2}} \ist (y \rmv-\rmv x_{0,k} )^{2} \rmv\bigg) \ist\ist dÊy \,. \nonumber
\end{align} 

\vspace{1.5mm}

\item
The minimum achievable variance at $\mathbf{x}_{0}$ is given by 
\be 
\label{equ_expr_min_achiev_var_diag_est_ISIT_SSNM}
\minachievvar_{\emph{SSNM}}(c(\cdot),\mathbf{x}_{0}) \eq v(\hat{x}_{k}(\cdot); \mathbf{x}_{0}) \,\phi(\mathbf{x}_{0}) \ist+\ist [ \phi(\mathbf{x}_{0}) \rmv-\rmv\rmv 1 ] \, 
  \gamma^{2}(\mathbf{x}_{0})Ê\,,
\ee 
with $\phi(\mathbf{x}_{0})$ as defined in \eqref{equ_factor_ISIT_SSNM_min_achiev_var}.

\vspace{1.5mm}

\item
The LMV estimator at $\mathbf{x}_{0}$ is given by  
\begin{equation}
\label{equ_expr_LMV_diag_est_ISIT_SSNM}
\hat{x}_{k}^{(c(\cdot),\mathbf{x}_{0})}(\mathbf{y})  \eq \hat{x}_{k}(y_{k})  \,\psi(\mathbf{y}, \mathbf{x}_{0}) \,,
\vspace{-1.5mm}
\end{equation}
with $\psi(\mathbf{y}, \mathbf{x}_{0})$ as defined 
\vspace{1.5mm}
in \eqref{equ_factor_ISIT_SSNM_LMV}.

\end{enumerate} 

\end{corollary} 

It follows from \eqref{equ_coeff_mean} and from Lemma \ref{lem_def_diag_finite_var_est_hilbert_space} that the given diagonal estimator $\hat{x}_{k}(\cdot)$
can be written as
\[
\hat{x}_{k}(\mathbf{y}) \ist\ist=\ist\ist \sigma^{2} \rmv\rmv \sum_{l\in\ist \mathbb{Z}_{+}} \!\frac{ m_{l} }{\sqrt{l!}} \ist\ist h^{(l)}(\mathbf{y}) \,.
\vspace{-.5mm}
\] 
Thus, the coefficients $m_{l}$ appearing in Theorem \ref{thm_diag_bias_min_achiev_var_LMV} have the 
interpretation of being (up to a factor of $1/\sqrt{l!}$) the expansion coefficients of the estimator $\hat{x}_{k}(\cdot)$---viewed as an element of 
$\mathcal{D}_{\text{SSNM}}$---with respect to 
the orthonormal basis $\big\{h^{(l)}(\mathbf{y})Ê\big\}_{l \in \mathbb{Z}_{+}}\!$.

Remarkably, as shown by \eqref{equ_expr_LMV_diag_est_ISIT_SSNM},
the LMV estimator 
can be obtained by multiplying the diagonal estimator $\hat{x}_{k}(\mathbf{y})$---which
is arbitrary except for the condition that its variance at $\mathbf{x}_{0}$ is finite---by the ``correction factor'' $\psi(\mathbf{y},\mathbf{x}_{0})$
in \eqref{equ_factor_ISIT_SSNM_LMV}. 
It can be easily verified that $\psi(\mathbf{y},\mathbf{x}_{0})$ does not depend on $y_{k}$. According to \eqref{equ_factor_ISIT_SSNM_LMV},
the following two cases have to be distinguished:

\begin{enumerate}

\vspace{1.5mm}

\item
For $k \rmv\in\rmv [N]$ such that $|\rmv\supp(\mathbf{x}_{0}) \cup \{ k \}| \le S$, we have $\psi(\mathbf{y},\mathbf{x}_{0}) = 1$, 
and therefore the LMV estimator is obtained from \eqref{equ_expr_LMV_diag_est_ISIT_SSNM} as 
$\hat{x}_{k}^{(c(\cdot),\mathbf{x}_{0})}(\mathbf{y}) = \hat{x}_{k}(y_{k}) = \hat{x}_{k}(\mathbf{y})$.
Thus, in that case,
it follows from Corollary \ref{cor_diag_est_min_achiev_var_LMV_ISIT} that every diagonal estimator 
$\hat{x}_{k}(\cdot) \rmv\rmv : \mathbb{R}^{N} \rmv\!\rightarrow\rmv \mathbb{R}$ for the SSNM that has finite variance 
at $\mathbf{x}_{0}$ is necessarily an LMV estimator. 
In particular, the variance $v(\hat{x}_{k}(\cdot);\mathbf{x}_{0})$ equals the minimum achievable variance 
$\minachievvar_{\text{SSNM}}(c(\cdot),\mathbf{x}_{0})$, i.e., the Barankin bound. Furthermore,
the sparsity information cannot be leveraged for improved MVE, because the estimator $\hat{x}_{k}(\cdot)$ is an LMV estimator for the parameter set 
$\mathcal{X}_{S}$ with arbitrary $S$, including the nonsparse case $\mathcal{X} \rmv\rmv=\rmv \mathbb{R}^{N}$. 

\vspace{1.5mm}

\item
For 
$k \rmv\in\rmv [N]$ such that $|\rmv\supp(\mathbf{x}_{0}) \cup \{ k \}| = S \rmv+\rmv 1$, 
it follows from Corollary \ref{cor_diag_est_min_achiev_var_LMV_ISIT} and 
\eqref{equ_factor_ISIT_SSNM_min_achiev_var} 
that there exist estimators (in particular, the LMV estimator $\hat{x}_{k}^{(c(\cdot),\mathbf{x}_{0})}(\mathbf{y})$) with the same bias function
as $\hat{x}_{k}(\cdot)$ but with a smaller variance at $\mathbf{x}_{0}$. Indeed, in this case, we have $\phi(\mathbf{x}_{0}) \rmv<\rmv 1$ 
in \eqref{equ_factor_ISIT_SSNM_min_achiev_var}, and by \eqref{equ_expr_min_achiev_var_diag_est_ISIT_SSNM} it thus follows 
that $\minachievvar_{\text{SSNM}}(c(\cdot),\mathbf{x}_{0}) 
< v(\hat{x}_{k}(\cdot); \mathbf{x}_{0})$. 

\vspace{1.5mm}

\end{enumerate}

Let us for the moment make the (weak) assumption that the given diagonal
estimator $\hat{x}_{k}(\cdot)$ has finite variance at every parameter vector $\mathbf{x} \rmv\in\rmv \mathbb{R}^{N}\rmv\rmv$. It can then be shown that 
the LMV estimator $\hat{x}_{k}^{(c(\cdot),\mathbf{x}_{0})}(\cdot)$
is robust to deviations from the nominal parameter $\mathbf{x}_{0}$ in the sense that its bias and variance depend continuously on $\mathbf{x}_{0}$. 
Furthermore, 
$\hat{x}_{k}^{(c(\cdot),\mathbf{x}_{0})}(\cdot)$ has finite bias and finite variance at any parameter vector $\mathbf{x} \rmv\in\rmv \mathbb{R}^{N}\rmv\rmv$, 
i.e., $\big| b\big(\hat{x}_{k}^{(c(\cdot),\mathbf{x}_{0})}(\cdot); \mathbf{x}\big) \big| < \infty$ and $v\big(\hat{x}_{k}^{(c(\cdot),\mathbf{x}_{0})}(\cdot);\mathbf{x}\big) < \infty$ 
for all $\mathbf{x} \rmv\in\rmv \mathbb{R}^{N}\rmv\rmv$. 

We finally note that Corollary \ref{cor_diag_est_min_achiev_var_LMV_ISIT} also applies to unbiased estimation, i.e., prescribed bias function $c(\cdot) \rmv\equiv\rmv 0$ 
(equivalently, $\gamma(\mathbf{x}) \rmv=\rmv x_{k}$). This is because $c(\cdot) \rmv\equiv\rmv 0$ is the actual bias function of the 
LS estimator $\hat{x}_{\text{LS},k}(\mathbf{y}) = y_{k}$. Clearly, the LS estimator is diagonal and has finite variance at $\mathbf{x}_{0}$.
Thus, it can be used as the given diagonal estimator $\hat{x}_{k}(\mathbf{y})$ in Corollary \ref{cor_diag_est_min_achiev_var_LMV_ISIT}.

\subsection{Lower Variance Bounds} 
\label{sec_RKHS_Lower_Bounds_SLGM_HCRB}

Finally, we complement the exact expressions of the minimum achievable variance $\minachievvar_{\text{SSNM}}(c(\cdot),\mathbf{x}_{0})$
presented above by simple lower bounds.
The following bound is obtained by specializing the sparse CRB in Theorem \ref{thm_CRB_SLM} to the SSNM ($\mathbf{H} \rmv=\rmv \mathbf{I}$).
 
\begin{corollary}
\label{cor_CRB_SSNM}
Consider the estimation problem $\mathcal{E}_{\emph{SSNM}} = \big(\mathcal{X}_{S},f_{\mathbf{I}}(\mathbf{y}; \mathbf{x}),g(\mathbf{x}) \!=\! x_{k} \big)$. 
Let $\mathbf{x}_{0} \rmv\in\rmv \mathcal{X}_{S}$.
If the prescribed bias function $c(\cdot) \rmv\rmv: \mathcal{X}_{S} \rmv\rightarrow\rmv \mathbb{R}$ is such that the partial derivatives 
$\frac{\partial c(\mathbf{x})}{\partial x_{l}} \big|_{\mathbf{x} = \mathbf{x}_{0}}$ exist for all $l \rmv\in\rmv [N]$, then 
\be 
\minachievvar_{\emph{SSNM}}(c(\cdot),\mathbf{x}_{0}) \,\geq\ist \begin{cases}
  \sigma^{2} \ist\ist \| \mathbf{b} \|_{2}^{2} \,, 
    & \mbox{if}\;\, {\| \mathbf{x}_{0} \|}_{0} < S \\[-1.5mm]Ê
  \sigma^{2} \ist\ist \| \mathbf{b}_{\mathbf{x}_{0}} \|^{2}_{2} \,, 
    & \mbox{if}\;\, {\| \mathbf{x}_{0} \|}_{0} = S \,. 
  \end{cases}
\label{equ_CRB_SSNM}
\ee 
Here, in the case ${\| \mathbf{x}_{0} \|}_{0} \rmv<\rmv S$, $\mathbf{b} \rmv\in\rmv \mathbb{R}^{N}\rmv$ is given 
by $b_{l} \triangleq \delta_{k,l} + \frac{\partial c(\mathbf{x})}{\partial x_{l}} \big|_{\mathbf{x} = \mathbf{x}_{0}}\ist$, $l \rmv\in\rmv [N]$,
and in the case ${\| \mathbf{x}_{0} \|}_{0} \rmv=\rmv S$,
$\mathbf{b}_{\mathbf{x}_{0}} \rmv\rmv\!\in\rmv \mathbb{R}^{S}\rmv$ consists of those entries of 
$\mathbf{b}$ that are indexed by $\supp(\mathbf{x}_{0})=\{k_{1},\ldots,k_{S}\}$, i.e., 
${(\mathbf{b}_{\mathbf{x}_{0}})}_{i} \rmv=\rmv b_{k_{i}}\ist$, $i \rmv\in\rmv [S]$. 
\end{corollary}

Specializing the alternative bound in Theorem \ref{thm_bound_asilomar} to the SSNM yields the following result.

\begin{corollary} 
\label{corr_bound_asilomar_SSNM}
Consider the estimation problem $\mathcal{E}_{\emph{SSNM}} = \big(\mathcal{X}_{S},f_{\mathbf{I}}(\mathbf{y}; \mathbf{x}),g(\mathbf{x}) \!=\! x_{k} \big)$. 
Let $\mathbf{x}_{0} \rmv\in\rmv \mathcal{X}_{S}$, and consider an arbitrary index set $\mathcal{K} =\{k_1,\ldots,k_{|\mathcal{K}|} \} \subseteq [N]$ 
consisting of no more than $S$ indices, i.e., $|\mathcal{K}| \leq S$.
If the prescribed bias function $c(\cdot) \rmv: \mathcal{X}_{S} \rmv\rmv\rightarrow\rmv \mathbb{R}$ is such that the partial derivatives 
$\frac{\partial c(\mathbf{x})}{\partial x_{k_i}} \big|_{\mathbf{x} = \mathbf{x}_{0}}\rmv$ exist for all $k_i \in \mathcal{K}$, 
\vspace{-1.5mm}
then 
\[ 
\minachievvar_{\emph{SSNM}}(c(\cdot),\mathbf{x}_{0}) \,\geq\, \exp\! \bigg( \!\!-\rmv\rmv \frac{1}{\sigma^{2}} \ist 
  \big\| \mathbf{x}^{[N] \setminus \mathcal{K}}_{0} \big\|^{2}_{2} \bigg) 
  \big[ \sigma^{2} \ist \| \bb_{\mathbf{x}_{0}} \|^{2}_{2} 
  \ist\ist+ \gamma^{2}(\mathbf{x}^{\mathcal{K}}_{0}) \big] -\ist \gamma^{2}(\mathbf{x}_{0}) \,.
\] 
Here, $\bb_{\mathbf{x}_{0}} \rmv\!\in\rmv \mathbb{R}^{|\mathcal{K}|}$ is defined elementwise as 
${(\bb_{\mathbf{x}_{0}})}_i \triangleq \delta_{k,k_i} \rmv+ \frac{\partial c(\mathbf{x})}{\partial x_{k_i}} \big|_{\mathbf{x} = \mathbf{x}^{\mathcal{K}}_{0}}\ist$ for $i \rmv\in\rmv [\ist |\mathcal{K}| \ist ]$, 
and $\gamma(\mathbf{x}) = c(\mathbf{x})+x_{k}$.
\end{corollary}

Furthermore, the modified bound in \eqref{equ_bound_asilomar_2} specialized to the SSNM reads as
\begin{equation}
\label{equ_CS_measur_matrix_bound_altern}
\SSNMMinAchVar  \,\geq\, \exp\! \bigg( \!\!-\rmv\rmv \frac{1}{\sigma^{2}} \ist 
  {\| (\mathbf{I} \!-\! \mathbf{P}) \ist \mathbf{x}_{0} \|}^{2}_{2} \bigg)
\, \sigma^{2} \ist {\| \bb_{\mathbf{x}_{0}} \|}_2^2
\,. 
\end{equation}  
Because $\mathbf{H} \rmv=\rmv \mathbf{I}$, we have
$\mathbf{P} = \mathbf{H}_{\mathcal{K}} (\mathbf{H}_{\mathcal{K}})^{\dagger} 
= {\mathbf{I}}_{\mathcal{K}} \ist ({\mathbf{I}}_{\mathcal{K}})^{\dagger}
= \sum_{l \in \mathcal{K}} \mathbf{e}_{l} \mathbf{e}_{l}^{T}$. 
Therefore, 
multiplying $\mathbf{x}_{0}$ by $\mathbf{I} \rmv-\rmv \mathbf{P}$ simply zeros
all entries of 
$\mathbf{x}_{0}$ whose indices belong to $\mathcal{K}$, i.e., 
$(\mathbf{I} \rmv\rmv-\rmv \mathbf{P}) \ist \mathbf{x}_{0} = \mathbf{x}_{0}^{\supp(\mathbf{x}_{0}) \setminus \mathcal{K}}\rmv\rmv$,
and thus \eqref{equ_CS_measur_matrix_bound_altern} becomes
\begin{equation}
\label{equ_CS_measur_matrix_bound_altern_1}
\SSNMMinAchVar  \,\geq\, \exp\! \bigg( \!\!-\rmv\rmv \frac{1}{\sigma^{2}} \ist 
  \big\| \mathbf{x}_{0}^{\supp(\mathbf{x}_{0}) \setminus \mathcal{K}} \big\|^{2}_{2} \bigg)
\, \sigma^{2} \ist {\| \bb_{\mathbf{x}_{0}} \|}_2^2
\,. 
\end{equation}  

For unbiased estimation ($c(\cdot) \rmv\equiv\rmv 0$), the following lower bound on $\minachievvar_{\text{SSNM}}(c(\cdot) \rmv\equiv\rmv 0, \mathbf{x}_{0})$ 
is based on the Hammersley-Chapman-Robbins bound (HCRB) \cite{RKHSExpFamIT2012,LC,GormanHero}.
This bound has been previously derived in a slightly different form in \cite{AlexZvikaJournal}.


\begin{theorem}
\label{thm_HCRB_SLM}
Consider the estimation problem $\mathcal{E}_{\emph{SSNM}} = \big(\mathcal{X}_{S},f_{\mathbf{I}}\rmv(\mathbf{y}; \mathbf{x}),g(\mathbf{x}) \!=\! x_{k} \big)$ with $k \rmv\in\rmv [N]$ 
and the prescribed bias function $c(\cdot) \equiv 0$.
Let $\mathbf{x}_{0} \rmv\in\rmv \mathcal{X}_{S}$. Then, 
\be 
\minachievvar_{\emph{SSNM}}(c(\cdot),\mathbf{x}_{0}) \,\geq\ist \begin{cases}
  \sigma^{2} \,, 
    & \mbox{if}\;\, |\rmv\supp(\mathbf{x}_{0}) \cup \{k\} | \le S \\[.5mm]Ê
  \displaystyle \sigma^{2} \, \frac{N \!-\rmv S \!-\! 1}{N \!-\rmv S} \ist \exp (-\xi_{0}^{2}/ \sigma^{2} ) \,, 
    & \mbox{if}\;\, |\rmv\supp(\mathbf{x}_{0}) \cup \{k\} | = S \rmv+\rmv 1 \,, 
  \end{cases}
\label{equ_HCRB_SLM}
\ee 
where $\xi_{0}$ denotes the value of the $S$-largest (in magnitude) entry of $\mathbf{x}_{0}$. 
\end{theorem} 

In \cite[Thm. 5.4.2]{JungPHD}, it is shown that the bound \eqref{equ_HCRB_SLM} for $|\rmv\supp(\mathbf{x}_{0}) \cup \{k\} | \le S$ 
is obtained from the generic bound \eqref{equ_projection_theorem_lower_bound_min_ach_variance} by using for the subspace $\mathcal{U}$ the limit of  
$\mathcal{U}^{(t)} \rmv\rmv\triangleq\ist \linspan\rmv\rmv \big\{ \genericfuncbound_{0}(\cdot), {\{ \genericfuncbound^{(t)}_{l}(\cdot) \}}_{l \in [N]}Ê\big\}$ as $t \!\to\! 0$. Here,
$\genericfuncbound_{0} (\cdot) \triangleq R_{\text{SSNM},\mathbf{x}_{0}} (\ist\cdot\ist\ist, \mathbf{x}_{0})$ and
\[
\genericfuncbound_{l}^{(t)}(\cdot) \,\triangleq\ist\ist \begin{cases}
  R_{\text{SSNM},\mathbf{x}_{0}} (\ist\cdot\ist\ist, \mathbf{x}_{0}+t\mathbf{e}_{l})-R_{\text{SSNM},\mathbf{x}_{0}} (\ist\cdot\ist\ist, \mathbf{x}_{0}) \,, 
    & \mbox{if}\;\, l \in \supp(\mathbf{x}_{0}) \\[-1mm]Ê
  R_{\text{SSNM},\mathbf{x}_{0}} (\ist\cdot\ist\ist, \mathbf{x}_{0} \rmv- \xi_{0}\mathbf{e}_{j_{0}}+t\mathbf{e}_{l})-R_{\text{SSNM},\mathbf{x}_{0}} (\ist\cdot\ist\ist, \mathbf{x}_{0}) \,, 
    & \mbox{if}\;\, l \in [N] \rmv\setminus\rmv \supp(\mathbf{x}_{0}) \,, 
  \end{cases}  \quad l \!\in\! [N] \,,
\] 
where $j_{0}$ denotes the index of the $S$-largest (in magnitude) entry of $\mathbf{x}_{0}$. 
Similarly, the bound \eqref{equ_HCRB_SLM} for $|\rmv\supp(\mathbf{x}_{0}) \cup \{k\} | = S \rmv+\rmv 1$ 
is obtained from \eqref{equ_projection_theorem_lower_bound_min_ach_variance} by using for $\mathcal{U}$ the limit of  
$\widetilde{\mathcal{U}}^{(t)} \rmv\rmv\triangleq\ist \linspan\rmv\rmv \big\{ \genericfuncbound_{0}(\cdot), u^{(t)}(\cdot)Ê\big\}$ as $t \!\to\! 0$, 
where $u^{(t)}(\cdot) \triangleq R_{\text{SSNM},\mathbf{x}_{0}}(\ist\cdot\ist\ist, \mathbf{x}_{0}+t\mathbf{e}_{k})-R_{\text{SSNM},\mathbf{x}_{0}} (\ist\cdot\ist\ist, \mathbf{x}_{0})$.
(An expression of $R_{\text{SSNM},\mathbf{x}_{0}} (\ist\cdot\ist\ist, \cdot)$ was given in \eqref{equ_R_SSNM_exp}.)
In \cite{AlexZvikaJournal}, an equivalent bound on the MSE (equivalently, on the variance, because $c(\cdot) \equiv 0$)
was formulated for a vector-valued estimator $\hat{\mathbf{x}}(\cdot)$; that bound can be obtained by summing 
\eqref{equ_HCRB_SLM} over all $k \rmv\in\rmv [N]$. 

It can be shown that the HCRB-type bound \eqref{equ_HCRB_SLM} is tighter (higher) 
than the CRB \eqref{equ_CRB_SSNM} specialized to $c(\cdot) \rmv\equiv\rmv 0$.
For $|\rmv\supp(\mathbf{x}_{0}) \cup \{k\} | = S \rmv+\rmv 1$ (which is true if both
${\| \mathbf{x} \|}_{0} \rmv\rmv=\rmv S$ and 
$k \rmv\not\in \supp(\mathbf{x}_{0})$), the HCRB-type bound \eqref{equ_HCRB_SLM} is a 
strictly upper semi-continuous function of $\mathbf{x}_{0}$, just as the CRB \eqref{equ_CRB_SSNM}.
Hence, it again follows from Corollary \ref{cor_lower_semi_cont_SLGM} that the bound cannot be tight, i.e., in general, we have a strict inequality in \eqref{equ_HCRB_SLM}. 
However, for $|\rmv\supp(\mathbf{x}_{0}) \cup \{k\} | \rmv\le\rmv S$ (which is true either if
${\| \mathbf{x} \|}_{0} \rmv\rmv<\rmv S$ or if both ${\| \mathbf{x} \|}_{0} \rmv\rmv=\rmv S$ and 
$k \rmv\in \supp(\mathbf{x}_{0})$),
the bound \eqref{equ_HCRB_SLM} is tight since it is achieved by the LS estimator $\hat{x}_{\text{LS},k}(\mathbf{y}) = y_{k}$.

\section{Exact versus Approximate Sparsity} 
\label{sec_strict_sparstiy_SLGM}

\begin{figure}[t]
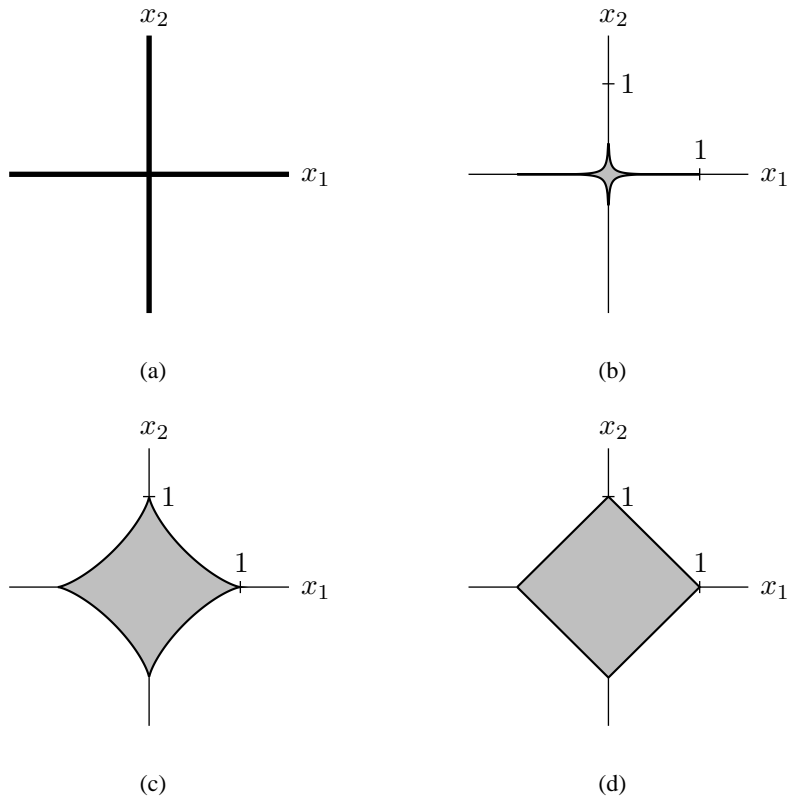

\psset{xunit=0.8cm,yunit=0.8cm}
\begin{center}
\subfigure[]
{\label{fig_B_q_0}
\hspace*{-0.5cm}
\pspicture(-2.5,-2.5)(2.5,2.5) 
\psline[linewidth=2pt](-2.3,0)(2.3,0)
\psline[linewidth=2pt](0,-2.3)(0,2.3) 
\rput[l](2.5,-0.05){$x_{1}$}
\rput[l](-0.15,2.6){$x_{2}$}
\endpspicture
}
\hspace*{2cm}
\subfigure[]
{\label{fig_B_q_0_25}
\hspace*{-0.5cm}
\pspicture(-2.5,-2.5)(2.5,2.5) 
\psline[linewidth=0.5pt](-2.3,0)(2.3,0)
\psline[linewidth=0.5pt](0,-2.3)(0,2.3) 
\pscustom[fillcolor=lightgray, fillstyle=solid]{
\psplot[linewidth=0.1pt,plotpoints=300]{-1.5}{1.5} {1.5 0.25 exp x abs 0.25 exp  sub 4 exp}
\psplot[linewidth=0.1pt,plotpoints=300]{1.5}{-1.5} {1.5 0.25 exp x abs 0.25 exp  sub 4 exp -1 mul}
}
\psline[linewidth=0.5pt](1.5,-0.1)(1.5,0.1)
\psline[linewidth=0.5pt](-0.1,1.5)(0.1,1.5)
\rput[l](1.4,0.4){$1$}
\rput[l](0.2,1.5){$1$}
\rput[l](2.5,-0.05){$x_{1}$}
\rput[l](-0.15,2.6){$x_{2}$}
\endpspicture
} \\[5mm]
\subfigure[]
{
\label{fig_B_q_0_75}
\hspace*{-0.5cm}
\pspicture(-2.5,-2.5)(2.5,2.5) 
\psline[linewidth=0.5pt](-2.3,0)(2.3,0)
\psline[linewidth=0.5pt](0,-2.3)(0,2.3) 
\pscustom[fillcolor=lightgray, fillstyle=solid]{
\psplot[linewidth=0.1pt,plotpoints=300]{-1.5}{1.5} {1.5 0.75 exp x abs 0.75 exp  sub 1.333 exp}
\psplot[linewidth=0.1pt,plotpoints=300]{1.5}{-1.5} {1.5 0.75 exp x abs 0.75 exp  sub 1.333 exp -1 mul}
}
\psline[linewidth=0.5pt](1.5,-0.1)(1.5,0.1)
\psline[linewidth=0.5pt](-0.1,1.5)(0.1,1.5)
\rput[l](1.4,0.4){$1$}
\rput[l](0.2,1.5){$1$}
\rput[l](2.5,-0.05){$x_{1}$}
\rput[l](-0.15,2.6){$x_{2}$}
\endpspicture
}
\hspace*{2cm}
\subfigure[]
{
\label{fig_B_q_1}
\hspace*{-0.5cm}
\pspicture(-2.5,-2.5)(2.5,2.5) 
\psline[linewidth=0.5pt](-2.3,0)(2.3,0)
\psline[linewidth=0.5pt](0,-2.3)(0,2.3) 
\pscustom[fillcolor=lightgray, fillstyle=solid]{
\psplot[linewidth=0.1pt,plotpoints=300]{-1.5}{1.5} {1.5 1 exp x abs 1 exp  sub 1 exp}
\psplot[linewidth=0.1pt,plotpoints=300]{1.5}{-1.5} {1.5 1 exp x abs 1 exp  sub 1 exp -1 mul}
}
\psline[linewidth=0.5pt](1.5,-0.1)(1.5,0.1)
\psline[linewidth=0.5pt](-0.1,1.5)(0.1,1.5)
\rput[l](1.4,0.4){$1$}
\rput[l](0.2,1.5){$1$}
\rput[l](2.5,-0.05){$x_{1}$}
\rput[l](-0.15,2.6){$x_{2}$}
\endpspicture
}
\end{center}
\renewcommand{\baselinestretch}{1.2}\small\normalsize
\caption{Examples of $\ell_{q}$-balls of radius $S \!=\!1$, $\mathcal{B}_{q}(1)$, in $\mathbb{R}^{2}$: 
\subref{fig_B_q_0} $q \!=\! 0$,
\subref{fig_B_q_0_25} $q \!=\! 0.25$,
\subref{fig_B_q_0_75} $q \!=\! 0.75$,
\subref{fig_B_q_1} $q \!=\! 1$.
}
\label{fig_approx_sparse_par_set}
\end{figure}

So far, the parameter set $\mathcal{X}$ 
has been the set $\mathcal{X}_{S}$ of $S$-sparse vectors.
In this section, we consider an approximate version of $S$-sparsity, which is modeled by a modified parameter set $\mathcal{X}$.
Following \cite{Raskutti2011,DonohoJohnstone94}, and \cite{Donoho94idealspatial}, we define this modified parameter set to be
the $\ell_{q}$-ball of radius $S$, i.e.,
\[
\mathcal{X} \ist\ist=\ist\ist \mathcal{B}_{q}(S)Ê\ist\ist\triangleq\ist\ist \big \{ \mathbf{x}' \!\rmv\in\rmv \mathbb{R}^{N} \big| \ist {\| \mathbf{x}' \|}_{q} \leq S \big\} \,, 
  \quad\; \text{with} \; 0 \rmv\leq\rmv q \rmv\leq\rmv 1 \,.
\vspace{-.5mm}
\]
The parameter set $\mathcal{X}_{S}$ of ``exactly'' $S$-sparse vectors is a special case obtained for
$q \rmv=\rmv 0$, i.e., $\mathcal{X}_{S} \rmv=\rmv \mathcal{B}_{0}(S)$. 
In Fig.~\ref{fig_approx_sparse_par_set}, we illustrate $\mathcal{B}_{q}(S)$ in $\mathbb{R}^2$ for 
$S \rmv=\rmv 1$ and various values of $q$. 
In contrast to $\mathcal{X}_{S} \rmv=\rmv \mathcal{B}_{0}(S)$, the parameter sets $\mathcal{B}_{q}(S)$ with $q \rmv>\rmv 0$ are bounded, 
i.e., for every $q \rmv>\rmv 0$ and $S \rmv\in\rmv [N]$, $\mathcal{B}_{q}(S)$ is contained in a finite ball about $\mathbf{0}$.
Thus, the set $\mathcal{X}_{S}$ of exactly $S$-sparse vectors
is not a subset of $\mathcal{B}_{q}(S)$ for any $q \rmv>\rmv 0$.

For a given system matrix $\mathbf{H} \rmv\in\rmv \mathbb{R}^{M \times N}\rmv\rmv$, sparsity degree $S \rmv\leq\rmv N$, and index $k \rmv\in\rmv [N]$, let us consider the estimation 
\vspace*{-1.5mm}
problem
\[
\mathcal{E}^{(q)} \ist\triangleq\ist\ist \big(\mathcal{B}_{q}(S),f_{\mathbf{H}}(\mathbf{y}; \mathbf{x}),g(\mathbf{x}) \rmv=\rmv x_{k} \big) \,.
\] 
Note that $\mathcal{E}^{(q)}$ differs from the SLGM-based estimation problem 
$\mathcal{E}_{\text{SLGM}} = \big(\mathcal{X}_{S},f_{\mathbf{H}}(\mathbf{y}; \mathbf{x}), g(\mathbf{x}) \!=\! x_{k} \big)$ only in the parameter set $\mathcal{X}$, which is 
$\mathcal{B}_{q}(S)$ instead of $\mathcal{X}_{S}$. Because $\mathcal{B}_{0}(S) \rmv=\rmv \mathcal{X}_{S}$, 
we have $\mathcal{E}^{(0)} \rmv\rmv=\rmv \mathcal{E}_{\text{SLGM}}$.
Furthermore, we consider a bias function $c(\cdot) \rmv\rmv: \mathbb{R}^{N} \!\rmv\rightarrow\rmv \mathbb{R}$ that is defined on all of $\mathbb{R}^{N}\rmv\rmv$,
and a parameter vector $\mathbf{x}_{0} \rmv\in\rmv \mathcal{B}_{q}(S) \cap \mathcal{X}_{S}$.
For $\mathcal{E}_{\text{SLGM}}$, as before, the bias function $c(\cdot)$ is prescribed on $\mathcal{X}_{S}$, i.e., we consider estimators $\hat{x}_k(\cdot)$ satisfying
(cf.\ \eqref{equ_bias_prescription})
\vspace*{-2mm}
\[
b(\hat{x}_k(\cdot); \mathbf{x}) \ist=\ist c(\mathbf{x}) \,, \quad\; \text{for all} \;\ist\ist \mathbf{x} \rmv\rmv\in\rmv\rmv \mathcal{X}_{S} \,. 
\]
Again as before, the minimum achievable variance at $\mathbf{x}_{0}$ is denoted as $\minachievvar_{\text{SLGM}}(c(\cdot),\mathbf{x}_{0})$.
On the other hand, for $\mathcal{E}^{(q)}\rmv$, the bias function $c(\cdot)$ is prescribed on $\mathcal{B}_{q}(S)$, i.e., we consider estimators $\hat{x}_k(\cdot)$ satisfying
\[
b(\hat{x}_k(\cdot); \mathbf{x}) \ist=\ist c(\mathbf{x}) \,, \quad\; \text{for all} \;\ist\ist \mathbf{x} \rmv\rmv\in\rmv\rmv \mathcal{B}_{q}(S) \,. 
\]
Here, the minimum achievable variance at $\mathbf{x}_{0}$ is denoted as $\minachievvar^{(q)}(c(\cdot),\mathbf{x}_{0})$. 

Evidently, because $\mathcal{B}_{0}(S) \rmv=\rmv \mathcal{X}_{S}$ and $\mathcal{E}^{(0)} \!=\rmv \mathcal{E}_{\text{SLGM}}$, we have
$\minachievvar^{(0)}(c(\cdot),\mathbf{x}_{0}) \rmv\rmv=\rmv \minachievvar_{\text{SLGM}}(c(\cdot),\mathbf{x}_{0})$.
It seems tempting to conjecture that $\minachievvar^{(q)}(c(\cdot),\mathbf{x}_{0}) \rmv\rmv\approx\rmv \minachievvar_{\text{SLGM}}(c(\cdot),\mathbf{x}_{0})$ for $q \rmv\rmv\approx\rmv 0$,
i.e., changing the parameter set $\mathcal{X}$ from $\mathcal{X}_{S} \rmv=\rmv \mathcal{B}_{0}(S)$ to $\mathcal{B}_{q}(S)$ with $q \rmv>\rmv 0$, and hence 
considering $\mathcal{E}^{(q)}\rmv$ instead of $\mathcal{E}_{\text{SLGM}}$, should not result in a significantly different
minimum achievable variance 
as long as $q$ is sufficiently small. However, the next result \cite[Thm. 5.6.1]{JungPHD} implies that there is a decisive difference, no matter how small $q$ is. 

\begin{theorem}
\label{thm_strict_sparsity_nec_SLM} 
Consider a subset $\mathcal{X} \rmv\subseteq\rmv \mathbb{R}^{N}\rmv$ that contains an open set,
and a function $c(\cdot) \rmv\rmv: \mathbb{R}^{N} \!\rmv\rightarrow\rmv \mathbb{R}$ that is valid at 
some
$\mathbf{x}_{0} \rmv\in\rmv \mathcal{X}$ 
for the LGM-based estimation problem $\mathcal{E}_{\emph{LGM}} = \big( \mathbb{R}^{N} \rmv\rmv,f_{\mathbf{H}}(\mathbf{y};\mathbf{x}),g(\mathbf{x}) \!=\! x_{k} \big)$,
with some system matrix $\mathbf{H}$ that does not necessarily satisfy condition \eqref{equ_spark_cond}. 
Let $\minachievvar_{\emph{LGM}}(c(\cdot), \mathbf{x}_{0})$ denote the minimum achievable variance at $\mathbf{x}_{0}$ for $\mathcal{E}_{\emph{LGM}}$
with bias function $c(\cdot)$ prescribed on $\mathbb{R}^{N}\rmv$. Furthermore let $\minachievvar'(c(\cdot), \mathbf{x}_{0})$ denote the minimum achievable variance at $\mathbf{x}_{0}$ 
for the estimation problem $\mathcal{E}' \rmv\triangleq \big(\mathcal{X},f_{\mathbf{H}} (\mathbf{y}; \mathbf{x}), g(\mathbf{x}) \!=\! x_{k} \big)$
with bias function $c(\cdot)$ prescribed on $\mathcal{X}$. 
\vspace{-1mm}
Then 
\[
\minachievvar'(c(\cdot), \mathbf{x}_{0}) \ist=\ist \minachievvar_{\emph{LGM}}(c(\cdot), \mathbf{x}_{0}) \,.
\vspace{-1mm}
\]
Moreover, the LMV 
estimator\footnote{This 
estimator is given 
by Part 3
of Theorem \ref{thm_Barankin_bound_LMV} specialized to $S \!=\! N$ (in which case the SLGM reduces to the 
LGM).} 
$\hat{g}_{\emph{LGM}}^{(c(\cdot),\mathbf{x}_{0})}(\cdot)$ for 
$\mathcal{E}_{\emph{LGM}}$ and bias function $c(\cdot)$ is simultaneously the LMV estimator for $\mathcal{E}'\rmv$ and bias function $c(\cdot) \big|_{\mathcal{X}}\ist$.
\vspace{1.5mm}
\end{theorem} 

Since for $q \rmv>\rmv 0$, the parameter set $\mathcal{X}=\mathcal{B}_{q}(S)$ contains an open set,
Theorem \ref{thm_strict_sparsity_nec_SLM} implies that 
\[
\minachievvar^{(q)}(c(\cdot), \mathbf{x}_{0}) \ist=\ist \minachievvar_{\text{LGM}}(c(\cdot), \mathbf{x}_{0}) \,, \quad\; \text{for all} \; q \rmv>\rmv 0 \,.
\]
Thus, 
the minimum achievable variance for $\mathcal{E}^{(q)}\rmv\rmv$, $q \rmv>\rmv 0$ with bias function $c(\cdot)$ prescribed on $\mathcal{B}_{q}(S)$
is always equal to the minimum achievable variance for $\mathcal{E}_{\text{LGM}}$ with bias function $c(\cdot)$ prescribed on $\mathbb{R}^{N} \rmv\rmv$. 
Furthermore, 
Theorem \ref{thm_strict_sparsity_nec_SLM} also implies that
the minimum achievable variance for $\mathcal{E}^{(q)} \rmv=\rmv \big(\mathcal{B}_{q}(S),f_{\mathbf{H}}(\mathbf{y}; \mathbf{x}),g(\mathbf{x}) \rmv\rmv=\rmv\rmv x_{k} \big)$, $q \rmv>\rmv 0$ 
is achieved by the LMV estimator for 
$\mathcal{E}_{\text{LGM}} = \big( \mathbb{R}^{N} \rmv\rmv,f_{\mathbf{H}}(\mathbf{y};\mathbf{x}),g(\mathbf{x}) \!=\! x_{k} \big)$.
But since in general\linebreak 
$\minachievvar_{\text{LGM}} (c(\cdot), \mathbf{x}_{0}) >  \minachievvar_{\text{SLGM}}(c(\cdot), \mathbf{x}_{0})$ 
(see \eqref{equ_diff_min_achiev_var_SSNM} for the special case given by the SSNM), 
it follows that\linebreak 
$\minachievvar^{(q)}(c(\cdot), \mathbf{x}_{0}) = \minachievvar_{\text{LGM}}(c(\cdot), \mathbf{x}_{0})$ does not generally converge to 
$\minachievvar_{\text{SLGM}}(c(\cdot), \mathbf{x}_{0})$ as $q$ approaches $0$.

For another interesting consequence of Theorem \ref{thm_strict_sparsity_nec_SLM}, consider an estimation problem $\mathcal{E}=\big( \mathcal{X} \rmv\rmv,f_{\mathbf{H}}(\mathbf{y};\mathbf{x}),g(\mathbf{x}) \!=\! x_{k} \big)$ whose parameter set $\mathcal{X}$ is the union of the set of exactly 
$S$-sparse vectors $\mathcal{X}_{S}$ and an open ball $\mathcal{B}(\mathbf{x}_{c},r) \triangleq \{ \mathbf{x} \in \mathbb{R}^{N} \,\, | \,\, \| \mathbf{x}- \mathbf{x}_{c} \|_{2} < r \}$), i.e., $\mathcal{X} = \mathcal{X}_{S} \cup \mathcal{B}(\mathbf{x}_{c},r)$. 
Then, it follows from Theorem 
\ref{thm_strict_sparsity_nec_SLM} that the minimum achievable variance for $\mathcal{E}$ at any sparse $\mathbf{x}_{0} \in \mathcal{X}_{S}$ coincides with $\minachievvar_{\text{LGM}} (c(\cdot), \mathbf{x}_{0})$. Since in general $\minachievvar_{\text{LGM}} (c(\cdot), \mathbf{x}_{0}) >  \minachievvar_{\text{SLGM}}(c(\cdot), \mathbf{x}_{0})$ this implies that the minimum achievable variance for $\mathcal{E}$ is in general strictly larger than the 
minimum achievable variance for the SLGM. Thus, no matter how small the radius $r$ is and how distant $\mathbf{x}_{c}$ is from $\mathcal{X}_{S}$, the inclusion of the open ball in $\mathcal{X}$ significantly affects the MVE of the $S$-sparse vectors in $\mathcal{X}_{S}$. 


The statement of Theorem \ref{thm_strict_sparsity_nec_SLM} is closely related to the facts that (i) the statistical model of the LGM belongs to an exponential family,
and (ii) the mean function $\gamma(\mathbf{x}) = \expect_{\mathbf{x}} \{ \hat{g}(\mathbf{y}) \}$ of any estimator $\hat{g}(\cdot)$ 
with finite bias and variance for an estimation problem whose 
statistical model belongs to an exponential family is an analytic function \cite[Lemma 2.8]{FundmentExpFamBrown}. 
Indeed, any analytic function is completely determined 
by its 
values on an arbitrary open set in its domain \cite{KranzPrimerAnalytic}. Therefore, because the mean function $\gamma(\mathbf{x})$ of any estimator
for the LGM is analytic, it is completely specified by its values for all $\mathbf{x} \rmv\in\rmv \mathcal{B}_{q}(S)$ with an arbitrary $q \rmv>\rmv 0$
(note that $\mathcal{B}_{q}(S)$ contains an open set).

\section{Numerical Results} 
\label{sec_numerical_SLGM}

In this section, we compare 
the lower variance bounds presented in Section \ref{sec_bounds_SLGM}
with the actual variance behavior of some well-known estimators. We consider the SLGM-based
estimation problem $\mathcal{E}_{\text{SLGM}} = \big(\mathcal{X}_{S},f_{\mathbf{H}}(\mathbf{y}; \mathbf{x}),g(\mathbf{x}) \!=\! x_{k} \big)$ for $k \rmv\in\rmv [N]$.
In what follows, we will denote the lower bounds \eqref{equ_CRB_SLM}, \eqref{equ_bound_asilomar_1}, and \eqref{equ_bound_asilomar_2} by 
$B_k^{(1)}(c(\cdot),\mathbf{x}_{0})$, $B_k^{(2)}(c(\cdot),\mathbf{x}_{0})$, and $B_k^{(3)}(c(\cdot),\mathbf{x}_{0})$, respectively. 
We recall that the latter two bounds depend on
an index set $\mathcal{K} \rmv\subseteq\rmv [N]$ with $|\mathcal{K}| \rmv\leq\rmv S$, which can be chosen freely.

Let $\hat{\mathbf{x}}(\cdot)$ be an estimator of $\mathbf{x}$ with bias function $\mathbf{c}(\cdot)$. Because of \eqref{equ_sum_var_scalar}, 
a lower bound on the estimator variance $v(\hat{\mathbf{x}}(\cdot);\mathbf{x}_{0})$ can be obtained by summing with respect to $k \rmv\in\rmv [N]$ the ``scalar bounds'' 
$B_k^{(1)}(c_k(\cdot),\mathbf{x}_{0})$ or $B_k^{(2)}(c_k(\cdot),\mathbf{x}_{0})$ or $B_k^{(3)}(c_k(\cdot),\mathbf{x}_{0})$,
where $c_{k}(\cdot) \triangleq \big(\mathbf{c}(\cdot)\big)_k$, i.e., 
\begin{equation}
\label{equ_lower_bound_var_vec_est_numerical_SLM}
v(\hat{\mathbf{x}}(\cdot);\mathbf{x}_{0}) \,\geq\, B^{(1/2/3)}(\mathbf{c}(\cdot),\mathbf{x}_{0})
\,\triangleq \sum_{k \in [N]} \rmv\rmv B_k^{(1/2/3)}(c_k(\cdot),\mathbf{x}_{0}) \,.
\vspace*{-.5mm}
\end{equation}
Here, the index sets $\mathcal{K}_k$ used in $B_k^{(2)}(c_k(\cdot),\mathbf{x}_{0})$ and $B_k^{(3)}(c_k(\cdot),\mathbf{x}_{0})$
can be chosen differently for different $k$.

\subsection{An SLGM View of Fourier Analysis}

Our first example is inspired by \cite[Example 4.2]{kay}.
We consider the 
SLGM with $N$ even, i.e., $N \rmv=\rmv 2L$, and $\sigma^{2} \rmv=\rmv 1$. The system matrix $\mathbf{H} \rmv\in\rmv \mathbb{R}^{M \times 2L}$ is given by
$H_{m,l} = \cos\!\big( \theta_{l} (m\!-\!1) \big)$ for $m \rmv\in\rmv [M]$ and $l \rmv\in\rmv [L]$ and 
$H_{m,l} = \sin\!\big( \theta_{l} (m\!-\!1) \big)$ for $m \rmv\in\rmv [M]$ and $l \rmv\in\rmv \{L+1,\ldots,2L\}$.
Here, the normalized angular frequencies $\theta_{l}$ are uniformly spaced according to $\theta_{l} = \theta_{0} + \big[ (l \!-\! 1) \,\, \mbox{mod} \,\, LÊ\big] \ist \Delta \theta$, $l \rmv\in\rmv [N]$.
The multiplication of $\mathbf{x}$ by $\mathbf{H}$ then
corresponds to an
inverse discrete Fourier transform that
maps $2L$ spectral samples
(the entries of $\mathbf{x}$) to $M$ temporal samples (the entries of $\mathbf{H}\mathbf{x}$).
In our simulation, we chose $M \rmv=\rmv 128$, $L \rmv=\rmv 8$ (hence, $N \rmv=\rmv 16$), $S \rmv=\rmv 4$, $\theta_{0} \rmv=\rmv 0.2$, and $\Delta \theta \rmv=\rmv 3.9 \cdot 10^{-3}\rmv$. 
The frequency spacing $\Delta \theta$ is about half the nominal DFT frequency resolution, which is $1/128 \approx 7.8 \times 10^{-3}\rmv$. 

\begin{figure}[t]
\vspace{-1mm}
\centering
\psfrag{SNR}[c][c][.9]{\uput{3.4mm}[270]{0}{\hspace{0mm}SNR [dB]}}
\psfrag{title}[c][c][.9]{\uput{2.5mm}[270]{0}{}}
\psfrag{x_m20}[c][c][.9]{\uput{0.3mm}[270]{0}{$-20$}}
\psfrag{x_m_10}[c][c][.9]{\uput{0.3mm}[270]{0}{$-10$}}
\psfrag{x_0}[c][c][.9]{\uput{0.3mm}[270]{0}{$0$}}
\psfrag{x_10}[c][c][.9]{\uput{0.3mm}[270]{0}{$10$}}
\psfrag{x_20}[c][c][.9]{\uput{0.3mm}[270]{0}{$20$}}
\psfrag{x_30}[c][c][.9]{\uput{0.3mm}[270]{0}{$30$}}
\psfrag{x_40}[c][c][.9]{\uput{0.3mm}[270]{0}{$40$}}
\psfrag{y_0}[c][c][.9]{\uput{0.1mm}[180]{0}{$0$}}
\psfrag{y_4}[c][c][.9]{\uput{0.1mm}[180]{0}{$4$}}
\psfrag{y_8}[c][c][.9]{\uput{0.1mm}[180]{0}{$8$}}
\psfrag{y_12}[c][c][.9]{\uput{0.1mm}[180]{0}{$12$}}
\psfrag{y_16}[c][c][.9]{\uput{0.1mm}[180]{0}{$16$}}
\psfrag{y_20}[c][c][.9]{\uput{0.1mm}[180]{0}{$20$}}
\psfrag{y_24}[c][c][.9]{\uput{0.1mm}[180]{0}{$24$}}
\psfrag{variance}[c][c][.9]{\uput{4mm}[90]{0}{\hspace*{-3mm}variance/bound}}
\psfrag{bML}[l][l][0.8]{bound on $v(\ML_est(\cdot);\mathbf{x}_0)$}
\psfrag{data1}[l][l][0.8]{\hspace*{1mm}$v(\hat{\mathbf{x}}_{\text{OMP}}(\cdot); \mathbf{x}_{0})$}
\psfrag{data2}[l][l][0.8]{\hspace*{1mm}$B^{(3)}(\mathbf{c}_{\text{OMP}}(\cdot),\mathbf{x}_{0})$}
\psfrag{data3}[l][l][0.8]{\hspace*{1mm}$B^{(2)}(\mathbf{c}_{\text{OMP}}(\cdot),\mathbf{x}_{0})$}
\psfrag{data4}[l][l][0.8]{\hspace*{1mm}$B^{(1)}(\mathbf{c}_{\text{OMP}}(\cdot),\mathbf{x}_{0})$}
\psfrag{CRB}[l][l][0.8]{Oracle CRB}
\psfrag{MLarrow}[l][l][0.8]{ML}
\psfrag{T3}[l][l][0.8]{HT \!($T \!\rmv=\! 3$)}
\psfrag{T4}[l][l][0.8]{HT \!($T \!\rmv=\! 4$)}
\psfrag{T5}[l][l][0.8]{HT \!($T \!\rmv=\! 5$)}
\centering
\hspace*{-0mm}\includegraphics[height=7cm,width=12cm]{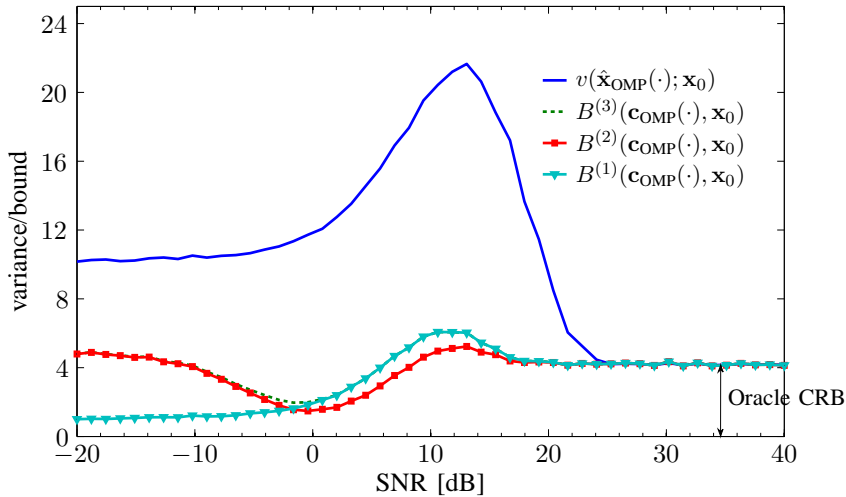}
\vspace*{-.5mm}
\renewcommand{\baselinestretch}{1.2}\small\normalsize
  \caption{Variance of the OMP estimator and corresponding lower bounds versus SNR,
  for the SLGM with $N \rmv=\rmv 16$, $M \rmv=\rmv 128$, $S \rmv=\rmv 4$, and $\sigma^{2} \rmv=\rmv 1$.} 
\label{fig_SLM_high_res}
\vspace*{3.5mm}
\end{figure}

We consider the OMP estimator $\hat{\mathbf{x}}_{\text{OMP}}(\cdot)$ that is obtained by applying the OMP \cite{GreedisGood,TroppGilbertOMP} with 
$S \rmv\rmv=\rmv\rmv 4$\linebreak 
iterations to the observation $\mathbf{y}$. We used Monte Carlo simulation with randomly generated noise $\mathbf{n} \rmv\rmv\sim\rmv\rmv \mathcal{N}(\mathbf{0}, \mathbf{I})$ 
to estimate the variance $v(\hat{\mathbf{x}}_{\text{OMP}}(\cdot); \mathbf{x}_{0})$ of $\hat{\mathbf{x}}_{\text{OMP}}(\cdot)$. 
The parameter vector was chosen as $\mathbf{x}_{0} = \sqrt{\text{SNR}} \,
\tilde{\mathbf{x}}_{0}$, 
where $\tilde{\mathbf{x}}_{0} \in \{0,1\}^{16}$, $\supp(\tilde{\mathbf{x}}_{0}) = \{3,6,11,14\}$, and $\text{SNR}$ varies between $10^{-2}$ and $10^{4}\rmv$. 
Thus, the observation $\mathbf{y}$ 
is a noisy superposition of four sinusoidal components with identical amplitudes; two of them are consine and sine components with frequency $\theta_{3} = \theta_{11} = \theta_{0}+2 \Delta \theta$, 
and two are cosine and sine components with frequency $\theta_{6}  = \theta_{14}  = \theta_{0} + 5 \Delta \theta$.
In Fig.\ \ref{fig_SLM_high_res}, we plot $v(\hat{\mathbf{x}}_{\text{OMP}}(\cdot); \mathbf{x}_{0})$ versus SNR. 
For comparison, we also plot the lower bounds 
$B^{(1)}(\mathbf{c}_{\text{OMP}}(\cdot),\mathbf{x}_{0})$, $B^{(2)}(\mathbf{c}_{\text{OMP}}(\cdot),\mathbf{x}_{0})$, and $B^{(3)}(\mathbf{c}_{\text{OMP}}(\cdot),\mathbf{x}_{0})$
in \eqref{equ_lower_bound_var_vec_est_numerical_SLM}, with 
$\mathbf{c}_{\text{OMP}}(\mathbf{x}) \triangleq \mathbf{b}(\hat{\mathbf{x}}_{\text{OMP}}(\cdot); \mathbf{x})$ being the actual bias function 
of the OMP estimator $\hat{\mathbf{x}}_{\text{OMP}}(\cdot)$. 
To evaluate these bounds, we computed the first-order partial derivatives of the bias functions $c_{\text{OMP},k}(\mathbf{x})$ 
(see Theorems \ref{thm_CRB_SLM} and \ref{thm_bound_asilomar})
by means of \eqref{equ_expr_par_der_CS_recovery_scheme}
and Monte Carlo simulation (see \cite{HeroUniformCRB} for details).
The index sets $\mathcal{K}_{k}$ in the bounds $B^{(2)}(\mathbf{c}_{\text{OMP}}(\cdot),\mathbf{x}_{0})$ and $B^{(3)}(\mathbf{c}_{\text{OMP}}(\cdot),\mathbf{x}_{0})$ 
were chosen as $\mathcal{K}_{k} = \supp(\mathbf{x}_{0})$ for $k \in \supp(\mathbf{x}_{0})$ and $\mathcal{K}_{k} = \{k\}$ for $k \notin \supp(\mathbf{x}_{0})$. 
This is the simplest nontrivial choice of the $\mathcal{K}_{k}$ for which $B^{(3)}(\mathbf{c}_{\text{OMP}}(\cdot),\mathbf{x}_{0})$ 
is tighter than the state-of-the-art bound $B^{(1)}(\mathbf{c}_{\text{OMP}}(\cdot),\mathbf{x}_{0})$ (the sparse CRB, which was originally presented
in \cite{ZvikaCRB}). 
Finally, Fig.~\ref{fig_SLM_high_res} also shows the ``oracle CRB,'' which is defined as the CRB for known $\supp(\mathbf{x}_{0})$. This is simply the 
CRB for a linear {G}aussian model with system matrix $\mathbf{H}_{\supp(\mathbf{x}_{0})}$ and is thus 
given by $\trace \!\big( \big( \mathbf{H}_{\supp(\mathbf{x}_{0})}^{T} \mathbf{H}_{\supp(\mathbf{x}_{0})} \big)^{\rmv\rmv -1} \big) \approx 4.19$ \cite{kay} 
for all values of SNR (recall that we set $\sigma^{2} \rmv=\rmv 1$).  

As can be seen from Fig.~\ref{fig_SLM_high_res}, for SNR below 20\,dB, $v(\hat{\mathbf{x}}_{\text{OMP}}(\cdot); \mathbf{x}_{0})$ is significantly higher than 
the
four lower bounds. This suggests that there might exist estimators with the same bias as that of the OMP estimator but a smaller variance; however,
a positive statement regarding the existence of such estimators 
cannot be based on our analysis.
For SNR larger than about 15\,dB,
the four lower bounds coincide. Furthermore, for SNR larger than about 11\,dB, $v(\hat{\mathbf{x}}_{\text{OMP}}(\cdot); \mathbf{x}_{0})$ quickly 
converges toward the lower bounds.
This is because for high SNR, the OMP estimator is able to detect $\supp(\mathbf{x}_{0})$ with very high probability. 
Note also that the results in Fig.\ \ref{fig_SLM_high_res} agree with our observation in Section \ref{sec_RKHS_Lower_Bounds_SLGM_improved},
around \eqref{equ_diff_asilomar_bounds}, that the bound $B^{(3)}(\mathbf{c}(\cdot),\mathbf{x}_{0})$ tends 
to be higher than $B^{(2)}(\mathbf{c}(\cdot),\mathbf{x}_{0})$.

\subsection{Minimum Variance Analysis for the SSNM}


Next, we consider the maximum likelihood (ML) estimator and the hard-thresholding (HT) estimator for the SSNM, i.e., for $M \rmv=\rmv N$ and $\mathbf{H} \rmv=\rmv \mathbf{I}$,
with $N \rmv=\rmv 50$, $S \rmv=\rmv 5$, and $\sigma^2 \rmv=\rmv 1$.
The ML estimator is given by 
\[
\ML_est(\mathbf{y}) \,\triangleq\, \argmax_{\mathbf{x}' \in \mathcal{X}_{S}} f ( \mathbf{y};  \mathbf{x}' ) \ist\ist=\, {\mathsf P}_{\!S} ( \mathbf{y}  ) \,,
\]
where the operator $\mathsf{P}_{\! S}$ retains the $S$ largest (in magnitude) entries and zeros all other entries. 
Closed-form expressions of the mean and variance of the ML estimator were derived in \cite{AlexZvikaJournal}.
The HT estimator $\hat{\mathbf{x}}_{\text{HT}}  (\cdot)$ is given 
\vspace{-1.5mm}
by
\begin{equation}
\label{equ_def_thr_func}
\hat{x}_{\text{HT},k}  (\mathbf{y}) \eq \hat{x}_{\text{HT},k}  (y_k) \,=\ist \begin{cases} y_k \,,  & |y_k| \geq T\\[-1mm]
  0 \,, & \text{else} \ist,
\end{cases} 
\qquad k \rmv\in\rmv [N] \,,
\end{equation} 
where $T$ is a fixed threshold. Note that in the limiting case 
$T=0$, the HT estimator coincides with the LS estimator 
$\hat{\mathbf{x}}_{\text{LS}}(\mathbf{y}) = \mathbf{y}$ \cite{kay,scharf91,LC}.
The mean and variance of the HT estimator are given by 
\begin{align}
\expect_{\mathbf{x}} \big\{ \hat{x}_{\text{HT},k}(\mathbf{y}) \big\} &\eq \frac{1}{\sqrt{2 \pi \sigma^{2}}} \int_{\mathbb{R} \setminus [-T,T]} y \,
  \exp \rmv\rmv\rmv\bigg( \!\!-\rmv\rmv  \frac{1}{2 \sigma^{2}} (y \rmv-\rmv x_{k})^{2}\rmv \bigg) \, dy
  \label{equ_mean_hard_th_est_ssnm}\\[1mm]
v(\hat{x}_{\text{HT},k}(\cdot); \mathbf{x}) &\eq
\frac{1}{\sqrt{2 \pi \sigma^{2}}} \int_{\mathbb{R} \setminus [-T,T]} y^{2} \,
  \exp \rmv\rmv\rmv\bigg( \!\!-\rmv\rmv  \frac{1}{2 \sigma^{2}} (y \rmv-\rmv x_{k})^{2}\rmv \bigg) \, dy
  \ist\ist-\ist\ist \big( \expect_{\mathbf{x}} \big\{ \hat{x}_{\text{HT},k}(\mathbf{y}) \big\} \big)^{2} .
  \label{equ_var_hard_th_est_ssnm}
\end{align}

\begin{figure}[t]
\vspace{-1mm}
\centering
\psfrag{SNR}[c][c][.9]{\uput{3.4mm}[270]{0}{\hspace{3mm}SNR [dB]}}
\psfrag{title}[c][c][.9]{\uput{2.5mm}[270]{0}{}}
\psfrag{x_5}[c][c][.9]{\uput{2.5mm}[270]{0}{}}
\psfrag{x_15}[c][c][.9]{\uput{2.5mm}[270]{0}{}}
\psfrag{x_m_5}[c][c][.9]{\uput{2.5mm}[270]{0}{}}
\psfrag{x_m_15}[c][c][.9]{\uput{2.5mm}[270]{0}{}}
\psfrag{x_0}[c][c][.9]{\uput{0.3mm}[270]{0}{$0$}}
\psfrag{x_m_10}[c][c][.9]{\uput{0.3mm}[270]{0}{$-10$}}
\psfrag{x_m_20}[c][c][.9]{\uput{0.3mm}[270]{0}{$-20$}}
\psfrag{x_10}[c][c][.9]{\uput{0.3mm}[270]{0}{$10$}}
\psfrag{x_20}[c][c][.9]{\uput{0.3mm}[270]{0}{$20$}}
\psfrag{y_0}[c][c][.9]{\uput{0.1mm}[180]{0}{$0$}}
\psfrag{y_5}[c][c][.9]{\uput{0.1mm}[180]{0}{$5$}}
\psfrag{y_10}[c][c][.9]{\uput{0.1mm}[180]{0}{$10$}}
\psfrag{y_15}[c][c][.9]{\uput{0.1mm}[180]{0}{$15$}}
\psfrag{y_20}[c][c][.9]{\uput{0.1mm}[180]{0}{$20$}}
\psfrag{y_25}[c][c][.9]{\uput{0.1mm}[180]{0}{$25$}}
\psfrag{y_30}[c][c][.9]{\uput{0.1mm}[180]{0}{$30$}}
\psfrag{y_35}[c][c][.9]{\uput{0.1mm}[180]{0}{$35$}}
\psfrag{y_40}[c][c][.9]{\uput{0.1mm}[180]{0}{$40$}}
\psfrag{y_45}[c][c][.9]{\uput{0.1mm}[180]{0}{$45$}}
\psfrag{y_50}[c][c][.9]{\uput{0.1mm}[180]{0}{$50$}}
\psfrag{y_55}[c][c][.9]{\uput{0.1mm}[180]{0}{$55$}}
\psfrag{y_60}[c][c][.9]{\uput{0.1mm}[180]{0}{$60$}}
\psfrag{variance}[c][c][.9]{\uput{3mm}[90]{0}{variance/bound}}
\psfrag{bML}[l][l][0.8]{bound on $v(\ML_est(\cdot);\mathbf{x}_0)$}
\psfrag{ML}[l][l][0.8]{ML}
\psfrag{data3}[l][l][0.8]{$v(\hat{\mathbf{x}}_{\text{HT}}(\cdot); \mathbf{x}_0)$, $T \!\rmv=\! 2$}
\psfrag{data5}[l][l][0.8]{$v(\hat{\mathbf{x}}_{\text{HT}}(\cdot); \mathbf{x}_0)$, $T \!\rmv=\! 3$}
\psfrag{data7}[l][l][0.8]{$v(\hat{\mathbf{x}}_{\text{HT}}(\cdot); \mathbf{x}_0)$, $T \!\rmv=\! 4$}
\psfrag{data4}[l][l][0.8]{$B^{(2/3)}(\mathbf{c}_{\text{HT}}(\cdot),\mathbf{x}_{0})$, $T \!\rmv=\! 2$}
\psfrag{data6}[l][l][0.8]{$B^{(2/3)}(\mathbf{c}_{\text{HT}}(\cdot),\mathbf{x}_{0})$, $T \!\rmv=\! 3$}
\psfrag{data8}[l][l][0.8]{$B^{(2/3)}(\mathbf{c}_{\text{HT}}(\cdot),\mathbf{x}_{0})$, $T \!\rmv=\! 4$}
\psfrag{data1}[l][l][0.8]{$v(\hat{\mathbf{x}}_{\text{LS}}(\cdot); \mathbf{x}_0) = v(\hat{\mathbf{x}}_{\text{HT}}(\cdot); \mathbf{x}_0)$, $T \!\rmv=\! 0$}
\psfrag{data2}[l][l][0.8]{$B^{(2/3)}(\mathbf{c}_{\text{HT}}(\cdot),\mathbf{x}_{0})$, $T \!\rmv=\! 0$}
\psfrag{data9}[l][l][0.8]{$v(\hat{\mathbf{x}}_{\text{ML}}(\cdot); \mathbf{x}_0)$}
\psfrag{data10}[l][l][0.8]{$B^{(2/3)}(\mathbf{c}_{\text{ML}}(\cdot),\mathbf{x}_{0})$}
\psfrag{MLarrow}[l][l][0.8]{ML}
\psfrag{T=3}[l][l][0.8]{\uput{0mm}[90]{0}{\hspace{0mm}$T \!\rmv=\! 3$}}
\psfrag{T=4}[l][l][0.8]{$T \!\rmv=\! 4$}
\psfrag{T=2}[l][l][0.8]{\uput{.5mm}[0]{0}{\vspace*{3mm} \!\!\!\!$T \!\rmv=\! 2$}}
\psfrag{T=0}[l][l][0.8]{\uput{0mm}[0]{0}{\hspace{-10mm}$T \!\rmv=\! 0\,\ist$(LS)}}
\psfrag{oracle}[l][l][0.8]{\uput{0mm}[0]{0}{$S \sigma^{2}$}}
\centering
\hspace*{-25mm}\includegraphics[height=8cm,width=15cm]{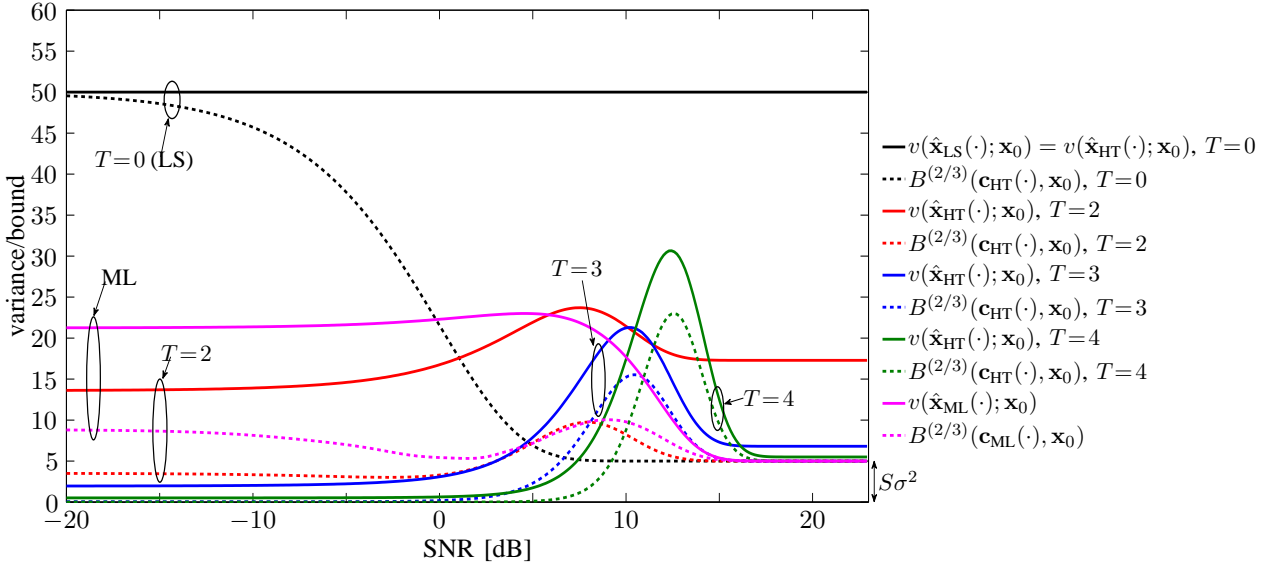}
\vspace*{1mm}
\renewcommand{\baselinestretch}{1.2}\small\normalsize
  \caption{Variance of the ML and HT estimators and corresponding lower bounds versus SNR,
  for the SSNM
  with $N \rmv=\rmv 50$, $S \rmv=\rmv 5$, and $\sigma^{2} \rmv=\rmv 1$.} 
\label{fig_bounds_1}
\vspace*{1mm}
\end{figure}

We calculated 
the variances $v(\hat{\mathbf{x}}_{\text{ML}}(\cdot); \mathbf{x}_{0})$ and $v(\hat{\mathbf{x}}_{\text{HT}}(\cdot); \mathbf{x}_{0})$
at parameter vectors $\mathbf{x}_{0} = \sqrt{\text{SNR}} \,
\tilde{\mathbf{x}}_{0}$, 
where $\tilde{\mathbf{x}}_{0} \in \{0,1\}^{50}$, 
$\supp(\tilde{\mathbf{x}}_{0}) = [S]$, and $\text{SNR}$ 
varies between $10^{-2}$ and $10^{2}\rmv$.
(The fixed choice $\supp(\mathbf{x}_{0}) = [S]$ is justified by the fact that neither the variances of the ML and HT estimators nor the corresponding variance bounds 
depend on the location of $\supp(\mathbf{x}_{0})$.)
In particular, $v(\hat{\mathbf{x}}_{\text{HT}}(\cdot); \mathbf{x}_{0})$ was calculated by numerical evaluation of the integrals \eqref{equ_var_hard_th_est_ssnm}
and \eqref{equ_mean_hard_th_est_ssnm}.
Fig.~\ref{fig_bounds_1} shows $v(\ML_est(\cdot); \mathbf{x}_{0})$ and $v( \hat{\mathbf{x}}_{\text{HT}} (\cdot); \mathbf{x}_0)$---the latter for four different choices of 
$T$ in \eqref{equ_def_thr_func}---versus SNR. 
Also shown are the lower bounds 
$B^{(2)}(\mathbf{c}_{\text{ML}}(\cdot),\mathbf{x}_{0})$ and $B^{(3)}(\mathbf{c}_{\text{ML}}(\cdot),\mathbf{x}_{0})$ as well as 
$B^{(2)}(\mathbf{c}_{\text{HT}}(\cdot),\mathbf{x}_{0})$ and $B^{(3)}(\mathbf{c}_{\text{HT}}(\cdot),\mathbf{x}_{0})$ (cf.\ \eqref{equ_lower_bound_var_vec_est_numerical_SLM}), 
with $\mathbf{c}_{\text{ML}}(\cdot)$ and $\mathbf{c}_{\text{HT}}(\cdot)$ being the actual bias functions 
of $\hat{\mathbf{x}}_{\text{ML}}(\cdot)$ and of $\hat{\mathbf{x}}_{\text{HT}}(\cdot)$, respectively. The index sets underlying the bounds were chosen
as $\mathcal{K}_{k} \rmv= \supp(\mathbf{x}_0)$ 
for $k \rmv\in\rmv \supp(\mathbf{x}_0)$ and $\mathcal{K}_{k} = \{k\} \cup \{ \supp(\mathbf{x}_{0}) \rmv\rmv\setminus\rmv\rmv \{j_{S} \} \}$ 
for $k \notin \supp(\mathbf{x}_{0})$, where $j_{S}$ denotes the index of the $S$-largest (in magnitude) entry of $\mathbf{x}_{0}$. 
For this choice of the $\mathcal{K}_{k}$, the two bounds are equal, i.e., 
$B^{(2)}(\mathbf{c}_{\text{ML}}(\cdot),\mathbf{x}_{0}) = B^{(3)}(\mathbf{c}_{\text{ML}}(\cdot),\mathbf{x}_{0})$ and
$B^{(2)}(\mathbf{c}_{\text{HT}}(\cdot),\mathbf{x}_{0}) = B^{(3)}(\mathbf{c}_{\text{HT}}(\cdot),\mathbf{x}_{0})$.
The first-order partial derivatives of the bias functions $c_{\text{ML},k}(\mathbf{x})$
involved in the bounds $B^{(2/3)}(\mathbf{c}_{\text{ML}}(\cdot),\mathbf{x}_{0})$ were approximated by a finite-difference quotient \cite{HeroUniformCRB}, 
i.e., $\frac{\partial c_{\text{ML},k}(\mathbf{x}) }{\partial x_{l}} = \delta_{k,l} \ist+ \frac{\partial \mathsf{E}_{\mathbf{x}} \{\hat{x}_{\text{ML},k}(\mathbf{y}) \} }{\partial x_{l}}$ with 
\[
\frac{\partial \mathsf{E}_{\mathbf{x}} \big\{\hat{x}_{\text{ML},k}(\mathbf{y})\big\} }{\partial x_{l}} 
  \,\approx\, \frac{ \mathsf{E}_{\mathbf{x}+ \Delta \mathbf{e}_{l}} \big\{\hat{x}_{\text{ML},k}(\mathbf{y})\big\} -\ist \mathsf{E}_{\mathbf{x}} \big\{\hat{x}_{\text{ML},k}(\mathbf{y})\big\} }{\Delta} \,, 
\]  
where $\Delta \rmv>\rmv 0$ is a small stepsize and the expectations were calculated using the closed-form expressions presented in \cite[Appendix I]{AlexZvikaJournal}.
The first-order partial derivatives of the bias functions $c_{\text{HT},k}(\mathbf{x})$
involved in the bounds $B^{(2/3)}(\mathbf{c}_{\text{HT}}(\cdot),\mathbf{x}_{0})$ were calculated by means of \eqref{equ_expr_par_der_CS_recovery_scheme}.


It can be seen in Fig.~\ref{fig_bounds_1} that for SNR larger than about 18 dB, 
the variances of the ML and HT estimators and the corresponding bounds are effectively equal (for the HT estimator, 
this is true if $T$ is not too small). Also, all bounds are close to $S \sigma^{2} \rmv=\rmv 4$;
this equals the variance of an oracle estimator that knows $\supp(\mathbf{x}_{0})$ and 
is given by $\hat{x}_{k}(\mathbf{y}) = y_{k}$ for $k \in \supp(\mathbf{x}_{0})$ and $\hat{x}_{k}(\mathbf{y}) = 0$ otherwise. 
However, in the medium-SNR
range, the variances of the ML and HT estimators are significantly higher than
the corresponding lower bounds. 
We can conclude that there might exist estimators with the same bias as that of the ML or HT estimator but a smaller variance; however, in general,
a positive statement regarding the existence of such estimators 
cannot be based on our analysis. 


\begin{figure}[t]
\vspace{-1mm}
\centering
\psfrag{SNR}[c][c][.85]{\uput{4mm}[270]{0}{\hspace{0mm}SNR\,[dB]}}
\psfrag{title}[c][c][.9]{\uput{2.5mm}[270]{0}{}}
\psfrag{x_m_20}[c][c][.9]{\uput{0.3mm}[270]{0}{$-20$}}
\psfrag{x_m_10}[c][c][.9]{\uput{0.3mm}[270]{0}{$-10$}}
\psfrag{x_0}[c][c][.8]{\uput{0.3mm}[270]{0}{\hspace{.3mm}$0$}}
\psfrag{x_10}[c][c][.8]{\uput{0.3mm}[270]{0}{\hspace{.2mm}$10$}}
\psfrag{x_20}[c][c][.8]{\uput{0.3mm}[270]{0}{\hspace{.2mm}$20$}}
\psfrag{y_0}[c][c][.9]{\uput{0.1mm}[180]{0}{$0$}}
\psfrag{y_5}[c][c][.9]{\uput{0.1mm}[180]{0}{$5$}}
\psfrag{y_10}[c][c][.9]{\uput{0.1mm}[180]{0}{$10$}}
\psfrag{y_15}[c][c][.9]{\uput{0.1mm}[180]{0}{$15$}}
\psfrag{y_20}[c][c][.9]{\uput{0.1mm}[180]{0}{$20$}}
\psfrag{y_25}[c][c][.9]{\uput{0.1mm}[180]{0}{$25$}}
\psfrag{y_30}[c][c][.9]{\uput{0.1mm}[180]{0}{$30$}}
\psfrag{y_35}[c][c][.9]{\uput{0.1mm}[180]{0}{$35$}}
\psfrag{y_40}[c][c][.9]{\uput{0.1mm}[180]{0}{$40$}}
\psfrag{y_45}[c][c][.9]{\uput{0.1mm}[180]{0}{$45$}}
\psfrag{y_50}[c][c][.9]{\uput{0.1mm}[180]{0}{$50$}}
\psfrag{y_55}[c][c][.9]{\uput{0.1mm}[180]{0}{$55$}}
\psfrag{y_60}[c][c][.9]{\uput{0.1mm}[180]{0}{$60$}}
\psfrag{variance}[c][c][.9]{\uput{3mm}[90]{0}{variance/Barankin bound}}
\psfrag{bML}[l][l][0.8]{bound on $v((\cdot);\mathbf{x}_0)$}
\psfrag{ML}[l][l][0.8]{$v((\cdot);\mathbf{x}_0)$}
\psfrag{data1}[l][l][0.8]{$v(\hat{\mathbf{x}}_{\text{LS}}(\cdot); \mathbf{x}_0) = v(\hat{\mathbf{x}}_{\text{HT}}(\cdot); \mathbf{x}_0)$, $T \!\rmv=\! 0$}
\psfrag{data2}[l][l][0.8]{$\minachievvar_{\text{HT}}(\mathbf{x}_{0})$, $T \!\rmv=\! 0$}
\psfrag{data3}[l][l][0.8]{$v(\hat{\mathbf{x}}_{\text{HT}}(\cdot); \mathbf{x}_0)$, $T \!\rmv=\! 2$}
\psfrag{data5}[l][l][0.8]{$v(\hat{\mathbf{x}}_{\text{HT}}(\cdot); \mathbf{x}_0)$, $T \!\rmv=\! 3$}
\psfrag{data7}[l][l][0.8]{$v(\hat{\mathbf{x}}_{\text{HT}}(\cdot); \mathbf{x}_0)$, $T \!\rmv=\! 4$}
\psfrag{data4}[l][l][0.8]{$\minachievvar_{\text{HT}}(\mathbf{x}_{0})$, $T \!\rmv=\! 2$}
\psfrag{data6}[l][l][0.8]{$\minachievvar_{\text{HT}}(\mathbf{x}_{0})$, $T \!\rmv=\! 3$}
\psfrag{data8}[l][l][0.8]{$\minachievvar_{\text{HT}}(\mathbf{x}_{0})$, $T \!\rmv=\! 4$}
\psfrag{data9}[l][l][0.8]{$v(\hat{\mathbf{x}}_{\text{ML}}(\cdot); \mathbf{x}_0)$}
\psfrag{MLarrow}[l][l][0.8]{ML}
\psfrag{T=3}[l][l][0.8]{\uput{0mm}[-10]{0}{\hspace{1mm}$T \!\rmv=\! 4$}}
\psfrag{T=4}[l][l][0.8]{$\!\!\!\!T \!\rmv=\! 3$}
\psfrag{T=2}[l][l][0.8]{\uput{.5mm}[0]{0}{\vspace*{3mm} \!\!\!\!\!\!\!$T \!\rmv=\! 2$}}
\psfrag{T=0}[l][l][0.8]{\uput{1mm}[270]{0}{\hspace{0mm}$T \!\rmv=\! 0\,\ist$(LS)}}
\psfrag{oracle}[l][l][0.8]{\uput{0mm}[0]{0}{$S \sigma^{2}$}}
\centering
\includegraphics[height=9cm,width=17cm]{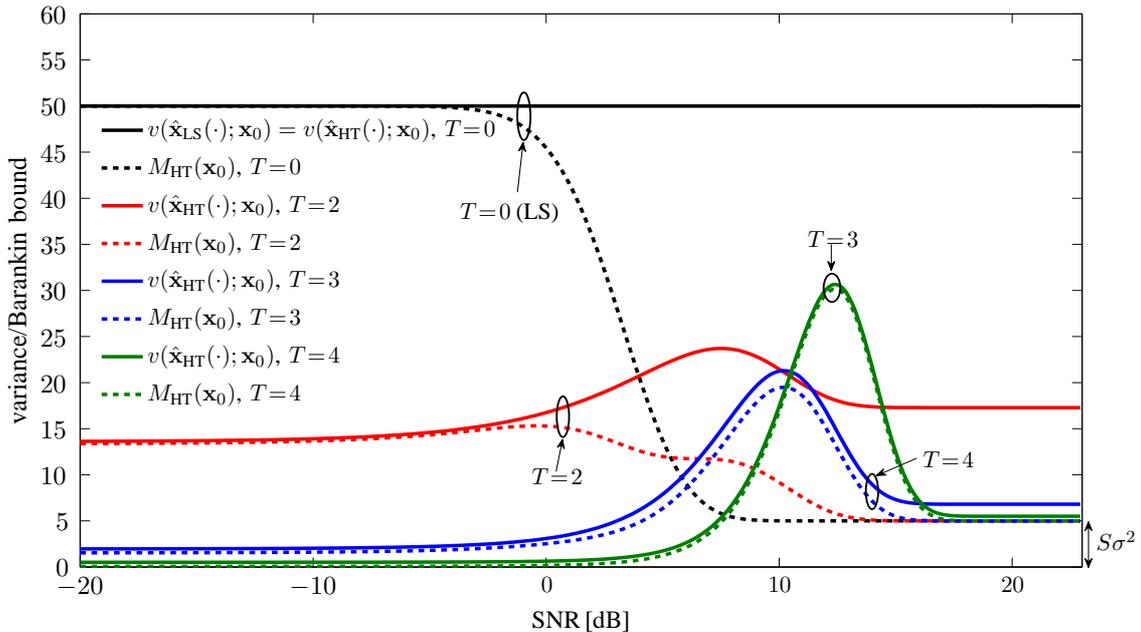}
\vspace*{1mm}
\renewcommand{\baselinestretch}{1.2}\small\normalsize
  \caption{Variance of the HT estimator, $v(\hat{\mathbf{x}}_{\text{HT}}(\cdot); \mathbf{x}_{0})$, for different $T$ (solid lines) 
  and corresponding minimum achievable variance (Barankin bound) $\minachievvar_{\text{HT}}(\mathbf{x}_{0})$ (dashed lines)
  versus SNR, for the SSNM with $N\!=\!50$, $S\!=\!5$, and $\sigma^2\!=\!1$.} 
\label{fig_barankin_bound_HT_ISIT}
\vspace*{1.5mm}
\end{figure}

On the other hand, for the special case of diagonal estimators,
such as the HT estimator, Theorem \ref{thm_diag_bias_min_achiev_var_LMV} and 
Corollary \ref{cor_diag_est_min_achiev_var_LMV_ISIT} make positive statements about the existence of estimators that 
have \emph{locally} a smaller variance than the HT estimator. In particular, we can use Corollary \ref{cor_diag_est_min_achiev_var_LMV_ISIT} to obtain the LMV 
estimator and corresponding minimum achievable variance at a parameter vector $\mathbf{x}_{0} \in \mathcal{X}_{S}$ for the given bias function of the HT estimator,
$\mathbf{c}_{\text{HT}}(\cdot)$. 
In Fig.~\ref{fig_barankin_bound_HT_ISIT}, we plot the variance $v(\hat{\mathbf{x}}_{\text{HT}}(\cdot); \mathbf{x}_{0})$ 
for four different choices of $T$ versus SNR. We also plot the
corresponding minimum achievable variance (Barankin bound) 
$\minachievvar_{\text{HT}}(\mathbf{x}_{0}) \triangleq \sum_{k \in [N]} \minachievvar_{\text{SSNM}}(c_{\text{HT},k}(\cdot),\mathbf{x}_{0})$.
Here, $\minachievvar_{\text{SSNM}}(c_{\text{HT},k}(\cdot),\mathbf{x}_{0})$
was obtained from \eqref{equ_expr_min_achiev_var_diag_est_ISIT_SSNM} in Corollary \ref{cor_diag_est_min_achiev_var_LMV_ISIT}.
(Note that \eqref{equ_expr_min_achiev_var_diag_est_ISIT_SSNM} is applicable because the estimator $\hat{x}_{\text{HT},k}(\mathbf{y})$ is diagonal and has 
finite variance at all $\mathbf{x}_{0} \rmv\in\rmv \mathcal{X}_{S}$.)
It is seen that for small $T$ (including $T\!=\!0$, where the HT estimator reduces to the LS estimator) and for SNR above $0\,$dB, 
$v(\hat{\mathbf{x}}_{\text{HT}}(\cdot); \mathbf{x}_{0})$ is significantly higher than 
$\minachievvar_{\text{HT}}(\mathbf{x}_{0})$.
However, as $T$ increases, the gap between the $v(\hat{\mathbf{x}}_{\text{HT}}(\cdot); \mathbf{x}_{0})$ and $\minachievvar_{\text{HT}}(\mathbf{x}_{0})$ curves becomes smaller;
in particular, the two curves are almost indistinguishable already for $T\!=\rmv 4$. 
For high SNR, 
$\minachievvar_{\text{HT}}(\mathbf{x}_{0})$ approaches the oracle variance $S \sigma^{2} \rmv=\rmv 4$ for any value of $T$.


\section{Conclusion}

We used RKHS theory to
analyze the MVE problem within the sparse linear Gaussian model (SLGM).
In the SLGM, the unknown parameter vector to be estimated is assumed to be sparse with a known sparsity degree, and the 
observed vector is a linearly transformed 
version of the parameter vector that is corrupted by i.i.d.\ Gaussian noise with a known variance.
The RKHS framework allowed us to establish a geometric interpretation of existing lower bounds on the estimator variance
and to derive novel lower bounds on the estimator variance, in both cases under a bias constraint. These bounds were obtained by an 
orthogonal projection of the prescribed mean function onto a subspace of the RKHS 
associated with the SLGM. Viewed as functions of the SNR,
the bounds 
were observed to vary between two extreme regimes. On the one hand, there is a low-SNR regime where the entries of the true parameter vector are small compared 
with the noise variance. Here, our bounds predict that if the estimator bias is approximately zero, the \emph{a priori} sparsity information 
does not help much in the estimation;
however, if the bias is allowed to be nonzero, the estimator variance can be reduced by the sparsity information. 
On the other hand, there is a high-SNR regime where the nonzero entries of the true parameter vector are 
large compared with the noise variance. 
Here, our bounds coincide with the \CRBfull
of an associated conventional linear Gaussian model 
in which
the support of the unknown parameter vector is supposed known. Our bounds exhibit a 
steep transition between these two regimes. In general, this transition has an exponential decay. 

For the special case of the SLGM that corresponds to the recovery problem in a linear compressed sensing 
scheme,
we expressed our lower bounds in terms of the restricted isometry and coherence parameters
of the measurement matrix. 
Furthermore, for the special case of the SLGM given by the sparse signal in noise model (SSNM), we 
derived closed-form expressions of the minimum achievable variance and the corresponding LMV estimator. 
These latter results include closed-form expressions of the (unbiased) Barankin bound and of the LMVU estimator for the SSNM. 
Simplified expressions of the minimum achievable variance and the LMV estimator were presented for the subclass of ``diagonal'' bias functions.

An analysis of the 
effects of exact and approximate sparsity information
from the MVE perspective showed that the minimum achievable variance under an 
exact sparsity constraint is not a limiting case of the minimum achievable variance under an approximate sparsity constraint.

Finally, a comparison of our bounds with the actual variance of established estimators for the SLGM and SSNM 
(maximum likelihood estimator, hard thresholding estimator, least squares estimator, and orthogonal matching pursuit) showed that 
there might exist estimators with the same bias but a smaller variance.

An interesting direction for future investigations is the search for (classes of) estimators that asymptotically approach our lower variance bounds 
when the estimation is based on an increasing number of
i.i.d.\ observation vectors $\mathbf{y}_i$.
In the unbiased case, the maximum likelihood estimator can be intuitively expected to achieve the variance bounds asymptotically. 
However, a rigorous proof of this conjecture seems to be nontrivial.
Indeed, most 
studies of the asymptotic behavior of maximum likelihood estimators assume that the parameter set is an open subset of $\mathbb{R}^{N}$ 
\cite{LC,IbragimovBook,AsympVanderVaartBook}, which is not the case for the parameter set $\mathcal{X}_{S}$. 
For the popular class of 
M-estimators or penalized maximum likelihood estimators, 
a characterization of the asymptotic behavior is available \cite{AsympVanderVaartBook,HuberRobustBook,RethinkingBiasedEldar}. 
Under mild conditions, 
M-estimators allow an efficient implementation via convex optimization techniques. 

Furthermore, it would be interesting to generalize our results 
to the case of block or group sparsity \cite{BlockSparsityEldarTSP,MishaliEldar2008,EldarRauhut2010}. This could be useful, e.g., 
for
sparse channel estimation in the case of clustered scatterers and delay-Doppler leakage \cite{EiwenSPAWC2010} 
and for the estimation of structured sparse spectra (extending sparsity-exploiting spectral estimation as proposed in \cite{jung-specesttit,CompWidBandSensThan,CompWidebandSensPolo,Tian2012}).

\bibliographystyle{IEEEtran}
\bibliography{/Users/ajung/Arbeit/LitAJ_ITC.bib,/Users/ajung/Arbeit/tf-zentral}
\end{document}